\documentclass[usegraphicx,usenatbib]{mn2e}
\usepackage{amsmath}
\usepackage{amssymb}
\usepackage{mathptmx}
\usepackage{subfigure}
\usepackage{epsfig}
\usepackage{bm}
\usepackage[usenames,dvipsnames]{color}
\definecolor{red}{rgb}{1,0,0}

\renewcommand{\descriptionlabel}[1]%
  {\hspace{\labelsep}\textbf{#1}}

\title[DIA: Methods for automatic kernel design]
      {Difference image analysis: Automatic kernel design using information criteria}

\author[D.M. Bramich et al.]
  {D. M. Bramich$^{1}$,    %\thanks{E-mail: dan.bramich@hotmail.co.uk},
   Keith Horne$^{2}$,
   K. A. Alsubai$^{1}$,
   E. Bachelet$^{1}$,
   D. Mislis$^{1}$
   \newauthor
   and N. Parley$^{1}$
  \medskip
  \\$^{1}$Qatar Environment and Energy Research Institute (QEERI), HBKU, Qatar Foundation, Doha, Qatar
  \\$^{2}$SUPA Physics \& Astronomy, North Haugh, St Andrews, KY16 9SS, Scotland, UK
  }

\begin{document}

\date{Accepted 2010 August ???. Received 2010 August ???; Submitted 2010 August ???}

\pagerange{\pageref{firstpage}--\pageref{lastpage}} \pubyear{2010}

\maketitle

\label{firstpage}

\begin{abstract} 

We present a selection of methods for automatically constructing an optimal kernel model for difference image analysis which require
very few external parameters to control the kernel design. Each method consists of two components; namely, a kernel design
algorithm to generate a set of candidate kernel models, and a model selection criterion to select the simplest kernel model
from the candidate models that provides a sufficiently good fit to the target image. We restricted our attention to
the case of solving for a spatially-invariant convolution kernel composed of delta basis functions, and we considered 19 different kernel
solution methods including six employing kernel regularisation. We tested these kernel solution methods
by performing a comprehensive set of image simulations and investigating how their performance in terms
of model error, fit quality, and photometric accuracy depends on the properties of the reference and target images.
We find that the irregular kernel design algorithm employing unregularised delta basis
functions, combined with either the Akaike or Takeuchi information criterion, is the best kernel solution method in terms of photometric accuracy.
Our results are validated by tests performed on two independent sets of real data. Finally, we provide some important
recommendations for software implementations of difference image analysis.

\end{abstract} 


\section{Introduction}
\label{sec:introduction}

In astronomy, the technique of difference image analysis (DIA) aims to measure changes, from one image to another, in the objects
(e.g. stars, galaxies, etc.) observed in a particular field. Typically these changes consist of variations in flux and/or position.
However, the variations in the object properties that we are interested in are entangled with the differences in the sky-to-detector
(or scene-to-image) transformation between pairs of images. Therefore, the DIA method must carefully model the changes in astrometry, throughput, background,
and blurring between an image pair in order to extract the required astronomical information.

The state of the art in DIA has evolved substantially over the last decade and a half. Possibly the most complicated part of DIA is the optimal modelling
of the convolution kernel describing the changes in point-spread function (PSF) between images. The seminal paper by \citet{ala1998}
set the current framework for doing this by detailing the expansion of the kernel as a linear combination of basis functions. \citet{ala2000} subsequently
showed how to model a spatially varying convolution kernel by modelling the coefficients of the kernel basis functions as polynomials of the image
coordinates. The most important ingredient then in constructing a kernel solution in the Alard DIA framework is the definition of the set of kernel
basis functions. The main developments in this area were achieved by \citet{ala1998}, who defined the Gaussian basis functions, \citet{bra2008} and \citet{mil2008}
who introduced the delta basis functions (DBFs), and \citet{bec2012} (hereafter Be12) who conceived of the regularised DBFs. A detailed
discussion of the kernel basis functions presented in the DIA literature may be found in \citet{bra2013} (hereafter Br13).

The traditional Gaussian basis functions require the specification of numerous parameters while demanding precise sub-pixel image registration for
optimal results, as do many other sets of kernel basis functions (e.g. the network of bicubic B-spline functions introduced by \citealt{yua2008}).
Consequently, the optimal choice of parameters for generating such sets of basis functions is not obvious, although some investigation into this
issue has been performed (\citealt{isr2007}). In contrast, the DBFs have the ultimate flexibility to represent a discrete kernel
of any form while requiring the absolute minimal user specification; namely the kernel size and shape (or equivalently the set of ``active'' kernel
pixels). They may even be used to model fractional pixel offsets between images, avoiding the need for image resampling in the absence
of other image misalignments (rotation, scale, shear and distortion). Unsurprisingly then, DIA photometry for kernels employing DBFs has been shown to
be better than that produced for kernels using Gaussian basis functions (\citealt{alb2009}).
However, the use of DBFs yields somewhat noisier kernel solutions than is desirable due to the relatively large number
of parameters in the kernel model. To tackle this weakness of the DBFs, Be12 developed the regularised DBFs
through the elegant application of Tikhonov regularisation to the kernel model. This refined approach produces very clean and low-noise
kernel solutions at the expense of introducing an extra parameter $\lambda$ into the kernel definition, where the value of $\lambda$ controls the strength
of the regularisation. Be12 recommend values of $\lambda$ between 0.1 and 1 for square kernels of size 19$\times$19 pixels although they caution that
the optimal value will likely be data set dependent.

The next logical step in the development of DIA is to investigate how the properties of the image pair under consideration
influence the composition of the optimal kernel model (i.e. the optimal set of DBFs, the optimal values of their coefficents,
and the optimal value of $\lambda$). In this context, ``optimality'' refers both to the Principle of Parsimony, in that the optimal kernel model
should constitute the simplest configuration of DBFs that provides a sufficiently good fit to the data, and to appropriate/relevant model
performance measure(s). The proposed investigation may be
accomplished both by generating and analysing a comprehensive set of simulated images, and by testing on a wide variety
of real image data. Neither of these tasks have yet been attempted. 

Various model selection criteria have been developed from different statistical view-points as implementations of the Principle of Parsimony
(e.g. the Aikaike information criterion - \citealt{aka1974}, the Bayesian information criterion - \citealt{sch1978}, etc.) and each one may
be used to automatically select a parsimonious model from a set of models\footnote{We note that the application of a model selection criterion
to model fitting may also be viewed as a regularisation technique.}.
Due to the sheer number of possible combinations of DBFs that may constitute the kernel model, the set of models that can be considered
will be limited to a set of feasible candidate kernel models defined via the adoption of an appropriate kernel design algorithm.
The performance of each model selection criterion may then be assessed by measuring the quality of the corresponding kernel solution with
respect to one or more desired metric(s). The final result will then be a recommendation,
dependent on the properties of the image pair under consideration, as to which model selection criterion should be adopted to consistently
yield the best kernel solutions for the specified kernel design algorithm.

In this paper, we report on the results of having carried out the proposed investigation for both the unregularised and
regularised DBFs (Section~\ref{sec:methods_dia}) using simulated images (Section~\ref{sec:simsec}) and real data (Section~\ref{sec:realsec}).
We restrict attention to the case of solving for a spatially-invariant convolution kernel.
The performance of three proposed kernel design algorithms (Section~\ref{sec:ker_design}) coupled with
up to eight model selection criteria (Section~\ref{sec:model_selection_criteria})
was assessed with regards to model error (simulations only), fit quality, and photometric accuracy. In total 19 methods were tested.
The conclusions and recommendations from our investigation are detailed in Section~\ref{sec:conclusions}.

\section{Modelling The Convolution Kernel}
\label{sec:methods_dia}

In this section, we briefly describe the methods used in this paper to solve for the spatially-invariant convolution
kernel matching the PSF between two images of the same field.

\subsection{Solving For A Spatially-Invariant Kernel: Recap}
\label{sec:solving_for_kernel}

Consider a pair of registered images of the same field with the same dimensions and sampled on the same pixel grid.
To avoid invalidating the assumption of a spatially-invariant kernel model, the image registration should be such that at most
there is a translational offset of a few pixels between the images, with no rotational (or other) image misalignments. Let the
images be referred to as the reference image $R$ and the target image $I$ with pixel values $R_{ij}$ and $I_{ij}$, respectively,
where $i$ and $j$ are pixel indices referring to the column $i$ and row $j$ of an image.

We model the target image $I$ as a model image $M$ formed by the convolution of the reference image $R$ with a spatially-invariant
discrete convolution kernel $K$ plus a spatially-invariant (constant\footnote{All of the results in this paper are easily
generalised to the case of a differential background that is a polynomial function of the image coordinates (e.g. see Br13).}) differential background $B$:
\begin{equation}
M_{ij} = [R \otimes K]_{ij} + B
\label{eqn:model1}
\end{equation}
where the $M_{ij}$ are the pixel values of the model image. As in \citet{ala1998}, we model $K$ as a linear
combination of basis functions:
\begin{equation}
K_{rs} = \sum_{q = 1}^{N_{\kappa}} a_{q} \, \kappa_{qrs}
\label{eqn:kernel}
\end{equation}
where the $K_{rs}$ are the kernel pixel values, $r$ and $s$ are pixel indices corresponding to the column $r$ and
row $s$ of the discrete kernel, $N_{\kappa}$ is the number of kernel basis functions, and the $\kappa_{qrs}$ are the pixel
values of the $q$th discrete kernel basis function $\kappa_{q}$ with corresponding coefficient $a_{q}$. Substitution of
equation~(\ref{eqn:kernel}) into equation~(\ref{eqn:model1}) yields:
\begin{equation}
M_{ij} = \sum_{q = 1}^{N_{\kappa}} a_{q} \, [R \otimes \kappa_{q}]_{ij} + B
\label{eqn:model2}
\end{equation}
with:
\begin{equation}
[R \otimes \kappa_{q}]_{ij} = \sum_{rs} R_{(i+r)(j+s)} \kappa_{qrs}
\label{eqn:basis_image}
\end{equation}
The image $[R \otimes \kappa_{q}]_{ij}$ is referred to as a {\it basis image}. The model image $M$ has
$N_{\mbox{\scriptsize par}} = N_{\kappa} + 1$ parameters. Note that equation~(\ref{eqn:model2})
may be derived as a special case of equation~(8) from Br13.

Assuming that the target image pixel values $I_{ij}$ are independent observations drawn from normal (or Gaussian) distributions $\mathcal{N}(M_{ij},\sigma_{ij})$
and that the parameters $a_{q}$ and $B$ of the model image have uniform Bayesian prior probability distribution functions (PDFs), then the
maximum likelihood estimator (MLE) of $a_{q}$ and $B$ may be found by minimising the chi-squared:
\begin{equation}
\chi^{2} = \sum_{ij} \left( \frac{I_{ij} - M_{ij}}{\sigma_{ij}} \right)^{2}
\label{eqn:chi_sqr}
\end{equation}
This is a general linear least-squares problem (see \citealt{pre2007}) with associated {\it normal equations} in matrix form:
\begin{equation}
\mathbf{H} \boldsymbol{\alpha} = \boldsymbol{\beta}
\label{eqn:matrix_normal_eqns}
\end{equation}
where the symmetric and positive-definite $(N_{\kappa}+1) \times (N_{\kappa}+1)$ matrix $\mathbf{H}$ is the least-squares matrix, the vector $\boldsymbol{\alpha}$ is the vector of
$N_{\kappa}+1$ model parameters, and $\boldsymbol{\beta}$ is another vector. For the vector of parameters:
\begin{equation}
\alpha_{q} = 
\begin{cases}
a_{q} & \mbox{for $1 \le q \le N_{\kappa}$} \\
B     & \mbox{for $q = N_{\kappa} + 1$}     \\
\end{cases}
\label{eqn:param_vector}
\end{equation}
the elements of $\mathbf{H}$ and $\boldsymbol{\beta}$ are given in terms of the basis images by:
\begin{equation}
H_{q q^{\,\prime}} = \sum_{ij} \, \frac{\psi_{qij} \, \psi_{q^{\,\prime} ij}}{\sigma_{ij}^{2}}
\label{eqn:least_squares_matrix}
\end{equation}
\begin{equation}
\beta_{q} = \sum_{ij} \, \frac{\psi_{qij} \, I_{ij}}{\sigma_{ij}^{2}}
\label{eqn:least_squares_vector}
\end{equation}
\begin{equation}
\psi_{qij} = \frac{\partial M_{ij}}{\partial \alpha_{q}} =
\begin{cases}
[R \otimes \kappa_{q}]_{ij} & \mbox{for $1 \le q \le N_{\kappa}$} \\
1                           & \mbox{for $q = N_{\kappa} + 1$}     \\
\end{cases}
\label{eqn:psi_pattern}
\end{equation}

Cholesky factorization of $\mathbf{H}$, followed by forward and back substitution is the most efficient and numerically stable method
(\citealt{gol1996}) for obtaining the solution $\boldsymbol{\alpha} = \boldsymbol{\widehat{\alpha}}$ to the normal equations
(i.e. $\boldsymbol{\widehat{\alpha}}$ is the vector of MLEs of the model parameters). The
inverse matrix $\mathbf{H}^{-1}$ is the covariance matrix of the parameter estimates 
$\operatorname{cov} \, (\widehat{\alpha}_{q},\widehat{\alpha}_{q^{\,\prime}}) = [ \mathbf{H}^{-1} ]_{q q^{\,\prime}}$
and consequently the uncertainty $\sigma_{q}$ in each $\widehat{\alpha}_{q}$ is given by:
\begin{equation}
\sigma_{q} = \sqrt{ [ \mathbf{H}^{-1}  ]_{q q} }
\label{eqn:sol_uncertainties}
\end{equation}

For the spatially-invariant kernel, the {\it photometric scale factor} $P$ between the reference and target image is a constant:
\begin{equation}
P = \sum_{rs} K_{rs}
\label{eqn:phot_scale}
\end{equation}
As noted by \citet{bra2008}, it is good practice to subtract an estimate of the sky background level from $R$ before solving for $K$
and $B$ in order to minimise any correlation between $P$ and $B$.

We adopt a noise model for the model-image pixel uncertainties $\sigma_{ij}$ of:
\begin{equation}
\sigma_{ij}^{2} = \frac{\sigma_{0}^{2}}{F_{ij}^{2}} + \frac{M_{ij}}{G \, F_{ij}}
\label{eqn:noise_model}
\end{equation}
where $\sigma_{0}$ is the CCD readout noise (ADU), $G$ is the CCD gain (e$^{-}$/ADU), and $F_{ij}$ is the
flat-field image. The $\sigma_{ij}$ depend on the $M_{ij}$ which renders our maximum likelihood problem as a non-linear problem and
also requires that the MLE of the model image parameters is obtained by minimising
$\chi^{2} + \sum_{ij} \ln ( \sigma_{ij}^{2} )$ instead of $\chi^{2}$. However, iterating the solution by considering the
$\sigma_{ij}$ and $M_{ij}$ in turn as fixed is an appropriate linearisation of the problem
that still allows for the model image parameters to be determined by minimising $\chi^{2}$ at each iteration as described above
(since the $\sigma_{ij}$ are considered as constant whenever the model image parameters are being estimated). For the first iteration, we estimate the $\sigma_{ij}$
by approximating $M_{ij}$ in equation~(\ref{eqn:noise_model}) with $I_{ij}$.
A $k$-sigma-clip algorithm is employed at the end of each iteration except for the first to prevent outlier target-image pixel values from 
influencing the solution (e.g. cosmic rays, variable stars, etc.). The criterion for pixel rejection is
$\left| \varepsilon_{ij} \right| = \left| (I_{ij} - M_{ij}) / \sigma_{ij} \right| \ge k$, and we use $k = 4$. Only 3-4 iterations are required for convergence and
the final solution is highly insensitive to the initial choice of $\sigma_{ij}$ (e.g. setting all of the $\sigma_{ij}$ to unity
for the first iteration gives exactly the same result as setting the $\sigma_{ij}$ by approximating $M_{ij}$ in equation~(\ref{eqn:noise_model}) with $I_{ij}$).
Finally, it should be noted that lack of iteration introduces a bias into the kernel and
differential background solution (see Br13 for a discussion and examples).

The difference image $D$ is defined by:
\begin{equation}
D_{ij} = I_{ij} - M_{ij}
\label{eqn:diff_image}  
\end{equation}
from which we may define a normalised difference image:
\begin{equation}
\varepsilon_{ij} = D_{ij} / \sigma_{ij}
\label{eqn:norm_diff_image}
\end{equation}
In the absence of varying objects, and for a reliable noise model, the distribution of the $\varepsilon_{ij}$ values provides an indication of
the quality of the difference image; namely, the $\varepsilon_{ij}$ should follow a Gaussian distribution with zero mean
and unit standard deviation. If the $\varepsilon_{ij}$ follow a Gaussian distribution with significant bias or standard deviation greater than unity,
then systematic errors are indicated, which may be due to under-fitting. If they follow a Gaussian distribution with standard deviation less than unity,
then over-fitting may be indicated. If they follow a non-Gaussian distribution, then an inappropriate noise
model may be at least part of the cause.

\subsection{The Delta Basis Functions}
\label{sec:delta_basis_functions}

The final ingredient required to construct a kernel solution is the definition of the set of kernel basis functions,
which in turn defines the set of basis images. In this paper we consider only the {\it delta basis functions}, which are defined by:
\begin{equation}
\kappa_{qrs} = \delta_{r\mu} \, \delta_{s\nu}
\label{eqn:delta_basis}
\end{equation}
where a one-to-one correspondence $q \leftrightarrow (\mu,\nu)$ associates the $q$th kernel basis function $\kappa_{q}$ with the discrete
kernel pixel coordinates $(\mu,\nu)$, and $\delta_{ij}$ is the Kronecker delta-function:
\begin{equation}
\delta_{ij} =   
\begin{cases}
1 & \mbox{if $i=j$} \\
0 & \mbox{if $i \ne j$} \\
\end{cases}
\label{eqn:kronecker_delta}
\end{equation}
As such, each DBF $\kappa_{q}$ and its corresponding coefficient $a_{q}$ represent a single kernel pixel and its value, respectively.
Note that this definition of the DBFs ignores the transformation that is required when the photometric scale factor is
spatially varying (Br13). 

The DBFs have a conveniently simple expression for the corresponding basis images:
\begin{equation}
[R \otimes \kappa_{q}]_{ij} = R_{(i+\mu)(j+\nu)}
\label{eqn:delta_basis_images}
\end{equation}

\subsection{Regularising The Delta Basis Functions}
\label{sec:reg_delta_basis}

For the DBFs, Be12 introduced a refinement to the normal equations to control the trade-off between noise and resolution in the
kernel solution. They used {\it Tikhonov regularisation} (see \citealt{pre2007}) to penalise kernel solutions that are
too noisy by adding a penalty term to the chi-squared that is derived from the second
derivative of the kernel surface and whose strength is parameterised by a tuning parameter $\lambda$. The addition of a penalty term
to the chi-squared is equivalent to adopting a non-uniform Bayesian prior PDF on the model parameters.
The corresponding maximum penalised likelihood estimator (MPLE) of $a_{q}$ and $B$ is obtained by minimising:
\begin{equation}
\chi^{2} + \lambda N_{\mbox{\scriptsize dat}} \, \boldsymbol{\alpha}^{\mbox{\scriptsize T}} \, \mathbf{L}^{\mbox{\scriptsize T}} \, \mathbf{L} \, \mathbf{\boldsymbol{\alpha}} =
\sum_{ij} \left( \frac{I_{ij} - M_{ij}}{\sigma_{ij}} \right)^{2} + \lambda N_{\mbox{\scriptsize dat}} \sum_{q = 1}^{N_{\kappa}} \, \sum_{u = 1}^{N_{\kappa}} \, \sum_{v = 1}^{N_{\kappa}} a_{q} \, L_{u q} \, L_{u v} \, a_{v}
\label{eqn:pen_chi_sqr}
\end{equation}
where $N_{\mbox{\scriptsize dat}}$ is the number of data values\footnote{Be12 accidentally omitted $N_{\mbox{\scriptsize dat}}$ from their equation~(12).} (i.e. target-image pixels)
and $\mathbf{L}$ is an $(N_{\kappa} + 1) \times (N_{\kappa} + 1)$ matrix with elements:
\begin{equation}
L_{u v} =
\begin{cases}
N_{\mbox{\scriptsize adj},u}   & \mbox{for $v = u \le N_{\kappa}$, and where $N_{\mbox{\scriptsize adj},u}$ is the number of} \\
                               & \;\;\; \mbox{DBFs adjacent to the DBF corresponding to $u$,}                                 \\
-1                             & \mbox{for $u \le N_{\kappa}$, $v \le N_{\kappa}$, $v \ne u$, and $u$ and $v$}                \\
                               & \;\;\; \mbox{corresponding to adjacent DBFs,}                                                \\
0                              & \mbox{otherwise.}                                                                            \\
\end{cases}
\label{eqn:L_elements}
\end{equation}
We consider two DBFs to be {\it adjacent} if they share a common kernel-pixel edge, {\it connected}
if they can be linked via any number of pairs of adjacent DBFs, and {\it disconnected} if they are not connected.
Note that the elements of the last row and column of $\mathbf{L}$, corresponding to the differential background parameter $B$, are all zero. 

The matrix $\mathbf{L}$ is the {\it Laplacian matrix} representing the connectivity graph of the set of DBFs (cf. graph theory). It is symmetric,
diagonally dominant, and positive-semidefinite. All of the eigenvalues of $\mathbf{L}$ are non-negative while $N_{\mbox{\scriptsize grp}} + 1$ of
them are equal to zero. Here, $N_{\mbox{\scriptsize grp}}$ is the number of disconnected sets of connected DBFs within the
full set of DBFs (i.e. the number of components of the connectivity graph). Consequently, the rank of $\mathbf{L}$ is $N_{\kappa} - N_{\mbox{\scriptsize grp}}$,
as is the rank of $\mathbf{L}^{\mbox{\scriptsize T}} \, \mathbf{L} = \mathbf{L} \mathbf{L}$, which are facts that we will use later in Section~\ref{sec:model_selection_criteria_dia}.
It is also useful to note that if all of the DBFs are connected to each other, then $\mathbf{L}$ and $\mathbf{L} \mathbf{L}$ are both of rank $N_{\kappa} - 1$.
In Appendix~A, we present a couple of example kernels with their corresponding $\mathbf{L}$ matrices.

The expression in equation~(\ref{eqn:pen_chi_sqr}) is at a minimum when its gradient with respect to each of
the parameters $a_{q}$ and $B$ is equal to zero. Performing the $N_{\kappa} + 1$ differentiations and rewriting the
set of linear equations in matrix form we obtain the {\it regularised normal equations}:
\begin{equation}
\mathbf{H_{\mbox{\scriptsize P}}} \, \boldsymbol{\alpha} = \boldsymbol{\beta}
\label{eqn:matrix_reg_normal_eqns}
\end{equation}
where:
\begin{equation}
\mathbf{H_{\mbox{\scriptsize P}}} = \mathbf{H} \, + \, \lambda N_{\mbox{\scriptsize dat}} \, \mathbf{L} \mathbf{L}
\label{eqn:reg_least_squares_matrix}
\end{equation}

Obtaining the solution to the regularised normal equations now proceeds as for the normal equations in Section~\ref{sec:solving_for_kernel}.
The covariance matrix of the parameter estimates $\boldsymbol{\alpha} = \boldsymbol{\widehat{\alpha}}_{\mbox{\scriptsize P}}$ is similarly given by
$\operatorname{cov} \, (\widehat{\alpha}_{\mbox{\scriptsize P}, q},\widehat{\alpha}_{\mbox{\scriptsize P}, q^{\,\prime}}) = [ \mathbf{H_{\mbox{\scriptsize P}}}^{-1} ]_{q q^{\,\prime}}$.

\section{Model Selection Criteria}
\label{sec:model_selection_criteria}

Here we describe our statistical tool-kit of model selection criteria that we will use for deciding on the
best set of DBFs to be employed in the modelling of the convolution kernel. The criteria are valid for
linear models, such as our model image $M$ in equation~(\ref{eqn:model2}), and for data drawn
from independent Gaussian distributions, which is a valid approximation to the Poissonian
statistics of photon detection for CCD image data $I$ that only breaks down at very low signal
levels ($\la$16~e$^{-}$). We direct the reader to \citet{kon2008} for an essential reference on
the information criteria presented below.

\subsection{Hypothesis Testing For Nested Models}
\label{sec:model_selection_criteria_nested}

\subsubsection{$\Delta \chi^{2}$-test}
\label{sec:dchi2}

The $\Delta \chi^{2}$-test may be used to compare two models A and B with parameter sets $P_{\mbox{\scriptsize A}}$
and $P_{\mbox{\scriptsize B}}$, respectively, that are nested (i.e. $P_{\mbox{\scriptsize A}} \subset P_{\mbox{\scriptsize B}}$).
The $\Delta \chi^{2}$-statistic is defined by:
\begin{equation}
\Delta \chi^{2} = \chi^{2}_{\mbox{\scriptsize A}} - \chi^{2}_{\mbox{\scriptsize B}}
\label{eqn:delta_chisq}
\end{equation}
where $\chi^{2}_{\mbox{\scriptsize A}}$ and $\chi^{2}_{\mbox{\scriptsize B}}$ are the chi-squared values of models A and B, respectively
(see equation~(\ref{eqn:chi_sqr})). Under the null hypothesis that model B does not provide a significantly better fit than model A,
the $\Delta \chi^{2}$-statistic follows a chi-squared distribution with $N_{\mbox{\scriptsize par},\mbox{\scriptsize B}} - N_{\mbox{\scriptsize par},\mbox{\scriptsize A}}$
degrees of freedom (DoF). We set our $\Delta \chi^{2}$ threshold for rejection of the null hypothesis at 1 per cent (e.g. $\Delta \chi^{2} \ga 6.63$ for DoF~=~1). We adopt the chi-squared
values of models A and B as those calculated during the first iteration of our kernel solution procedure to enable a fair comparison
between models since they are both computed using the same pixel uncertainties
(i.e. the $\sigma_{ij}$ estimated by approximating $M_{ij}$ in equation~(\ref{eqn:noise_model}) with $I_{ij}$). However, the values of the 
model image parameters are still taken as those calculated in the final iteration of the kernel solution procedure.

Model selection using the $\Delta \chi^{2}$-test applies only to models A, B, ..., Z with sequentially nested parameter sets
$P_{\mbox{\scriptsize A}} \subset P_{\mbox{\scriptsize B}} \subset ... \subset P_{\mbox{\scriptsize Z}}$. Starting with models A and B, the $\chi^{2}$
is minimised for each model and the $\Delta \chi^{2}$-test is used to determine whether or not model B provides a significantly better fit than model A.
If it does not, then model A is accepted as the correct model and the procedure terminates, otherwise the next pair of models B and C are evaluated using
the same method. The procedure continues by evaluating sequential model pairs in this fashion until either the $\Delta \chi^{2}$-test indicates that the
next model does not provide a significantly better fit or until there are no more models to test.

\subsubsection{$F$-test}
\label{sec:f}

The $F$-test may also be used to compare two nested models A and B. The $F$-statistic is defined by:
\begin{equation}
F = \frac{\Delta \chi^{2} / \left( N_{\mbox{\scriptsize par},\mbox{\scriptsize B}} - N_{\mbox{\scriptsize par},\mbox{\scriptsize A}} \right)}
         {\chi^{2}_{\mbox{\scriptsize B}} / \left( N_{\mbox{\scriptsize dat}} - N_{\mbox{\scriptsize par},\mbox{\scriptsize B}} \right)}
\label{eqn:fstat}
\end{equation}
where $N_{\mbox{\scriptsize dat}}$ is the number of data values. Again, under the null hypothesis that model B does not provide a significantly better fit than model A,
$F$ follows an $F$-distribution with
DoF~$=~\left( N_{\mbox{\scriptsize par},\mbox{\scriptsize B}} - N_{\mbox{\scriptsize par},\mbox{\scriptsize A}}, \, N_{\mbox{\scriptsize dat}} - N_{\mbox{\scriptsize par},\mbox{\scriptsize B}} \right)$.
We set our $F$ threshold for rejection of the null hypothesis at 1 per cent (e.g. $F \ga 4.63$ for DoF~=~(2,1000)) and we compute the $F$-statistic using the chi-squared
values of models A and B calculated during the first iteration of our kernel solution procedure. Model selection with the $F$-test applies to models A, B, ..., Z with sequentially
nested parameter sets and proceeds in the same way as model selection with the $\Delta \chi^{2}$-test.

\subsection{Information Criteria For Maximum Likelihood}
\label{sec:model_selection_criteria_ic_ml}

The principal of maximum likelihood assumes a uniform prior PDF on the model parameters. A
consequence of this is that as parameters are added to a model, the maximum likelihood always increases, rendering it
useless for the purpose of model selection between models with different dimensionality. Information criteria are used
as an alternative for evaluating models with different numbers of parameters. They may be applied regardless of whether
the models under consideration are nested or non-nested.

\subsubsection{AIC$_{\mbox{\scriptsize C}}$}
\label{sec:aicc}

The Akaike information criterion (AIC; \citealt{aka1974}) is derived as an asymptotic approximation to the Kullback-Leibler divergence 
(\citealt{kul1951})\footnote{Use of the AIC as a model selection criterion is also equivalent to
assuming a prior PDF on the model parameters that is proportional to $\exp \left( -N_{\mbox{\scriptsize par}} \right)$,
hence favouring models with smaller numbers of parameters.}, which measures the distance
of a candidate model from the true underlying model under the assumption that the true model
is of infinite dimension and is therefore not represented in the set of candidate models. The aim of the AIC is to
evaluate models based on their prediction accuracy.

A version of the AIC for Gaussian linear regression problems that corrects for the small-sample bias while being asymptotically the same as the AIC for $N_{\mbox{\scriptsize dat}} \gg N_{\mbox{\scriptsize par}}$
was derived by \citet{sug1978}:
\begin{equation}
\mbox{AIC}_{\mbox{\scriptsize C}} = -2 \ln \mathcal{L} ( \boldsymbol{\widehat{\theta}} ) \, + \, 2 N_{\mbox{\scriptsize par}} \left( \frac{N_{\mbox{\scriptsize dat}}}{N_{\mbox{\scriptsize dat}} - N_{\mbox{\scriptsize par}} - 1} \right)
\label{eqn:AIC}
\end{equation}
where $\mathcal{L} ( \boldsymbol{\theta} )$ is the likelihood function for the vector of model parameters $\boldsymbol{\theta}$, and $\boldsymbol{\widehat{\theta}}$
is a vector of MLEs for the model parameters. Model selection with the AIC$_{\mbox{\scriptsize C}}$ is performed
by minimising $-2 \ln \mathcal{L} ( \boldsymbol{\theta} )$ for each model, and then minimising AIC$_{\mbox{\scriptsize C}}$ over the full set of models under consideration.

\subsubsection{TIC}
\label{sec:tic}

The Takeuchi information criterion (TIC; \citealt{tak1976}) is a generalisation of the AIC (\citealt{kon2008}) given by:
\begin{equation}
\mbox{TIC} = -2 \ln \mathcal{L} ( \boldsymbol{\widehat{\theta}} ) + 2 \operatorname{tr} \left( \mathbf{I} ( \boldsymbol{\widehat{\theta}} ) \,\, \mathbf{J}^{-1} ( \boldsymbol{\widehat{\theta}} ) \right)
\label{eqn:TIC}
\end{equation}
where tr is the matrix trace operator. The matrices $\mathbf{I}$ and $\mathbf{J}$ are defined as:
\begin{equation}
\mathbf{I} ( \boldsymbol{\theta} ) = \frac{1}{N_{\mbox{\scriptsize dat}}} \sum_{i = 1}^{N_{\mbox{\tiny dat}}} 
                                     \frac{\partial \ln l_{i} ( \boldsymbol{\theta} )}{\partial \boldsymbol{\theta}} \,
                                     \frac{\partial \ln l_{i} ( \boldsymbol{\theta} )}{\partial \boldsymbol{\theta}^{\mbox{\scriptsize T}}}
\label{eqn:I} 
\end{equation}
\begin{equation}
\mathbf{J} ( \boldsymbol{\theta} ) = - \left( \frac{1}{N_{\mbox{\scriptsize dat}}} \right) \frac{\partial^{2} \ln \mathcal{L} ( \boldsymbol{\theta} )}{\partial \boldsymbol{\theta} \, \partial \boldsymbol{\theta}^{\mbox{\scriptsize T}}}
\label{eqn:J} 
\end{equation}
\begin{equation}
\ln \mathcal{L} ( \boldsymbol{\theta} ) = \sum_{i = 1}^{N_{\mbox{\tiny dat}}} \ln l_{i} ( \boldsymbol{\theta} )
\label{eqn:lnlike}
\end{equation}
where $l_{i} ( \boldsymbol{\theta} )$ is the likelihood function for the $i$th (single) data point.
Model selection with the TIC proceeds as for the AIC$_{\mbox{\scriptsize C}}$.

\subsubsection{BIC}
\label{sec:bic} 

The Bayesian approach to model selection is to choose the model with the largest Bayesian posterior probability. By approximating the posterior probability of each model,
\citet{sch1978} derived the Bayesian information criterion (BIC) for model selection:
\begin{equation}
\mbox{BIC} = -2 \ln \mathcal{L} ( \boldsymbol{\widehat{\theta}} ) \, + \, N_{\mbox{\scriptsize par}} \ln N_{\mbox{\scriptsize dat}} \, - \, N_{\mbox{\scriptsize par}} \ln 2 \pi
\label{eqn:BIC}
\end{equation}
The BIC generally includes a heavier penalty than the AIC$_{\mbox{\scriptsize C}}$ for more complicated models
(e.g. in the regime $N_{\mbox{\scriptsize par}} < 20$ and $N_{\mbox{\scriptsize dat}} > 100$), therefore
favouring models with fewer parameters than those favoured by the AIC$_{\mbox{\scriptsize C}}$. Model selection with the BIC 
proceeds as for the AIC$_{\mbox{\scriptsize C}}$.

\subsubsection{BIC$_{\mbox{\scriptsize I}}$}
\label{sec:bici}

\citet{kon2004} performed a deeper Bayesian analysis to derive an improved BIC:
\begin{equation}
\mbox{BIC}_{\mbox{\scriptsize I}} = -2 \ln \mathcal{L} ( \boldsymbol{\widehat{\theta}} ) \, + \, N_{\mbox{\scriptsize par}} \ln N_{\mbox{\scriptsize dat}}
                                    \, + \, \ln ( \det ( \mathbf{J} ( \boldsymbol{\widehat{\theta}} ) ) ) \, - \, N_{\mbox{\scriptsize par}} \ln 2 \pi
\label{eqn:BICI}
\end{equation}
Model selection with the BIC$_{\mbox{\scriptsize I}}$ proceeds as for the AIC$_{\mbox{\scriptsize C}}$.

It is worth mentioning that the BIC and BIC$_{\mbox{\scriptsize I}}$ are {\it consistent} model selection criteria in that they select with high probability
the true model from the set of candidate models whenever the true model is represented in the set of candidate models.

\subsection{Information Criteria For Maximum Penalised Likelihood}
\label{sec:model_selection_criteria_ic_mpl}

The AIC, AIC$_{\mbox{\scriptsize C}}$, TIC, BIC and BIC$_{\mbox{\scriptsize I}}$ apply only to models estimated by maximum likelihood.

\subsubsection{GIC$_{\mbox{\scriptsize P}}$}
\label{sec:gicp}

\citet{kon1996} derived a further generalisation of the AIC and TIC, called the generalised information criterion (GIC), that can be
applied to model selection for models with parameters estimated by maximum penalised likelihood:
\begin{equation}
\mbox{GIC}_{\mbox{\scriptsize P}} (\lambda) = -2 \ln \mathcal{L} ( \boldsymbol{\widehat{\theta}}_{\mbox{\scriptsize P}} ) 
          + 2 \operatorname{tr} \left( \mathbf{I_{\mbox{\scriptsize P}}} ( \boldsymbol{\widehat{\theta}}_{\mbox{\scriptsize P}} ) \,\, \mathbf{J_{\mbox{\scriptsize P}}}^{-1} ( \boldsymbol{\widehat{\theta}}_{\mbox{\scriptsize P}} ) \right) 
\label{eqn:GICP}
\end{equation}
where $\boldsymbol{\widehat{\theta}}_{\mbox{\scriptsize P}}$ is a vector of MPLEs for the model parameters, and:
\begin{equation}
\mathbf{I_{\mbox{\scriptsize P}}} ( \boldsymbol{\theta} ) = \mathbf{I} ( \boldsymbol{\theta} ) \, - \, \frac{\lambda}{N_{\mbox{\scriptsize dat}}} \mathbf{L}^{\mbox{\scriptsize T}} \, \mathbf{L} \, \boldsymbol{\theta}
                                                                                               \, \frac{\partial \ln \mathcal{L} ( \boldsymbol{\theta} )}{\partial \boldsymbol{\theta}^{\mbox{\scriptsize T}}}
\label{eqn:Idash}
\end{equation}
\begin{equation}
\mathbf{J_{\mbox{\scriptsize P}}} ( \boldsymbol{\theta} ) = \mathbf{J} ( \boldsymbol{\theta} ) + \lambda \mathbf{L}^{\mbox{\scriptsize T}} \, \mathbf{L}
\label{eqn:Jdash} 
\end{equation}
Here $\mathbf{L}^{\mbox{\scriptsize T}} \, \mathbf{L}$ is an $N_{\mbox{\scriptsize par}} \times N_{\mbox{\scriptsize par}}$ matrix and we have used the fact that it is symmetric
to slightly simplify the \citet{kon1996} expressions for $\mathbf{I_{\mbox{\scriptsize P}}} ( \boldsymbol{\theta} )$ and $\mathbf{J_{\mbox{\scriptsize P}}} ( \boldsymbol{\theta} )$ (their equation~(21)).
Model selection with the GIC$_{\mbox{\scriptsize P}} (\lambda)$ is performed by minimising GIC$_{\mbox{\scriptsize P}} (\lambda)$ over $\lambda$ for each model,
and then selecting the model for which GIC$_{\mbox{\scriptsize P}} (\lambda)$ is minimised over the full set of models under consideration.

\subsubsection{BIC$_{\mbox{\scriptsize P}}$}
\label{sec:bicp}

Using the same Bayesian analysis as for the derivation of the BIC$_{\mbox{\scriptsize I}}$, \citet{kon2004} also extended the BIC$_{\mbox{\scriptsize I}}$
to apply to model selection for models with parameters estimated by maximum penalised likelihood.
For $\mathbf{L}^{\mbox{\scriptsize T}} \, \mathbf{L}$ of rank $N_{\mbox{\scriptsize par}} - d$,
and denoting the product of the $N_{\mbox{\scriptsize par}} - d$ non-zero eigenvalues of $\mathbf{L}^{\mbox{\scriptsize T}} \, \mathbf{L}$ by $\Lambda_{+}$,
they derived:
\begin{equation}
\begin{aligned}
\mbox{BIC}_{\mbox{\scriptsize P}} (\lambda) = & \, -2 \ln \mathcal{L} ( \boldsymbol{\widehat{\theta}}_{\mbox{\scriptsize P}} ) \, + \, d \ln N_{\mbox{\scriptsize dat}} \, 
                                                + \, \ln ( \det ( \mathbf{J_{\mbox{\scriptsize P}}} ( \boldsymbol{\widehat{\theta}}_{\mbox{\scriptsize P}} ) ) ) \, - \, d \ln 2 \pi  \\
  & \, + \, \lambda N_{\mbox{\scriptsize dat}} \, \boldsymbol{\widehat{\theta}}^{\mbox{\scriptsize T}}_{\mbox{\scriptsize P}} \, \mathbf{L}^{\mbox{\scriptsize T}} \, \mathbf{L} \, \boldsymbol{\widehat{\theta}}_{\mbox{\scriptsize P}} \,
       - \, \ln \Lambda_{+} \, - \, (N_{\mbox{\scriptsize par}} - d) \ln \lambda \\
\end{aligned}
\label{eqn:BICP}
\end{equation}
Model selection with the BIC$_{\mbox{\scriptsize P}} (\lambda)$ proceeds as for the GIC$_{\mbox{\scriptsize P}} (\lambda)$.

\subsection{Information Criteria For DIA}
\label{sec:model_selection_criteria_dia}

We may adapt the various information criteria from Sections~\ref{sec:model_selection_criteria_ic_ml}~and~\ref{sec:model_selection_criteria_ic_mpl}
to our problem of solving for the kernel and differential background in DIA.
The model image $M$ has $N_{\mbox{\scriptsize par}} = N_{\kappa} + 1$ parameters and we use the notation $\boldsymbol{\theta} \equiv \boldsymbol{\alpha}$,
$\boldsymbol{\widehat{\theta}} \equiv \boldsymbol{\widehat{\alpha}}$ and $\boldsymbol{\widehat{\theta}}_{\mbox{\scriptsize P}} \equiv \boldsymbol{\widehat{\alpha}}_{\mbox{\scriptsize P}}$.

Firstly, we compute the log-likelihood function for data drawn from Gaussian distributions $\mathcal{N}(M_{ij},\sigma_{ij})$ as:
\begin{equation}
-2 \ln \mathcal{L} ( \boldsymbol{\alpha} ) = \chi^2 \, + \, \sum_{ij} \ln ( \sigma_{ij}^{2} ) \, + \, N_{\mbox{\scriptsize dat}} \, \ln 2 \pi
\label{eqn:ln_L}
\end{equation}
For model selection purposes, the last term $N_{\mbox{\scriptsize dat}} \, \ln 2 \pi$ is constant and can be ignored.
Secondly, we note that since the $\sigma_{ij}$ are considered as constant at each iteration of the maximum likelihood problem in Section~\ref{sec:solving_for_kernel},
the matrices $\mathbf{I} ( \boldsymbol{\alpha} )$ and $\mathbf{J} ( \boldsymbol{\alpha} )$ evaluated at $\boldsymbol{\alpha} = \boldsymbol{\widehat{\alpha}}$
are given by:
\begin{equation}
[ \mathbf{I} ( \boldsymbol{\widehat{\alpha}} ) ]_{q q^{\,\prime}} = \frac{1}{N_{\mbox{\scriptsize dat}}} \sum_{ij} \, \varepsilon_{ij}^{2} \, \frac{\psi_{qij} \, \psi_{q^{\,\prime} ij}}{\sigma_{ij}^{2}}
\label{eqn:Ihat}
\end{equation}
\begin{equation}
\mathbf{J} ( \boldsymbol{\widehat{\alpha}} ) = \mathbf{H} / N_{\mbox{\scriptsize dat}}
\label{eqn:Jhat}
\end{equation}
For computational purposes it is useful to note that $\mathbf{I} ( \boldsymbol{\widehat{\alpha}} )$ is symmetric.
From these two expressions, we may derive the following results:
\begin{equation}
\operatorname{tr} \left( \mathbf{I} ( \boldsymbol{\widehat{\alpha}} ) \,\, \mathbf{J}^{-1} ( \boldsymbol{\widehat{\alpha}} ) \right) =
                  \, N_{\mbox{\scriptsize dat}} \sum_{q = 1}^{N_{\kappa} + 1} \; \sum_{q^{\,\prime} = 1}^{N_{\kappa} + 1} [ \mathbf{I} ( \boldsymbol{\widehat{\alpha}} ) ]_{q q^{\,\prime}} \, [ \mathbf{H}^{-1} ]_{q q^{\,\prime}}
\label{eqn:trIJ}
\end{equation}
\begin{equation}
\ln ( \det ( \mathbf{J} ( \boldsymbol{\widehat{\alpha}} ) ) ) = \ln ( \det ( \mathbf{H} )) - ( N_{\kappa} + 1 ) \ln N_{\mbox{\scriptsize dat}}
\label{eqn:detJ}
\end{equation}
Finally, we consider that the solution of the normal equations requires the computation of the Cholesky factorisation
$\mathbf{H} = \mathbf{G} \mathbf{G}^{\mbox{\scriptsize T}}$, where $\mathbf{G}$
is a lower triangular matrix with positive diagonal entries $g_{qq}$, from which we may immediately calculate the determinant of $\mathbf{H}$ as
$\det ( \mathbf{H} ) = \prod_{q = 1}^{N_{\kappa} + 1} g_{qq}^{2}$. Hence, with minimal extra computation, the Cholesky factorisation of $\mathbf{H}$ yields:
\begin{equation}
\ln ( \det ( \mathbf{H} )) = 2 \sum_{q = 1}^{N_{\kappa} + 1} \ln g_{qq}
\label{eqn:detH}
\end{equation}

Therefore, using equations~(\ref{eqn:ln_L}) and (\ref{eqn:trIJ})~-~(\ref{eqn:detH}) for the maximum likelihood problem
in Section~\ref{sec:solving_for_kernel}, we have the following formulae for the relevant information criteria from
Section~\ref{sec:model_selection_criteria_ic_ml}:
\begin{equation}
\mbox{AIC}_{\mbox{\scriptsize C}} = \chi^2 \, + \, \sum_{ij} \ln ( \sigma_{ij}^{2} ) \, + \, 2 (N_{\kappa} + 1) \left( \frac{N_{\mbox{\scriptsize dat}}}{N_{\mbox{\scriptsize dat}} - N_{\kappa} - 2} \right)
\label{eqn:AIC_use}
\end{equation}
\begin{equation}
\mbox{TIC} = \chi^2 \, + \, \sum_{ij} \ln ( \sigma_{ij}^{2} ) 
                    \, + \, 2 N_{\mbox{\scriptsize dat}} \sum_{q = 1}^{N_{\kappa} + 1} \; \sum_{q^{\,\prime} = 1}^{N_{\kappa} + 1} [ \mathbf{I} ( \boldsymbol{\widehat{\alpha}} ) ]_{q q^{\,\prime}} \, [ \mathbf{H}^{-1} ]_{q q^{\,\prime}}
\label{eqn:TIC_use}
\end{equation}
\begin{equation}
\mbox{BIC} = \chi^2 \, + \, \sum_{ij} \ln ( \sigma_{ij}^{2} ) \, + \, (N_{\kappa} + 1) ( \ln N_{\mbox{\scriptsize dat}} \, - \, \ln 2 \pi )
\label{eqn:BIC_use}
\end{equation}
\begin{equation}
\mbox{BIC}_{\mbox{\scriptsize I}} = \chi^2 \, + \, \sum_{ij} \ln ( \sigma_{ij}^{2} ) \, + \, 2 \sum_{q = 1}^{N_{\kappa} + 1} \ln g_{qq} \, - \, (N_{\kappa} + 1) \ln 2 \pi
\label{eqn:BICI_use}
\end{equation}

Considering now the maximum penalised likelihood problem, for constant $\sigma_{ij}$ we have
$\partial \ln \mathcal{L} ( \boldsymbol{\alpha} ) / \partial \boldsymbol{\alpha}^{\mbox{\scriptsize T}} = \boldsymbol{\beta}^{\mbox{\scriptsize T}} - \boldsymbol{\alpha}^{\mbox{\scriptsize T}} \mathbf{H}$,
which is equal to $\lambda N_{\mbox{\scriptsize dat}} \, ( \mathbf{L} \mathbf{L} \boldsymbol{\widehat{\alpha}}_{\mbox{\scriptsize P}} )^{\mbox{\scriptsize T}}$ when evaluated at $\boldsymbol{\alpha} = \boldsymbol{\widehat{\alpha}}_{\mbox{\scriptsize P}}$
(using equations~(\ref{eqn:matrix_reg_normal_eqns})~and~(\ref{eqn:reg_least_squares_matrix})). Then, using $\mathbf{L}^{\mbox{\scriptsize T}} \, \mathbf{L} = \mathbf{L} \mathbf{L}$,
the matrices $\mathbf{I_{\mbox{\scriptsize P}}} ( \boldsymbol{\alpha} )$ and $\mathbf{J_{\mbox{\scriptsize P}}} ( \boldsymbol{\alpha} )$ evaluated
at $\boldsymbol{\alpha} = \boldsymbol{\widehat{\alpha}}_{\mbox{\scriptsize P}}$ are given by:
\begin{equation}
\mathbf{I_{\mbox{\scriptsize P}}} ( \boldsymbol{\widehat{\alpha}}_{\mbox{\scriptsize P}} ) = \mathbf{I} ( \boldsymbol{\widehat{\alpha}}_{\mbox{\scriptsize P}} )
                 \, - \, \lambda^{2} \, ( \mathbf{L} \, \mathbf{L} \, \boldsymbol{\widehat{\alpha}}_{\mbox{\scriptsize P}} ) ( \mathbf{L} \, \mathbf{L} \, \boldsymbol{\widehat{\alpha}}_{\mbox{\scriptsize P}} )^{\mbox{\scriptsize T}}
\label{eqn:Idashhat}
\end{equation}
\begin{equation}
\mathbf{J_{\mbox{\scriptsize P}}} ( \boldsymbol{\widehat{\alpha}}_{\mbox{\scriptsize P}} ) = \mathbf{H_{\mbox{\scriptsize P}}} / N_{\mbox{\scriptsize dat}}
\label{eqn:Jdashhat}
\end{equation}
Writing $\Omega_{q} = \sum_{u = 1}^{N_{\kappa}} \sum_{v = 1}^{N_{\kappa}} L_{qu} \, L_{uv} \, \widehat{\alpha}_{\mbox{\scriptsize P}, v}$, then, from these two expressions, we may derive the following results:
\begin{equation}
\operatorname{tr} \left( \mathbf{I_{\mbox{\scriptsize P}}} ( \boldsymbol{\widehat{\alpha}}_{\mbox{\scriptsize P}} ) \,\, \mathbf{J_{\mbox{\scriptsize P}}}^{-1} ( \boldsymbol{\widehat{\alpha}}_{\mbox{\scriptsize P}} ) \right) =
                  \, N_{\mbox{\scriptsize dat}} \sum_{q = 1}^{N_{\kappa} + 1} \; \sum_{q^{\,\prime} = 1}^{N_{\kappa} + 1}
                  \left( [ \mathbf{I} ( \boldsymbol{\widehat{\alpha}}_{\mbox{\scriptsize P}} ) ]_{q q^{\,\prime}} \, - \, \lambda^{2} \, \Omega_{q} \, \Omega_{q^{\,\prime}} \right) \,
                  [ \mathbf{H_{\mbox{\scriptsize P}}}^{-1} ]_{q q^{\,\prime}}
\label{eqn:trIJdash}
\end{equation}
\begin{equation}
\ln ( \det ( \mathbf{J_{\mbox{\scriptsize P}}} ( \boldsymbol{\widehat{\alpha}}_{\mbox{\scriptsize P}} ) ) ) = \ln ( \det ( \mathbf{H_{\mbox{\scriptsize P}}} )) - ( N_{\kappa} + 1 ) \ln N_{\mbox{\scriptsize dat}}
\label{eqn:detJdash}
\end{equation}
Also, the Cholesky factorisation of $\mathbf{H_{\mbox{\scriptsize P}}} = \mathbf{G_{\mbox{\scriptsize P}}} \mathbf{G_{\mbox{\scriptsize P}}^{\mbox{\scriptsize T}}}$ yields:
\begin{equation}
\ln ( \det ( \mathbf{H_{\mbox{\scriptsize P}}} )) = 2 \sum_{q = 1}^{N_{\kappa} + 1} \ln g_{\mbox{\scriptsize P}, qq}
\label{eqn:detHdash}
\end{equation}
Finally, we note that the matrix $\mathbf{L} \mathbf{L}$ is of rank $N_{\mbox{\scriptsize par}} - d = N_{\kappa} - N_{\mbox{\scriptsize grp}}$,
and hence $d = N_{\mbox{\scriptsize grp}} + 1$.

Therefore, using equations~(\ref{eqn:ln_L}) and (\ref{eqn:trIJdash})~-~(\ref{eqn:detHdash}) for the maximum penalised likelihood problem
in Section~\ref{sec:reg_delta_basis}, 
we have the following formulae for the relevant information criteria
from Section~\ref{sec:model_selection_criteria_ic_mpl}:
\begin{equation}
\begin{aligned}
\mbox{GIC}_{\mbox{\scriptsize P}} (\lambda) = & \, \chi^2 \, + \, \sum_{ij} \ln ( \sigma_{ij}^{2} ) \\
                                              & \, + \, 2 N_{\mbox{\scriptsize dat}} \sum_{q = 1}^{N_{\kappa} + 1} \; \sum_{q^{\,\prime} = 1}^{N_{\kappa} + 1}                                                   
                                                \left( [ \mathbf{I} ( \boldsymbol{\widehat{\alpha}}_{\mbox{\scriptsize P}} ) ]_{q q^{\,\prime}} \, - \, \lambda^{2} \, \Omega_{q} \, \Omega_{q^{\,\prime}} \right) \,
                                                [ \mathbf{H_{\mbox{\scriptsize P}}}^{-1} ]_{q q^{\,\prime}}
\end{aligned}
\label{eqn:GICP_use}
\end{equation}
\begin{equation}
\begin{aligned} 
\mbox{BIC}_{\mbox{\scriptsize P}} (\lambda) = & \, \chi^2 \, + \, \sum_{ij} \ln ( \sigma_{ij}^{2} ) \, + \, 2 \sum_{q = 1}^{N_{\kappa} + 1} \ln g_{\mbox{\scriptsize P}, qq}
                                                \, - \, ( N_{\kappa} - N_{\mbox{\scriptsize grp}} ) \ln \lambda N_{\mbox{\scriptsize dat}} \\
                                              & \, - \, (N_{\mbox{\scriptsize grp}} + 1) \ln 2 \pi
                                                \, + \, \lambda N_{\mbox{\scriptsize dat}} \sum_{q = 1}^{N_{\kappa}} \widehat{\alpha}_{\mbox{\scriptsize P}, q} \, \Omega_{q}
                                                \, - \, \ln \Lambda_{+} \\
\end{aligned}
\label{eqn:BICP_use}
\end{equation}

\section{Kernel Design Algorithms}
\label{sec:ker_design}

Let us introduce the concept of a {\it kernel design}, which we define
as a specific choice of DBFs (or, equivalently, kernel pixels) to be employed in the modelling of the
convolution kernel. From a master set of $N$ DBFs, the model selection criteria will each select a single ``best'' kernel design,
which requires the evaluation of the criteria via the estimation of the model image parameters for each of the $2^{N}$ possible
kernel designs\footnote{This number includes the kernel design with zero DBFs, i.e. a model image with the differential background as the only parameter.}.
This computational problem is formidable and currently infeasible for values of $N$ that are required for typical kernel models
(e.g. a relatively small 9x9 kernel pixel grid yields $\sim$2.4$\times$10$^{24}$ potential kernel designs!). Furthermore, 
branch-and-bound algorithms (e.g. \citealt{fur1974}) for speeding up this exhaustive search are only applicable to some of our model
selection criteria in Section~\ref{sec:model_selection_criteria_dia}.

It is well known that by not considering all of the possible combinations of predictor variables in a linear regression problem (e.g.
by using stepwise regression for variable selection), the optimal set of predictors may be misidentified.
However, in our case, we know from the nature/purpose of the kernel (and copious amounts of prior experience!) that the
true kernel model has a peak signal at the kernel coordinates corresponding to the translational offset between the reference
and target images (which is at the kernel origin when they are properly registered) and that this signal decays away from the peak.
There may be other peaks (e.g. due to a telescope jump in the target image), but again these also have profiles that decay away
from the peak(s). The best kernel designs are therefore generally limited to sets of DBFs in close proximity that form relatively compact and
regular shapes. Based on these observations, we have devised two algorithms for automatic kernel design that compare a manageable
number of sensible kernel models; the circular kernel design algorithm
(Section~\ref{sec:simim_ckda}) and the irregular kernel design algorithm (Section~\ref{sec:simim_ikda}).

\subsection{The Circular Kernel Design Algorithm}
\label{sec:simim_ckda}

\begin{table}
\caption{The number of DBFs in a circular kernel design for different ranges of the kernel radius $r_{\kappa}$. The ranges are defined
         by $r_{\kappa , \mbox{\tiny lo}} \le r_{\kappa} < r_{\kappa , \mbox{\tiny hi}}$. The table may be extended as appropriate
         for larger values of $r_{\kappa}$.}
\centering
\begin{tabular}{cccccccc}
\hline
$r_{\kappa , \mbox{\tiny lo}}$ (pix) & $r_{\kappa , \mbox{\tiny hi}}$ (pix) & $N_{\kappa}$ & & & $r_{\kappa , \mbox{\tiny lo}}$ (pix) & $r_{\kappa , \mbox{\tiny hi}}$ (pix) & $N_{\kappa}$ \\
\hline
0.000 & 1.000 &   1 & & &    4.472 &    5.000 &       69 \\
1.000 & 1.414 &   5 & & &    5.000 &    5.099 &       81 \\
1.414 & 2.000 &   9 & & &    5.099 &    5.385 &       89 \\
2.000 & 2.236 &  13 & & &    5.385 &    5.657 &       97 \\
2.236 & 2.829 &  21 & & &    5.657 &    5.831 &      101 \\
2.829 & 3.000 &  25 & & &    5.831 &    6.000 &      109 \\
3.000 & 3.162 &  29 & & &    6.000 &    6.083 &      113 \\
3.162 & 3.606 &  37 & & &    6.083 &    6.325 &      121 \\
3.606 & 4.000 &  45 & & &    6.325 &    6.403 &      129 \\
4.000 & 4.123 &  49 & & &    6.403 &    6.708 &      137 \\
4.123 & 4.243 &  57 & & &    6.708 &    7.000 &      145 \\
4.243 & 4.472 &  61 & & & $\hdots$ & $\hdots$ & $\hdots$ \\
\hline
\end{tabular}
\label{tab:ker_rad}
\end{table}

One very simple way to greatly reduce the number of candidate kernel designs that is in line with the
expected kernel properties is to restrict the kernel designs to those that correspond to a circularly shaped pixel grid centred
at the origin of the kernel pixel coordinates. We therefore define a circular kernel design of radius $r_{\kappa}$
as the set of DBFs corresponding to the kernel pixels whose centres lie at or within $r_{\kappa}$ pixels of the kernel
origin, which is taken to be at the centre of the $(r,s) = (0,0)$ kernel pixel. As $r_{\kappa}$ is increased,
the circular kernel design includes progressively more DBFs leading to a set of nested kernel designs. In Table~\ref{tab:ker_rad}, we list the
number of DBFs in a circular kernel design for a range of values of $r_{\kappa}$.

The circular kernel design algorithm (CKDA) works for a pair of images and an adopted model selection criterion.
The algorithm sequentially evaluates a set of nested model images. It finishes when the current
model image under consideration fails the selection criterion, and consequently the previously considered
model image is selected. For the $\Delta \chi^{2}$ and $F$-tests, this means that the current model
image does not provide a significantly better fit than the previous one. For the information criteria,
this means that the current model image yields a larger value of the criterion than the previous one, 
where $\lambda$ has already been optimised individually for each model if appropriate.
The algorithm proceeds as follows:
\begin{enumerate}
\item Fit the model image with the differential background $B$ as the only parameter and calculate the
      desired model selection criterion.
\item Adopt a circular kernel design of radius $r_{\kappa} = 0.5$~pix, which defines a kernel model with a single DBF.
      Fit the model image consisting of the differential background and the kernel model,
      and calculate the desired model selection criterion.
      If the model image from (i) is selected, then finish.
\item Increment $r_{\kappa}$ until the kernel model includes a new (larger) set of DBFs.
      Fit the model image consisting of the differential background and the new kernel model,
      and calculate the desired model selection criterion.
      If the model image from the previous iteration is selected, then finish.
\item Repeat step (iii) until the algorithm terminates.
\end{enumerate}
Note that the CKDA is intended to be applied to reference and target images that are registered to within a single pixel
(but without requiring sub-pixel alignment necessitating image resampling).

Special care must be taken when using sigma-clipping during the fitting of the model images in the CKDA. Since each model image fit
within the algorithm has the potential to clip different sets of target-image pixel values, the calculation of the model selection criterion
may end up employing different sets of pixels at each step, which leads to undesirable jumps in its value that
are unrelated to the properties of the fits. If sigma-clipping is required due to the presence of outlier pixel values, then,
to avoid this problem, it is recommended to run the CKDA to conclusion without using sigma-clipping and to use the selected model
image to identify and clip the outliers. The CKDA may then be re-run ignoring this same set of clipped pixel values at each step,
and the whole process may be iterated more than once if necessary. This issue with sigma-clipping applies to all kernel design   
algorithms, and also whenever a fair comparison between algorithms is required (see Section~\ref{sec:realsec}).

In the early phases of testing the CKDA, we ran the algorithm past the finishing point to check that kernel designs with
larger radii than the radius of the selected design do not yield smaller values of the information criterion,
which, if this was the case, would indicate that the algorithm is terminating too early at a local minimum.  
We found that only in a relatively small proportion of the simulations (Section~\ref{sec:simsec}) a slightly smaller value of the information criterion
is achieved for a kernel design with a larger radius than the selected design (usually 2-3 steps larger in Table~\ref{tab:ker_rad}),
and that when this occurs, the values of the model performance metrics (Section~\ref{sec:mod_perf_metrics})
for the two designs are very similar with no systematic improvement for the kernel design with a larger radius.
Given that running the CKDA to larger radii comes at considerable
cost in terms of processing power, the termination criterion of the CKDA was fixed at the first minimum of the information criterion.
The same conclusions were also found to apply to the irregular kernel design algorithm (Section~\ref{sec:simim_ikda}).

\subsection{The Irregular Kernel Design Algorithm}
\label{sec:simim_ikda}

Another way to limit the number of candidate kernel designs is to ``grow'' the kernel model as a
connected set of DBFs from a single ``seed'' DBF by including one new DBF at each iteration. We 
call this the irregular kernel design algorithm (IKDA), and it works for a pair of images and   
an adopted model selection criterion as follows:
\begin{enumerate}
\item Fit the model image with the differential background $B$ as the only parameter and calculate the
      desired model selection criterion.
\item Define a master set of $N$ DBFs by taking an appropriately large grid of kernel pixels centred on the pixel at
      the kernel origin. For each DBF in the master set, fit the model image consisting of the differential
      background and a kernel model with the single DBF, and calculate the desired model selection criterion. If the model image from (i)
      is selected in all $N$ cases, then finish. Otherwise, accept the DBF from the master set that gives the best model image
      (according to the selection criterion) as the first DBF to be included in the kernel model (referred to as the seed DBF).
      Remove the seed DBF from the master set.
\item Find the subset of DBFs from the master set that are adjacent to at least one of the DBFs in the kernel model from
      the previous iteration. For each candidate DBF in this subset, fit the model image consisting of the differential 
      background and a new kernel model with a set of DBFs that is the union of the set of DBFs in the kernel model from
      the previous iteration with the candidate DBF. If the model image from the previous iteration
      is selected in all cases, then finish. Otherwise, accept the candidate DBF that gives the best model image
      as the next DBF to be included in the kernel model. Remove the accepted candidate DBF from the master set.
\item Repeat step (iii) until the algorithm terminates.
\end{enumerate}
Note that the IKDA may be applied to reference and target images that are not registered to within a single pixel since
step (ii) is effectively a form of image registration. Again, special care must be taken with the
application of sigma-clipping within the IKDA (see Section~\ref{sec:simim_ckda}).

The IKDA may generate different sequences of DBFs during the growth of the kernel model for different model selection criteria.
However, for the $\Delta \chi^{2}$ and $F$-statistics, the IKDA follows the same sequence of DBFs since
both statistics are maximised at each iteration of the IKDA by minimising $\chi^{2}_{\mbox{\scriptsize B}}$.
For similar reasons, the IKDA follows the same sequence of DBFs for the AIC$_{\mbox{\scriptsize C}}$ and the BIC.
In these cases, the different model selection criteria simply terminate the IKDA at different points in the sequence.
Still, regardless of the actual model selection criterion used, we find that the IKDA always grows the kernel solution outwards
from the selected seed DBF.

There are various alternative ways in which the kernel model may be grown within the IKDA. We have experimented
with dropping the constraint that each new DBF must be adjacent to at least one DBF in the previous kernel model. However,
this produced similar kernel solutions to those produced by the IKDA with the adjacency constraint but with an extra
scattering of isolated DBFs arbitrarily far from the peak signal in the kernel. We also experimented
with relaxing the definition of ``adjacent'' to include more nearby kernel pixels, but the results from these versions of the IKDA are
virtually indistinguishable in terms of the model performance metrics from the results for the IKDA described above (both for the simulated and real data).
Hence we have not considered these variations on the IKDA any further.

Finally, we mention that the IKDA may be modified to generate multiple seed DBFs (possibly as part of step (ii) or by generating a new seed DBF
after the algorithm terminates for the first time). This modification would be useful for adapting to situations similar to when
the telescope has jumped during a target image exposure, and consequently the true kernel model consists of two or more disconnected peaks.

\section{Testing Automatic Kernel Design Algorithms On Simulated Images}
\label{sec:simsec}

The main aim of this paper is to find out which combination of kernel design algorithm and model selection
criterion consistently selects a kernel model that provides the best performance in terms
of model error, fit quality, and photometric accuracy. The conclusions drawn from our investigation will likely
depend on the properties of the reference and target images, and hence we must systematically map out the performance of
each method accordingly. This task is most efficiently performed by generating and analysing simulated images.
Furthermore, with respect to model error, the performance of each method may only be measured through the
use of simulations where the true model image is known. Simulations also provide a setting in which the noise model is precisely known
since it is used to generate the simulated data. Thus simulations allow for the degree of under- or over-fitting to be assessed accurately.
For these reasons, we have performed detailed DIA simulations for a wide range of image properties.

\subsection{Generating Simulated Images}
\label{sec:simim}

We employed a Monte Carlo method for our investigation. We adopted reasonable values for the CCD readout noise and gain of
$\sigma_{0} = 5$~ADU and $G = 1$~e$^{-}$/ADU, respectively. For each simulation,
we generated both noiseless and noisy versions of a reference and target image pair, along with the noise maps
used for generating the noisy images, via the following procedure:
\begin{enumerate}
\item The size of the reference image was set to 141$\times$141 pixels.
\item The sky background level for the reference image $S_{\mbox{\scriptsize ref}}$ was drawn from a uniform
      distribution on the interval $[16,1000]$~ADU.
\item The log of the field star density, parameterised as the number of stars per 100$\times$100 pixel image region, was drawn from
      a uniform distribution on the interval $[0,3]$, and this density was used to calculate the number of stars $N_{\mbox{\scriptsize star}}$
      to be generated in the reference image.
\item The pixel coordinates of each star in the reference image were drawn from a uniform distribution over the image
      area. Also, for each star, the value of $\mathcal{F}^{-3/2}$, where $\mathcal{F}$ is the star flux (ADU), was drawn from a uniform
      distribution on the interval $[10^{-9},10^{-9/2}]$. This flux distribution is appropriate when imaging to a fixed depth through
      a uniform space density of stars (e.g. a good approximation for certain volumes in our Galaxy). For the purposes of performing PSF
      photometry on the difference image, and without loss of generality, the pixel coordinates of the brightest star were modified by an
      integer pixel shift to lie within the central pixel of the reference image.
\item The same normalised two-dimensional Gaussian profile of full-width at half-maximum (FWHM) $f_{\mbox{\scriptsize ref}}$ pixels
      was adopted for the profile of each star in the reference image. The value of $f_{\mbox{\scriptsize ref}}$ was drawn from a uniform distribution on the interval $[1,6]$~pix,
      adequately covering the under- to over-sampled regimes.
\item A square image pixel array $R_{\mbox{\scriptsize noiseless}}$ of size 141$\times$141 pixels was created with all of the
      pixel values set to the sky background level $S_{\mbox{\scriptsize ref}}$.
\item For each star in the reference image, the Gaussian profile was centred at the star pixel coordinates and sampled at 7 times the image resolution
      over the image area. The over-sampled Gaussian profile was then binned (by averaging) to match the image resolution and re-normalised to a sum of unity.
      Finally, it was scaled by the star flux and added to the image $R_{\mbox{\scriptsize noiseless}}$.
\item An image of standard deviations $\sigma_{\mbox{\scriptsize in},\mbox{\scriptsize ref}}$ (i.e. a noise map) corresponding to
      $R_{\mbox{\scriptsize noiseless}}$ was created via:
      \begin{equation}
      \sigma_{\mbox{\scriptsize in},\mbox{\scriptsize ref},ij} = \sqrt{\sigma_{0}^{2} + R_{\mbox{\scriptsize noiseless},ij} / G}
      \label{eqn:noise_map_ref}
      \end{equation}
      which may be derived from equation~(\ref{eqn:noise_model}) by setting the $F_{ij} = 1$. A 141$\times$141 pixel image $W$ of values
      drawn from a Gaussian distribution with zero mean and unit standard deviation was also generated and used to construct a noisy reference image $R_{\mbox{\scriptsize noisy}}$ via:
      \begin{equation}
      R_{\mbox{\scriptsize noisy},ij} = R_{\mbox{\scriptsize noiseless},ij} + W_{ij} \, \sigma_{\mbox{\scriptsize in},\mbox{\scriptsize ref},ij}
      \label{eqn:noisy_ref}
      \end{equation}
\item The size of the target image was set to 141$\times$141 pixels.
\item For simplicity, the sky background level for the target image $S_{\mbox{\scriptsize tar}}$ was set to $S_{\mbox{\scriptsize ref}}$, which is equivalent
      to assuming a differential background of zero.
\item A single sub-pixel shift in each of the $x$ and $y$ image coordinate directions was drawn from a uniform distribution on the interval $[-0.5,0.5]$~pix and applied
      to the pixel coordinates of the stars in the reference image to generate the coordinates of the same stars in the target image. The fluxes of the stars
      in the target image were assumed to be the same as their fluxes in the reference image, which is equivalent to assuming non-variable stars and a 
      photometric scale factor of unity.
\item The convolution kernel matching the PSF between the reference and target images was assumed to be a normalised two-dimensional Gaussian profile
      of FWHM $| f_{\mbox{\scriptsize ker}} |$ pixels, where a non-negative or negative value of $f_{\mbox{\scriptsize ker}}$ indicates that the kernel
      convolves the reference or target image PSF, respectively. The value of $f_{\mbox{\scriptsize ker}}$ was drawn from a uniform
      distribution on the interval $[-1,5]$~pix and the FWHM $f_{\mbox{\scriptsize tar}}$ of the Gaussian profile for the stars in the target image was
      then calculated from:
      \begin{equation}
      f_{\mbox{\scriptsize tar}}^2 =
      \begin{cases}
      f_{\mbox{\scriptsize ref}}^2 + f_{\mbox{\scriptsize ker}}^2 & \mbox{for $f_{\mbox{\scriptsize ker}} \ge 0$} \\
      f_{\mbox{\scriptsize ref}}^2 - f_{\mbox{\scriptsize ker}}^2 & \mbox{for $f_{\mbox{\scriptsize ker}} < 0$}   \\
      \end{cases}
      \label{eqn:fwhm_tar}
      \end{equation}
\item A square image pixel array $I_{\mbox{\scriptsize noiseless}}$ of size 141$\times$141 pixels was created with all of the
      pixel values set to the sky background level $S_{\mbox{\scriptsize tar}}$.
\item The flux profiles of the stars in the target image were added to $I_{\mbox{\scriptsize noiseless}}$ using the same method as that used in step~(vii)
      for the reference image.
\item An image of standard deviations $\sigma_{\mbox{\scriptsize in},\mbox{\scriptsize tar}}$ corresponding to $I_{\mbox{\scriptsize noiseless}}$
      was created via:
      \begin{equation}
      \sigma_{\mbox{\scriptsize in},\mbox{\scriptsize tar},ij} = \sqrt{\sigma_{0}^{2} + I_{\mbox{\scriptsize noiseless},ij} / G}
      \label{eqn:noise_map_tar}
      \end{equation}
      A new 141$\times$141 pixel image $W$ of values drawn from a Gaussian distribution with zero mean and unit standard deviation was also generated and
      used to construct a noisy target image $I_{\mbox{\scriptsize noisy}}$ via:
      \begin{equation}
      I_{\mbox{\scriptsize noisy},ij} = I_{\mbox{\scriptsize noiseless},ij} + W_{ij} \, \sigma_{\mbox{\scriptsize in},\mbox{\scriptsize tar},ij}
      \label{eqn:noisy_tar}
      \end{equation}
\item The images $I_{\mbox{\scriptsize noiseless}}$, $I_{\mbox{\scriptsize noisy}}$ and $\sigma_{\mbox{\scriptsize in},\mbox{\scriptsize tar}}$ were each trimmed to a size of 101$\times$101 pixels.
      Hence the number of data values in each simulation is $N_{\mbox{\scriptsize dat}}~=~10201$.
\item The signal-to-noise (S/N) ratio of the noisy target image SN$_{\mbox{\scriptsize tar}}$ was calculated as:
      \begin{equation}
      \mbox{SN}_{\mbox{\scriptsize tar}} = \frac{\sum_{ij} ( I_{\mbox{\scriptsize noiseless},ij} - S_{\mbox{\scriptsize tar}} ) }{\sqrt{\sum_{ij} \sigma_{\mbox{\scriptsize in},\mbox{\scriptsize tar},ij}^{2}}}
      \label{eqn:sn_tar}
      \end{equation}
      The value of $\operatorname{log}$(SN$_{\mbox{\scriptsize tar}}$) is distributed approximately uniformly due to the way the field star density is generated in step~(iii).
      It is important to note that it does not necessarily follow that a high S/N target image has a bright high-S/N star in the centre. The high S/N in the target image may
      be the consequence of the presence of a reasonable number of faint stars. In this case, the star at the centre of the image is of low S/N, even though it is the brightest star in the image.
\end{enumerate}

In total, we generated the reference and target images for 548,392 simulations. We call this set of images ``Simulation set S1''.

The above method for generating reference and target images represents the case where the reference image has approximately the same S/N ratio as the target image.
However, it is common practice in DIA to create a high-S/N ratio reference image by stacking images or integrating longer. We therefore also repeated the whole procedure
of generating simulated images for reference images with ten times less variance in each pixel value than the corresponding target images (or $\sim 3.16$ times better S/N). This was
achieved by scaling the $\sigma_{\mbox{\scriptsize in},\mbox{\scriptsize ref},ij}$ in step (viii) by $10^{-1/2} \sim 0.316$. This
second set of reference and target images, ``Simulation set S10'', was generated for a total of 529,571 simulations. Figure~\ref{fig:example_sim} shows an example
noisy reference and target image pair from one of these simulations.

\subsection{Model Performance Metrics}
\label{sec:mod_perf_metrics}

We used the following metrics to measure the quality of each kernel and differential background solution:
\begin{itemize}
\item {\bf Model error:}
      The mean squared error MSE measures how well the fitted model image $M$ matches the true model image
      $I_{\mbox{\scriptsize noiseless}}$. It is defined by:
      \begin{equation}
      \mbox{MSE} = \frac{1}{N_{\mbox{\scriptsize dat}}} \sum_{ij} (M_{ij} - I_{\mbox{\scriptsize noiseless},ij})^{2}
      \label{eqn:mod_err}
      \end{equation}
      Kernel and differential background solutions with the smallest values of MSE exhibit the best performance
      in terms of model error.

      We also consider the photometric scale factor $P$ and the differential
      background $B$ as supplementary measures of model error. For our simulations,
      the closer to unity the value of $P$ and the closer to zero the value of $B$,
      the better the performance of a kernel and differential background solution with respect to model error.
      Systematic errors in the photometric scale factor are especially important since
      a fractional error E$_{P}$ in $P$ introduces a fractional error of E$_{P}$ into the photometry (\citealt{bra2015}).

\item {\bf Fit quality:}
      The bias and excess variance in the fitted model image may be measured by the following statistics:
      \begin{equation}
      \mbox{MFB} = \frac{1}{N_{\mbox{\scriptsize dat}}} \sum_{ij} \frac{(I_{\mbox{\scriptsize noisy},ij} - M_{ij})}{\sigma_{\mbox{\scriptsize in},\mbox{\scriptsize tar},ij}}
      \label{eqn:fit_bias}
      \end{equation}  
      \begin{equation}
      \mbox{MFV} = \frac{1}{N_{\mbox{\scriptsize dat}}-1} \sum_{ij} \left(\frac{(I_{\mbox{\scriptsize noisy},ij} - M_{ij})}{\sigma_{\mbox{\scriptsize in},\mbox{\scriptsize tar},ij}} - \mbox{MFB} \right)^{2}
      \label{eqn:fit_var}
      \end{equation}
      MFB is the mean fit bias and MFV is the mean fit variance with units of sigma and sigma-squared, respectively.
      The closer to zero the value of MFB, and the closer to unity the value of MFV, the better the performance of a kernel and
      differential background solution with respect to fit quality.

\item {\bf Photometric accuracy:}
      To assess the photometric accuracy, we perform PSF fitting on the difference image at the position of the brightest star
      in the reference image under the assumption that the reference image PSF is perfectly known. In detail, we generate a
      normalised two-dimensional Gaussian profile of FWHM $f_{\mbox{\scriptsize ref}}$ pixels centred at the pixel coordinates
      of the brightest star in the reference image (guaranteed by construction to be within half a pixel of the image centre)
      and sampled at 7 times the image resolution. The over-sampled Gaussian is then
      binned (by averaging) to match the image resolution, convolved with the kernel solution, trimmed in extent to
      a circularly shaped pixel grid of radius $\lceil 2 f_{\mbox{\scriptsize tar}} \rceil$ pixels around the star coordinates, and renormalised.
      This model PSF for the target image is then optimally scaled to the difference image at the position of the brightest star
      by simultaneously fitting a scaling factor $\mathcal{F}_{\mbox{\scriptsize diff}}$ and an additive constant, and
      using the known pixel variances in the target image $\sigma_{\mbox{\scriptsize in},\mbox{\scriptsize tar}}^2$.
%      The uncertainty on the scaling factor $\sigma_{\mathcal{F}_{\mbox{\scriptsize diff}}}$ is also recorded.
      The difference flux $\mathcal{F}_{\mbox{\scriptsize meas}}$ of the brightest star on the photometric scale of
      the reference image is then computed using $\mathcal{F}_{\mbox{\scriptsize meas}} = \mathcal{F}_{\mbox{\scriptsize diff}} / P$.
%      along with the corresponding uncertainty $\sigma_{\mathcal{F}_{\mbox{\scriptsize meas}}} = \sigma_{\mathcal{F}_{\mbox{\scriptsize diff}}} / P$.

      The theoretical minimum variance $\sigma_{\mbox{\scriptsize min}}^2$ in the difference flux $\mathcal{F}_{\mbox{\scriptsize meas}}$
      for PSF fitting with a scaling factor only is given by:
      \begin{equation}
      \sigma_{\mbox{\scriptsize min}}^2 = \frac{1}{P_{\mbox{\scriptsize true}}^{2}} 
                                          \left( \sum_{ij} \frac{\mathcal{P}_{\mbox{\scriptsize tar},ij}^2}{\sigma_{\mbox{\scriptsize in},\mbox{\scriptsize tar},ij}^2} \right)^{-1}
      \label{eqn:minsig}
      \end{equation}
      where $P_{\mbox{\scriptsize true}}$ is the true photometric scale factor ($P_{\mbox{\scriptsize true}} = 1$ in our simulations)
      and $\mathcal{P}_{\mbox{\scriptsize tar}}$ is the true PSF for the brightest star in the target image
      (a normalised two-dimensional Gaussian profile of FWHM $f_{\mbox{\scriptsize tar}}$
      pixels in our simulations). Since all of the stars in the simulations are non-variable,
      the best kernel and differential background solutions should yield a distribution of values of 
      $\mathcal{F}_{\mbox{\scriptsize meas}} / \sigma_{\mbox{\scriptsize min}}$ with zero mean and unit variance.
      Hence, for a set of $N_{\mbox{\scriptsize set}}$ simulations indexed by $k$, appropriate measures for assessing the photometric accuracy are:
      \begin{equation}
      \mbox{MPB} = \frac{1}{N_{\mbox{\scriptsize set}}} \sum_{k} \frac{\mathcal{F}_{\mbox{\scriptsize meas},k}}{\sigma_{\mbox{\scriptsize min},k}}
      \label{eqn:mean_fdiff}
      \end{equation}
      \begin{equation}
      \mbox{MPV} = \frac{1}{N_{\mbox{\scriptsize set}} - 1} \sum_{k} \left( \frac{\mathcal{F}_{\mbox{\scriptsize meas},k}}{\sigma_{\mbox{\scriptsize min},k}} - \mbox{MPB} \right)^{2}
      \label{eqn:var_fdiff}
      \end{equation}
      MPB is the mean photometric bias and MPV is the mean photometric variance with units of $\sigma_{\mbox{\scriptsize min}}$ and $\sigma_{\mbox{\scriptsize min}}^2$, respectively.
      We note that even though MPV is normalised by the theoretical minimum variance in the difference flux, it may still achieve values that
      are less than unity when the target image is over-fit and/or when the model PSF for the target image differs from the true PSF.
\end{itemize}

\subsection{Results}   
\label{sec:sim_results}

\begin{figure*}
\centering
\epsfig{file=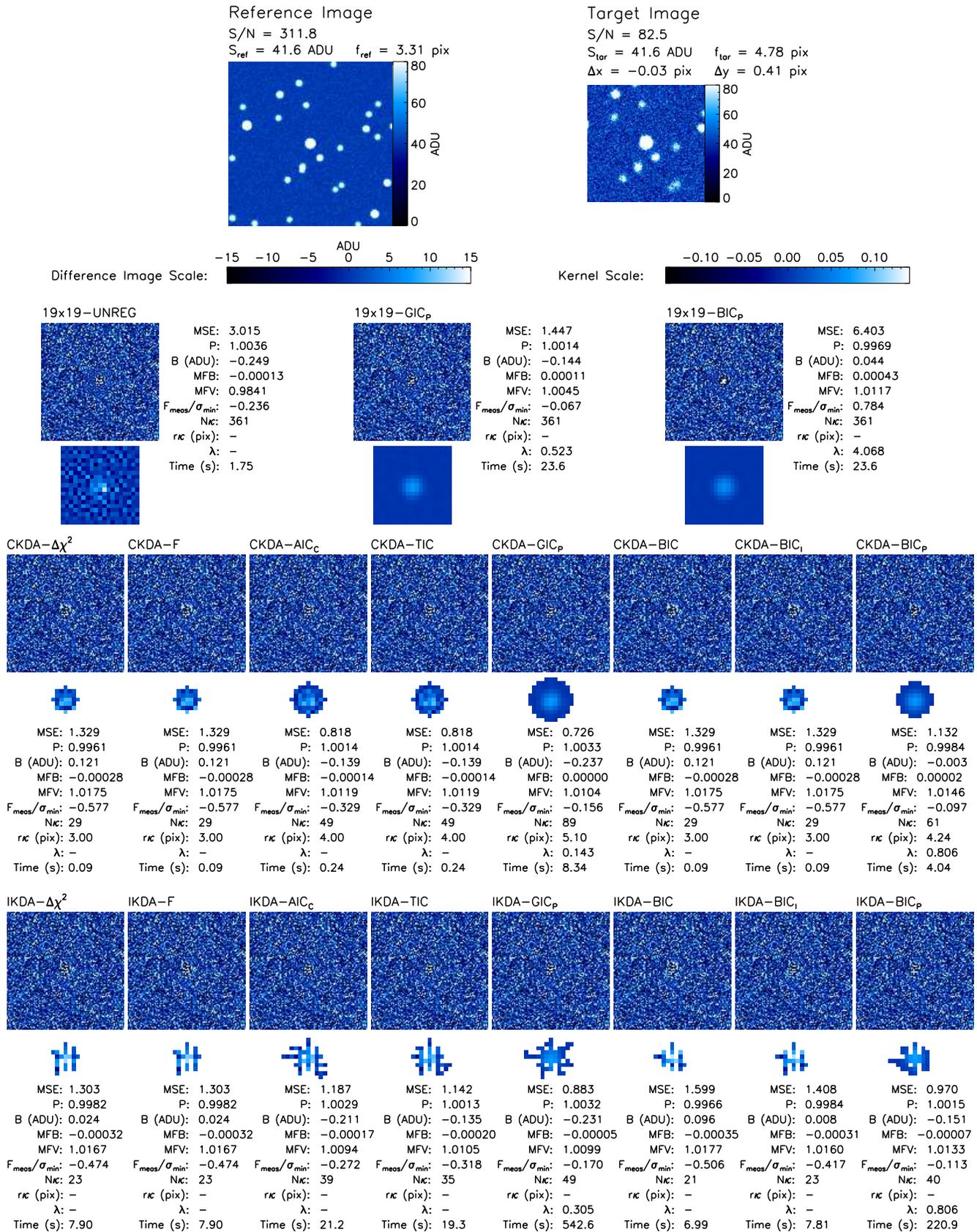,angle=0.0,width=\linewidth}
\caption{A reference and target image pair from simulation set S10 are shown at the top. The corresponding results for each of the 19 kernel solution methods
         are shown below. For each method, the difference image, kernel solution, and model performance metrics are displayed.
         The difference images and kernels are all displayed using the same linear scales of $[-15,15]$~ADU and $[-0.14,0.14]$, respectively.
         Processing times were measured for non-optimised code running on a desktop computer with an Intel Core i7-2600 CPU (3.40~GHz) and 16~Gb RAM.
         \label{fig:example_sim}}
\end{figure*}

\begin{figure*}
\centering
\epsfig{file=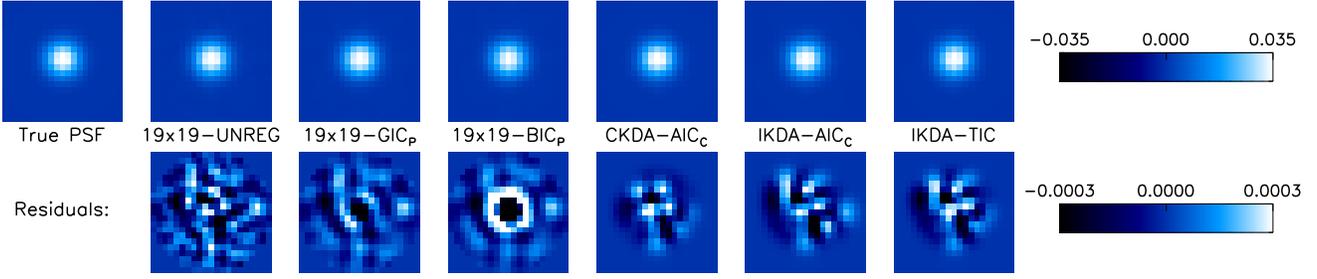,angle=0.0,width=\linewidth}
\caption{The true PSF for the brightest star in the target image from Figure~\ref{fig:example_sim} is shown at the top left. The true PSF is a normalised
         two-dimensional Gaussian of FWHM 4.78~pix centred in the image stamp using the sub-pixel coordinates of the brightest star.
         A selection of six model PSFs for the target image are shown in the top row and
         labelled with the corresponding kernel solution methods. The residuals of these model PSFs from the true PSF
         are shown in the bottom row. Each row of plots uses the linear scale reproduced at the right-hand end of the row.
         \label{fig:example_psfs}}
\end{figure*}

\begin{figure}
\centering
\begin{tabular}{c}
\subfigure[]{\epsfig{file=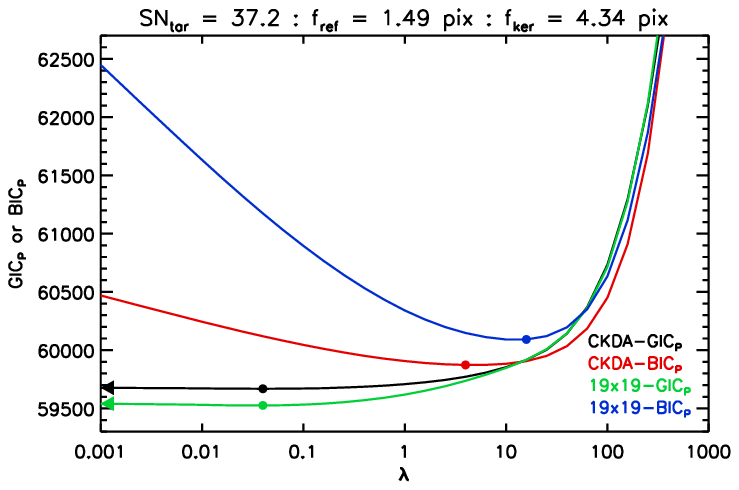,angle=0.0,width=\linewidth} \label{fig:lam_ex1}} \\
\subfigure[]{\epsfig{file=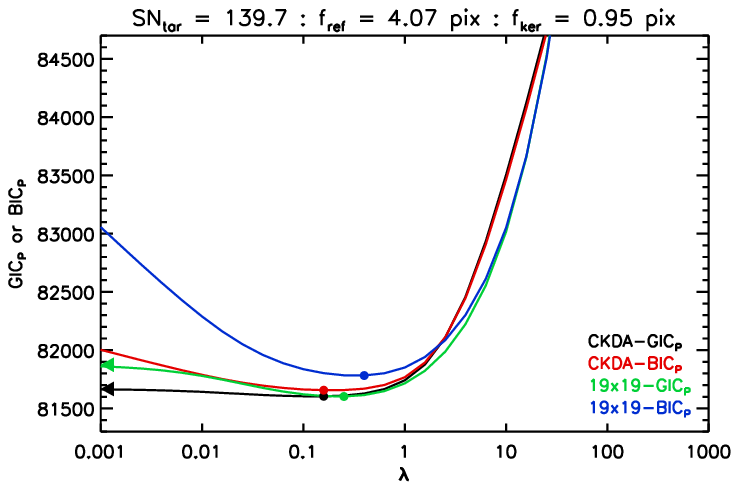,angle=0.0,width=\linewidth} \label{fig:lam_ex2}} \\
\end{tabular}
\caption{Examples from the simulations of the variation of the GIC$_{\mbox{\scriptsize P}}$ and BIC$_{\mbox{\scriptsize P}}$ criteria as a function of $\lambda$
         for the kernel solution methods CKDA-GIC$_{\mbox{\scriptsize P}}$ (black), CKDA-BIC$_{\mbox{\scriptsize P}}$ (red),
         19x19-GIC$_{\mbox{\scriptsize P}}$ (green), and 19x19-BIC$_{\mbox{\scriptsize P}}$ (blue). For the CKDA-GIC$_{\mbox{\scriptsize P}}$
         and CKDA-BIC$_{\mbox{\scriptsize P}}$ methods, the curves correspond to the selected kernel radius. For each curve, the minimum
         is marked with a solid circle. For the CKDA-GIC$_{\mbox{\scriptsize P}}$ and 19x19-GIC$_{\mbox{\scriptsize P}}$ methods, the value of the curve
         at $\lambda = 0$ is marked on the left-hand side of the plot with a triangle. Note that while GIC$_{\mbox{\scriptsize P}} (\lambda)$ converges to TIC
         for $\lambda \rightarrow 0$, BIC$_{\mbox{\scriptsize P}} (\lambda)$
         diverges as $\lambda \rightarrow 0$ because of the divergence of the term involving $\ln \lambda$ in equation~\ref{eqn:BICP_use}. Hence no triangles
         are plotted for the CKDA-BIC$_{\mbox{\scriptsize P}}$ and 19x19-BIC$_{\mbox{\scriptsize P}}$ methods.
         \label{fig:lam_ex}}
\end{figure}

For each possible combination of kernel design algorithm and model selection criterion, we computed kernel and differential background solutions for all of
the reference and target image pairs in both of the simulation sets S1 and S10. Furthermore, for comparison purposes,
for each simulation we solved for a model image employing a square 19x19-pixel kernel design
which corresponds to the unregularised kernel analysed in Be12. We also solved for the same 19x19-pixel kernel design with
regularised DBFs where the optimal choice of $\lambda$ was determined using either GIC$_{\mbox{\scriptsize P}} (\lambda)$ or BIC$_{\mbox{\scriptsize P}} (\lambda)$
(equations~\ref{eqn:GICP_use}~and~\ref{eqn:BICP_use}) which corresponds to the regularised kernel analysed in Be12. In all cases,
we used three iterations for each solution, but without employing sigma-clipping since the simulated images do not suffer from outlier pixel values (see Section~\ref{sec:solving_for_kernel}).
The optimisation of $\lambda$ for the GIC$_{\mbox{\scriptsize P}}$ and BIC$_{\mbox{\scriptsize P}}$ model selection criteria was performed using
a binary search algorithm in $\log(\lambda)$ for the range $-3~\le~\log(\lambda)~\le~3$ with a final resolution in $\lambda$ of 15\%, while also considering the limit $\lambda~=~0$.
Finally, the corresponding model performance metrics from Section~\ref{sec:mod_perf_metrics} were calculated for each solution.

Hereafter we use a string of the form \\ $<$ALGORITHM$>$-$<$CRITERION$>$ to refer to a specific combination of kernel design algorithm
(CKDA or IKDA) and model selection criterion ($\Delta \chi^{2}$, $F$, AIC$_{\mbox{\scriptsize C}}$, TIC, GIC$_{\mbox{\scriptsize P}}$,
BIC, BIC$_{\mbox{\scriptsize I}}$ or BIC$_{\mbox{\scriptsize P}}$). For the 19x19-pixel kernel design, we use 19x19-UNREG, 19x19-GIC$_{\mbox{\scriptsize P}}$,
and 19x19-BIC$_{\mbox{\scriptsize P}}$. Each of these combinations constitutes a kernel solution method, and hence we have 19 methods to consider.

In Figure~\ref{fig:example_sim}, we show the difference images, kernel solutions, and model performance metrics for each of the 19 kernel solution methods applied to
the reference and target image pair displayed at the top (taken from simulation set S10). The target image is of medium S/N, and the reference and target images are
both over-sampled (with $f_{\mbox{\scriptsize ker}} > 2.35$~pix). Notice how the regularisation in the 19x19-GIC$_{\mbox{\scriptsize P}}$ and 19x19-BIC$_{\mbox{\scriptsize P}}$ methods drastically
reduces the noise in the kernel compared to the 19x19-UNREG method as demonstrated previously in Be12. Notice also how, as expected, the BIC-type criteria
(BIC, BIC$_{\mbox{\scriptsize I}}$ and BIC$_{\mbox{\scriptsize P}}$) select kernel designs with fewer DBFs than the kernel designs selected by
the AIC-type criteria (AIC$_{\mbox{\scriptsize C}}$, TIC and GIC$_{\mbox{\scriptsize P}}$). Somewhat surprising is the ``spidery'' form of
the kernel solutions generated by the IKDA. A selection of model PSFs for this target image, used to perform PSF fitting on the difference images,
are displayed in the top row of Figure~\ref{fig:example_psfs} alongside the true PSF for the brightest star. The residuals of these model PSFs from the true PSF
(bottom row) demonstrate that the spidery form of the IKDA kernel solutions has no discernable detrimental effect, when compared to the other
kernel solutions, on the convolution of the reference image PSF to obtain the target image PSF.

To provide an idea of what the functional forms of GIC$_{\mbox{\scriptsize P}} (\lambda)$ and BIC$_{\mbox{\scriptsize P}} (\lambda)$ look like, we plot these quantities
as a function of $\lambda$ for two example simulations in Figure~\ref{fig:lam_ex}. Each plot shows the curves for the CKDA-GIC$_{\mbox{\scriptsize P}}$,
CKDA-BIC$_{\mbox{\scriptsize P}}$, 19x19-GIC$_{\mbox{\scriptsize P}}$ and 19x19-BIC$_{\mbox{\scriptsize P}}$ methods. Clear minima exist indicating the
optimal values of $\lambda$. All of the simulations yield similar functional forms for GIC$_{\mbox{\scriptsize P}} (\lambda)$ and BIC$_{\mbox{\scriptsize P}} (\lambda)$,
and while the minima of the GIC$_{\mbox{\scriptsize P}} (\lambda)$ curves may sometimes lie at $\lambda = 0$, they very rarely lie at values of $\lambda$ that are greater than 10 for GIC$_{\mbox{\scriptsize P}}$,
or that are greater than 100 for BIC$_{\mbox{\scriptsize P}}$.
Note that for the example shown in Figure~\ref{fig:lam_ex2}, the optimal value of $\lambda$ for each method lies in the range 0.1-1.0 which
matches with the recommendation for $\lambda$ from Be12. However, for the other example shown in Figure~\ref{fig:lam_ex1}, the GIC$_{\mbox{\scriptsize P}}$ and BIC$_{\mbox{\scriptsize P}}$
criteria yield optimal values of $\lambda$ that are $<$0.1 and $>$1.0, respectively.

\begin{figure*}
\centering
\epsfig{file=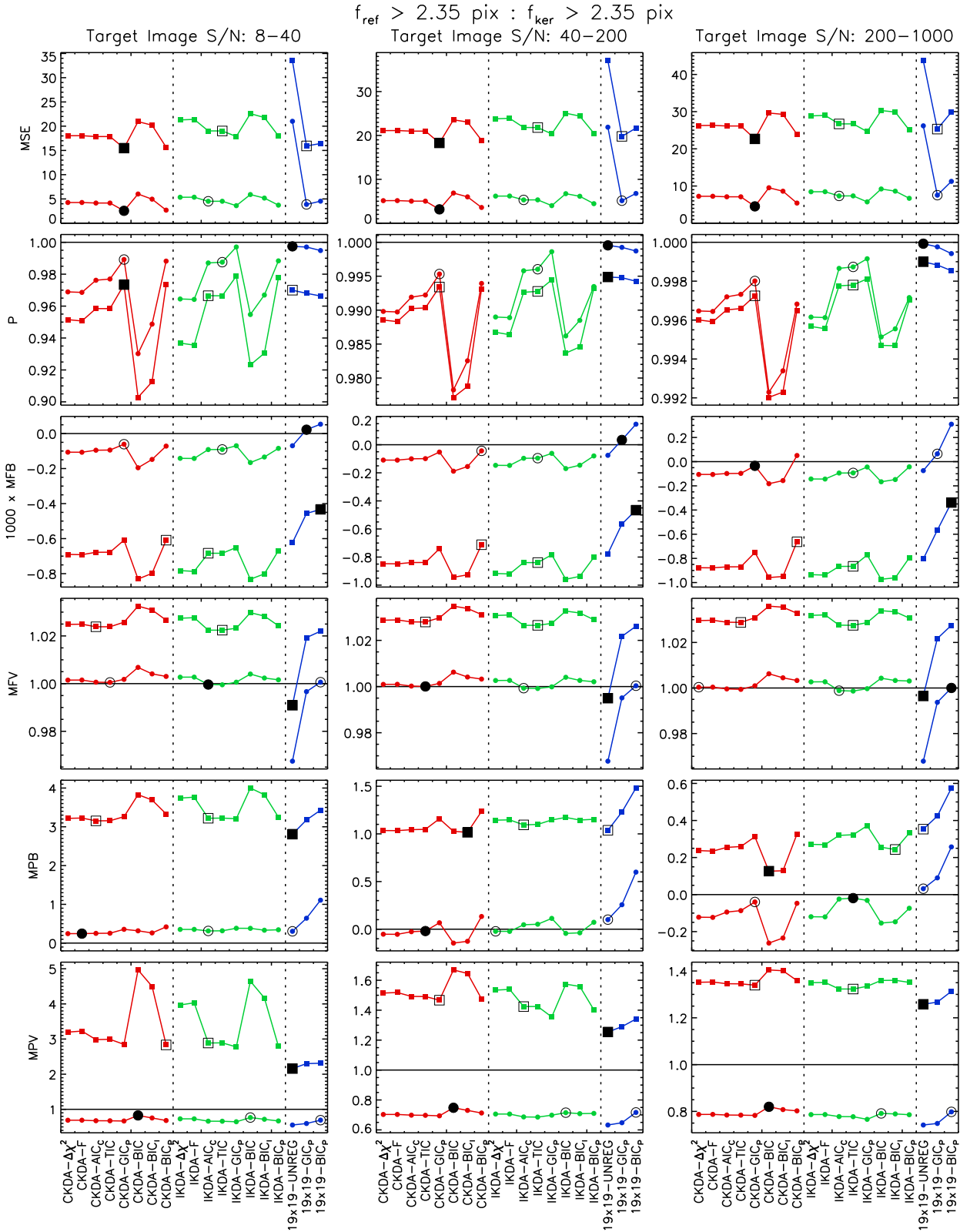,angle=0.0,width=\linewidth}
\caption{Plots of the median MSE, $P$, MFB, and MFV values (equations~\ref{eqn:mod_err}, \ref{eqn:phot_scale}, \ref{eqn:fit_bias} and \ref{eqn:fit_var}),
         and the MPB and MPV measures (equations~\ref{eqn:mean_fdiff}~and~\ref{eqn:var_fdiff}), for each kernel solution method for $f_{\mbox{\scriptsize ref}} \ge 2.35$~pix
         and $f_{\mbox{\scriptsize ker}} \ge 2.35$~pix. The results in each plot have been calculated from $\sim$60000 simulations for each of the simulation sets S1 and S10.
         {\bf Layout:} The three columns of plots correspond to low (8-40), medium (40-200) and high (200-1000) S/N target images. Each row of plots corresponds
         to a different model performance metric.
         {\bf Individual plots:} Square and circular symbols represent the results for simulation sets S1 and S10, respectively. Red, green, and blue colours correspond to the
         kernel design algorithms CKDA, IKDA, and 19x19, respectively. For each algorithm, the kernel solution method with the best
         value of the relevant model performance metric is also plotted with an open black symbol. The method with the overall best metric value is plotted with a solid black symbol.
         The IKDA-GIC$_{\mbox{\scriptsize P}}$ and IKDA-BIC$_{\mbox{\scriptsize P}}$ methods are excluded when determining the best metric values since
         their results are noisier having been determined from many fewer simulations, and because they are too computationally intensive to be of   
         practical use with currently available computing equipment.
         \label{fig:results_reg5}}
\end{figure*}

In Figure~\ref{fig:results_reg5}, for each kernel solution method we plot the median MSE, $P$, MFB, and MFV values, and the MPB and MPV measures,
for a subset of our simulations corresponding to over-sampled reference images ($f_{\mbox{\scriptsize ref}} \ge 2.35$~pix) and kernels with $f_{\mbox{\scriptsize ker}} \ge 2.35$~pix.
The corresponding plots for $B$ are not presented because the results are very similar to those for $P$ since the photometric scale factor and differential
background are correlated. We have further split the simulations into subsets based on target image S/N (low: 8-40, medium: 40-200, high: 200-1000; three columns of plots)
and reference image S/N (simulation sets S1 and S10; square or circular symbols). Similar style figures representing the results for different subsets of simulations chosen based
on image sampling are presented in Appendix~B (Figures~\ref{fig:results_reg1}~-~\ref{fig:results_reg4}).

Within each subset of simulations, the distributions of the various model performance metrics are single-peaked bell-shapes with rapidly falling wings and
they are not far-off being Gaussian in some cases. Skewness affects some of the distributions as do a few outlier points.
However, for each simulation subset and model performance metric, the shape and width of the distributions are very similar between the kernel solution methods.
The differences in the distributions lie in their central values. Consequently we have used the median of the model performance metrics MSE, $P$, MFB and MFV in
Figures~\ref{fig:results_reg5} and~\ref{fig:results_reg1}~-~\ref{fig:results_reg4} to compare the kernel solution methods since the median is a robust
estimator of the central value. Given the Gaussian-like shape of the distributions of $\mathcal{F}_{\mbox{\scriptsize meas}} / \sigma_{\mbox{\scriptsize min}}$,
our choice of measures MPB and MPV (equations~\ref{eqn:mean_fdiff}~and~\ref{eqn:var_fdiff}) for assessing the photometric accuracy is justified.

The processing time to run the IKDA-GIC$_{\mbox{\scriptsize P}}$ and IKDA-BIC$_{\mbox{\scriptsize P}}$ methods is prohibitive (see the timings noted in Figure~\ref{fig:example_sim}).
Hence we only ran these kernel solution methods on 25,410 and 25,320 reference and target image pairs from simulation sets S1 and S10, respectively. The results from these methods
are plotted in Figures~\ref{fig:results_reg5} and~\ref{fig:results_reg1}~-~\ref{fig:results_reg4}, although they suffer from more noise than the other methods because they
are derived from many fewer simulations. Consequently we do not consider these two kernel solution methods any further.

\subsection{Discussion}
\label{sec:sim_disc}

Unless otherwise stated, the discussion in this section refers to the results plotted in all of the Figures~\ref{fig:results_reg5} and~\ref{fig:results_reg1}~-~\ref{fig:results_reg4},
while Figure~\ref{fig:results_reg5} alone is sufficient to demonstrate the points raised.

In preparation for our discussion, it is worth considering how closely the candidate model images generated by our kernel design algorithms
are able represent the true underlying model image. In each simulation, the Gaussian PSF profile in the reference (or target) image is convolved with a Gaussian kernel
to obtain a Gaussian PSF profile in the target (or reference) image. In the limits of a noiseless reference image with infinitely fine image sampling, and for a kernel that
convolves the reference image, a kernel of DBFs of infinite extent is sufficient to allow for a full representation of the true underlying model image (i.e. the noiseless target image).
In practice, the reference image is noisy, the reference and target images are sampled at a finite scale with a spatial offset between them,
the target image may be sharper than the reference image, and the kernel model
employs a finite number of DBFs. It is clear therefore that none of the candidate model images will actually represent the true model image. However,
for reference images with higher S/N and better sampling, and for kernel models employing more DBFs (without over-fitting), the candidate model images will include
models that are closer to the true model. Referring back to Section~\ref{sec:model_selection_criteria_ic_ml}, it seems then that the model selection criteria derived
considering the Kullback-Leibler divergence (i.e. the AIC-type criteria) should perform the best for DIA (especially in terms of model error), and that all of the criteria
should perform better with improved reference image S/N and sampling.

Unsurprisingly then, the first major conclusion that can be drawn from the results of the simulations is that with very few exceptions it is vastly advantageous,
as demonstrated by all of the model performance metrics, to use a reference image with a higher S/N than the target image regardless of the target image S/N,
the reference or target image FWHM, or the kernel solution method employed. Our discussion will therefore focus on the results for simulation set S10. Furthermore,
the best estimates for the photometric scale factor are achieved for higher S/N target images, and that in general $P$ is under-estimated.
Since an accurate estimate of the photometric scale factor is crucial for performing accurate photometry (\citealt{bra2015}),
our discussion will further focus on the results for target-image S/N ranges of 40-200 and 200-1000 where $P$ is estimated to better than 1\% for simulation set S10.

We observe that the smallest median MSE values for simulation set S10 are always achieved by a kernel solution method employing an AIC-type criterion.
What differs between the S/N and FWHM regimes is which algorithm combined with an AIC-type criterion performs the best in terms of model error. This result validates our discussion at the
beginning of this section about the fact that the set of candidate model images generated by our kernel design algorithms does not include the true underlying model.

We find that the median MFB values, which have units of sigma, are very small negative numbers with absolute values less than $\sim$3$\times$10$^{-3}~\sigma_{\mbox{\scriptsize min}}$
(simulation set~S10). Hence in terms of fit quality, we give more weight to the results for the MFV metric.

The results for the photometric accuracy in Figure~\ref{fig:results_reg5} reveal the surprising fact that the variance in the photometry of the brightest star for simulation set S10
is smaller than the theoretical minimum $\sigma_{\mbox{\scriptsize min}}^2$. This can also be seen in Figures~\ref{fig:results_reg2}~-~\ref{fig:results_reg4}.
We have investigated how this might be possible. Firstly we checked the photometry of the
faintest star in each simulation. We did this by modifying step~(iv) of the image simulation procedure in Section~\ref{sec:simim}
to shift the pixel coordinates of the faintest star (instead of the brightest) to lie within the central pixel of the reference image.
We then generated 60,000 reference and target image pairs with the $\sigma_{\mbox{\scriptsize in},\mbox{\scriptsize ref},ij}$ in step (viii) scaled by $10^{-1/2} \sim 0.316$
as was done for simulation set~S10, and for each simulated image pair, we computed $\mathcal{F}_{\mbox{\scriptsize meas}} / \sigma_{\mbox{\scriptsize min}}$ for the faintest
star for each kernel solution method. We found that the MPV measures for the faintest star are greater than unity with values in the range 
$\sim$1.1-1.6~$\sigma_{\mbox{\scriptsize min}}^{2}$ for Figures~\ref{fig:results_reg5} and~\ref{fig:results_reg2}~-~\ref{fig:results_reg4}.
Hence the variance in the photometry of the faintest stars does not achieve the theoretical minimum $\sigma_{\mbox{\scriptsize min}}^2$.

We believe that these facts may be explained by considering that all of the kernel solution methods are over-fitting the brightest star(s) and under-fitting the faintest star(s)
in most of our simulations. Careful inspection of the difference images in Figure~\ref{fig:example_sim} reveals that the pixel noise in the area around the brightest star is
suppressed\footnote{The effect is more easily discernable on a digital display than on a printed copy.}, and one can see
that this effect is visible to various extents for all of the kernel solution methods including the methods employing regularisation. Furthermore, the noise suppression around the
brightest star is clearly visible in figures~2~and~3 of Be12. We experimented with using a noise model in our simulations with equal pixel variances calculated using the sky background level
only. In this case, the target-image pixel values for the brightest star are given even more weight relative to the other pixels when solving for the kernel and differential background,
and the MPV measures for the brightest star are found to be even smaller. However, if we increase the size of the images
in our simulations, the MPV measures that are smaller than unity in Figures~\ref{fig:results_reg5} and~\ref{fig:results_reg2}~-~\ref{fig:results_reg4}
are found to increase to values that are closer to unity. The same effect may also be achieved by considering only those simulations
with higher star densities.

From this we may conclude that the kernel solution methods which yield MPV measures that are closest to unity, regardless of whether they
are greater than or less than unity, are those that achieve the best balance between under- or over-fitting the target image for the brightest stars. Consequently
it is these methods that produce the most reliable photometry whenever the corresponding MPB measures are also closest to zero. Also, it is clear that in practice
the image regions used to derive the kernel solutions should be as large as possible while satisfying the assumption of a spatially-invariant kernel,
and that ideally they should each contain at least a few bright objects. This helps to avoid the situation where a single bright object dominates the kernel solution in
each region.

Based on the above general observations and discussion, we have attempted a detailed analysis of the results presented in Figures~\ref{fig:results_reg5} and~\ref{fig:results_reg1}~-~\ref{fig:results_reg4}.
However, it has proven impossible to identify any single kernel solution method, or even an individual algorithm or criterion, that consistently performs the best for all of the
model performance metrics. Even breaking the analysis down into each of the five image sampling regimes does not help much.
Since we are unable to reach a clear conclusion from the way the results have been analysed and presented so far, further investigation is required.

Finally, we checked how well the kernel solutions recover the input sub-pixel shifts between the reference and target images. To do this,
we computed the centroid of each kernel solution and compared the centroid coordinates to the appropriate sub-pixel shifts. Reassuringly, we find
that the residuals are scattered around zero with decreasing scatter for higher S/N target images. Furthermore, we note that while all of the CKDA and IKDA methods
perform equally well in recovering the shifts (e.g. $\sim$0.015~pix rms at SN$_{\mbox{\scriptsize tar}} \sim 300$), the 19x19-UNREG, 19x19-GIC$_{\mbox{\scriptsize P}}$, and
19x19-BIC$_{\mbox{\scriptsize P}}$ methods all perform considerably worse in this respect (e.g. $\sim$0.043~pix rms at SN$_{\mbox{\scriptsize tar}} \sim 300$).

\subsection{Further Investigation}
\label{sec:sim_fur_inv}

\begin{figure*}
\centering
\epsfig{file=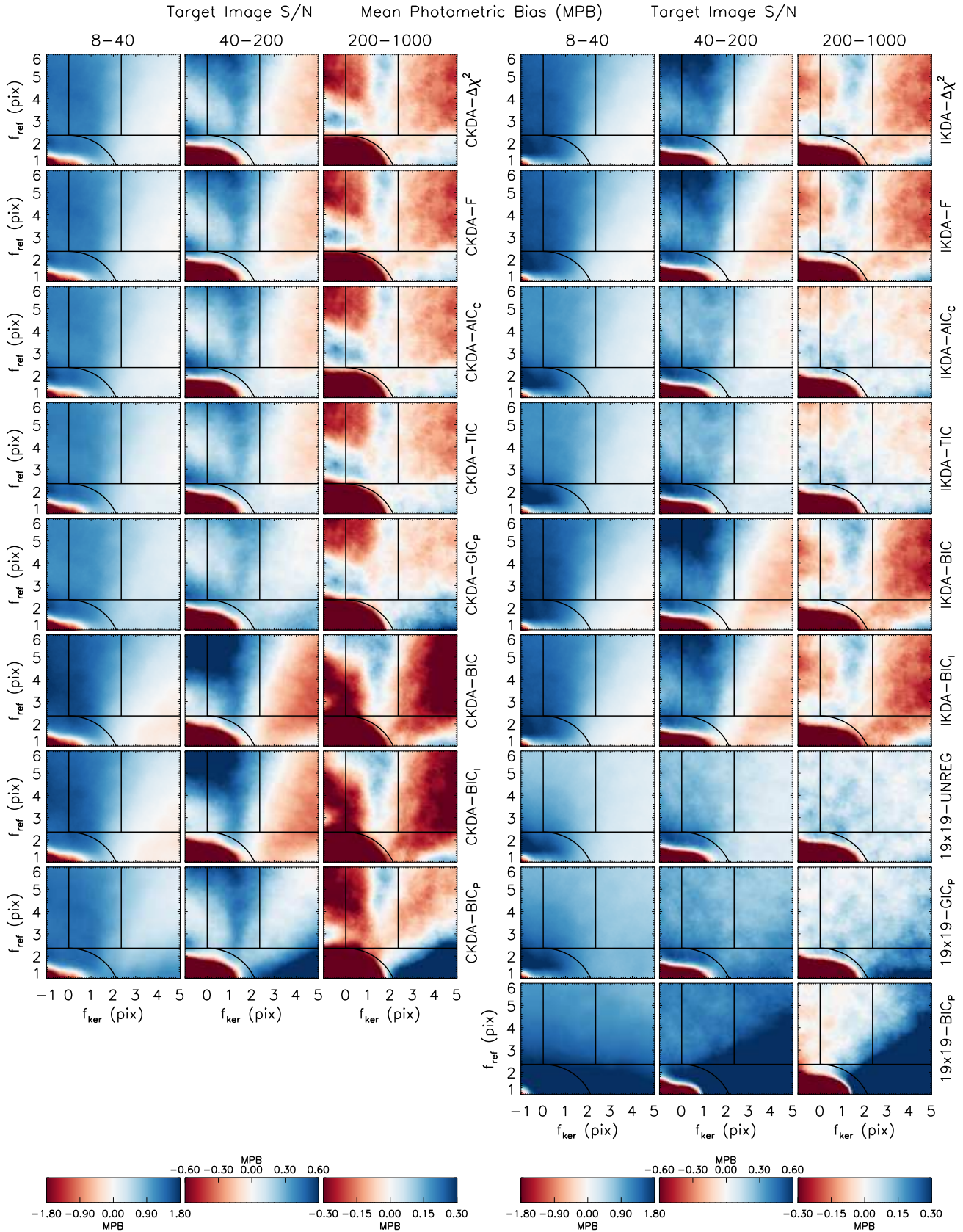,angle=0.0,width=0.99\linewidth}
\caption{Plots of surfaces representing the MPB measure (equation~\ref{eqn:mean_fdiff}) for simulation set S10 as a function of the reference image and kernel FWHM.
         Each plot corresponds to a specific kernel solution method (labelled on the right-hand side of each row of three plots)
         and target-image S/N regime (labelled at the top of each column of plots).
         The colour-intensity bar corresponding to each column of plots is reproduced at the bottom of the figure.
         In each plot, the surface values are calculated using a circular smoothing region of radius 0.33~pix and the
         image sampling regimes corresponding to Figures~\ref{fig:results_reg5} and~\ref{fig:results_reg1}~-~\ref{fig:results_reg4}
         are delimited by continuous black lines. Specifically, the curved line corresponds to a critically sampled target image for
         an under-sampled reference image (i.e. $f_{\mbox{\scriptsize tar}} = 2.35$ and $f_{\mbox{\scriptsize ref}} \le 2.35$~pix).
         \label{fig:results_mpb_map}}
\end{figure*}

\begin{figure*}
\centering
\epsfig{file=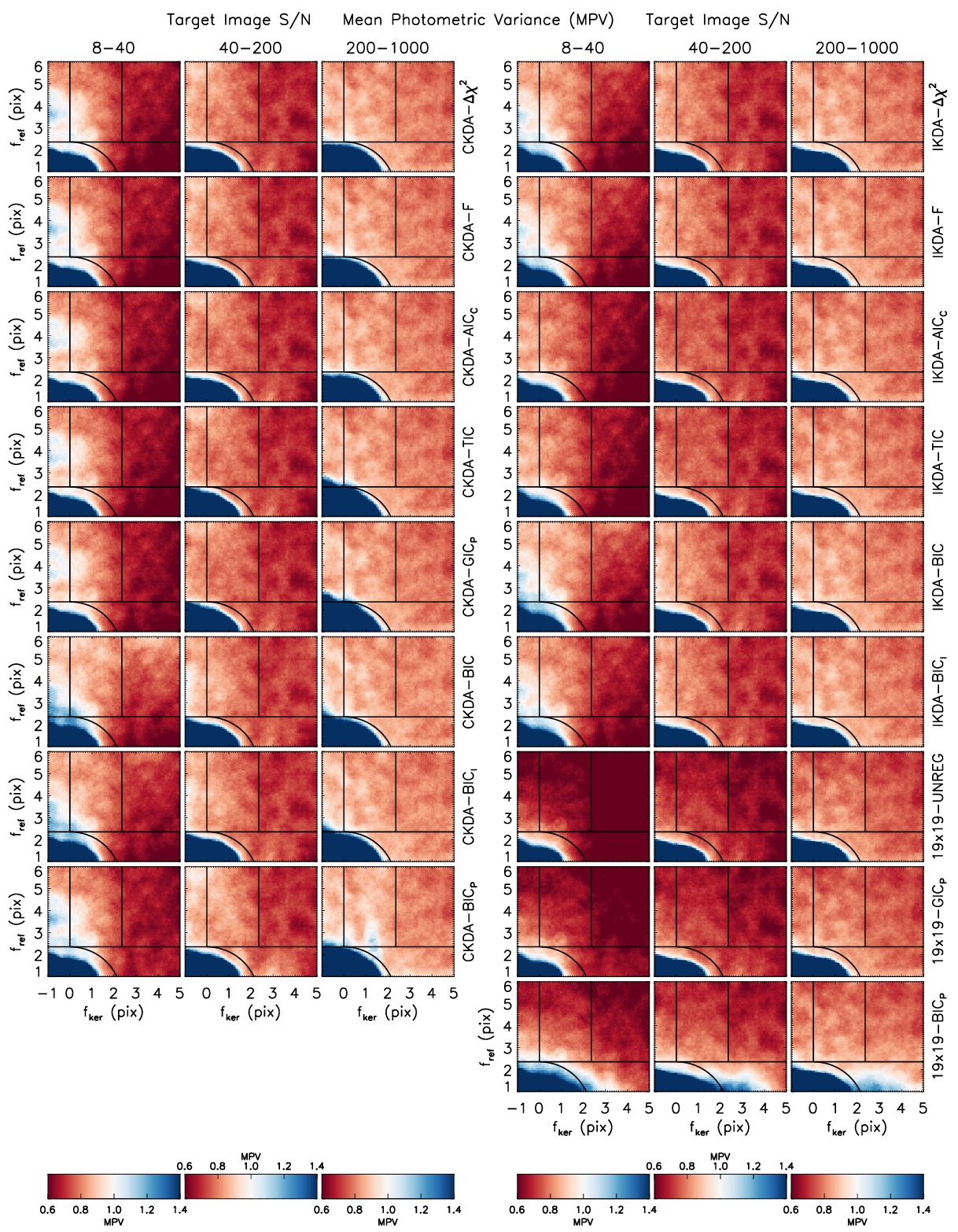,angle=0.0,width=0.99\linewidth}
\caption{Plots of surfaces representing the MPV measure (equation~\ref{eqn:var_fdiff}) for simulation set S10 as a function of the reference image and kernel FWHM.
         The format of the figure is the same as in Figure~\ref{fig:results_mpb_map}.
         \label{fig:results_mpv_map}}
\end{figure*}

One of the most important applications of DIA is for precision photometry. Therefore, we are highly motivated in developing a kernel solution method
that provides the best possible photometry in the sense that the chosen method should deliver the smallest photometric bias while also
striking the optimal balance between under- and over-fitting. In photometric applications, DIA is typically used to obtain photometry for
the objects in a set of time-series images. The properties of these images, such as the PSF FWHM and S/N, will likely vary substantially
during the course of the observations (e.g. due to the atmosphere).
Therefore, a further aspect on which we may assess the kernel solution methods studied in this paper
is on the uniformity of the photometric bias and variance as a function of FWHM and S/N.

In Figure~\ref{fig:results_mpb_map}, for each kernel solution method and target-image S/N regime, we plot
surfaces representing the MPB measure for simulation set S10 as a function of the reference image and kernel FWHM.
A circular smoothing region of radius 0.33~pix, which encompasses
the results from $\sim$2000 simulations, is used to calculate the MPB surface values.
Blue and red colours in the plot panels indicate positive and negative mean photometric biases, respectively,
while white indicates zero bias. The image sampling regimes corresponding to Figures~\ref{fig:results_reg5} and~\ref{fig:results_reg1}~-~\ref{fig:results_reg4}
are delimited in each plot by continuous black lines. These plots are complimentary to those in Figures~\ref{fig:results_reg5} and~\ref{fig:results_reg1}~-~\ref{fig:results_reg4}
in the sense that they reveal considerably more detail about the dependence of the MPB measure on image sampling, even though it is more difficult to assess
the exact MPB values in each case.

Since the DIA photometry for a set of time-series images is extracted using a single reference image, the uniformity of the MPB surfaces in
Figure~\ref{fig:results_mpb_map} should be assessed via horizontal cuts (i.e. at fixed $f_{\mbox{\scriptsize ref}}$), and by comparing the cuts between
the plot columns (for S/N dependence). Immediately it is clear
to the eye that the IKDA-AIC$_{\mbox{\scriptsize C}}$, IKDA-TIC, 19x19-UNREG and 19x19-GIC$_{\mbox{\scriptsize P}}$ methods produce by far the
most uniform and least biased MPB surfaces in the S/N$>$40 regime with the IKDA-AIC$_{\mbox{\scriptsize C}}$ and IKDA-TIC methods yielding nearly identical results.
Specifically, in the over-sampled reference image regime and for S/N$>$200, each of these four methods suffers from only a small photometric bias
(MPB~$\sim \pm 0.1 \; \sigma_{\mbox{\scriptsize min}}$), with the two IKDA methods showing both positive and negative biases, and the two 19x19-pixel kernel designs
showing just a positive bias. Between S/N regimes, the IKDA-AIC$_{\mbox{\scriptsize C}}$, IKDA-TIC and 19x19-UNREG methods are relatively uniform whereas
the 19x19-GIC$_{\mbox{\scriptsize P}}$ method exhibits more noticeable non-uniformity.
In the under-sampled reference image regime with over-sampled target images, the IKDA-AIC$_{\mbox{\scriptsize C}}$, IKDA-TIC and 19x19-UNREG
methods produce very similar uniform MPB surfaces with a slight positive bias of $\sim 0.1 \; \sigma_{\mbox{\scriptsize min}}$, while the 19x19-GIC$_{\mbox{\scriptsize P}}$ method
shows significant non-uniformity. In the under-sampled reference and target image regime, the MPB surfaces for all 17 kernel solution methods have large gradients indicating that
the time-series photometry will suffer from large systematic variations whenever there are small variations in $f_{\mbox{\scriptsize ker}}$ between the images (unsurprisingly).
For S/N$>$200, the four kernel solution methods identified here provide uniform MPB surfaces down to $f_{\mbox{\scriptsize ref}} \sim 2.1$ and $f_{\mbox{\scriptsize tar}} \sim 2.1$ pix (slightly
below critical sampling).

The surfaces representing the MPV measure for simulation set S10 as a function of the reference image and kernel FWHM, created in the same fashion as the MPB surfaces,
are displayed in Figure~\ref{fig:results_mpv_map}. They clearly show that all of the kernel solution methods yield a mean photometric variance that
is smaller than the theoretical minimum $\sigma_{\mbox{\scriptsize min}}^{2}$ (except in the under-sampled reference and target image regime).
In Section~\ref{sec:sim_disc}, we came to the conclusion that this is because the kernel solution methods are over-fitting the brightest star(s).
The 19x19-UNREG method is the worst performer in this respect, followed closely by 19x19-GIC$_{\mbox{\scriptsize P}}$ and then 19x19-BIC$_{\mbox{\scriptsize P}}$.
Otherwise, the MPV surfaces for the CKDA and IKDA are all very similar with the IKDA providing more uniform photometric variance near the locus of critically sampled
target images. The MPV surfaces for the IKDA-AIC$_{\mbox{\scriptsize C}}$ and IKDA-TIC methods are virtually indistinguishable.

Hence, we may conclude that the IKDA-AIC$_{\mbox{\scriptsize C}}$ and IKDA-TIC methods are equally the best kernel solution methods in terms of the photometry that they yield,
with the 19x19-GIC$_{\mbox{\scriptsize P}}$ method coming in as a close second best. The plots of the equivalent surfaces for the remaining model performance metrics for simulation set S10,
reproduced in Appendix~C, also support this conclusion. Briefly, Figure~\ref{fig:results_mse_map} clearly demonstrates the very poor performance of the 19x19-UNREG method in terms of model error.
The uniformity and accuracy of the estimated photometric scale factor as a function of FWHM and S/N is also important for obtaining time-series photometry that is free from systematic errors,
and the best methods in this respect are 19x19-UNREG and
19x19-GIC$_{\mbox{\scriptsize P}}$, followed closely by IKDA-AIC$_{\mbox{\scriptsize C}}$, IKDA-TIC, and 19x19-BIC$_{\mbox{\scriptsize P}}$ (Figure~\ref{fig:results_p_map}).
In fact, the MPB surfaces (Figure~\ref{fig:results_mpb_map}) show considerable correlation with the $P$ surfaces, which further highlights the importance of obtaining a precise and unbiased estimate of the 
photometric scale factor in order to obtain precise and unbiased photometry (\citealt{bra2015}). For the MFB surfaces (Figure~\ref{fig:results_mfb_map}), which also correlate
somewhat with the MPB surfaces, the most uniform and least biased methods
are 19x19-UNREG and 19x19-GIC$_{\mbox{\scriptsize P}}$, followed closely by IKDA-AIC$_{\mbox{\scriptsize C}}$, IKDA-TIC, and 19x19-BIC$_{\mbox{\scriptsize P}}$.
The MFV surfaces in Figure~\ref{fig:results_mfv_map} further demonstrate the target-image over-fitting by the 19x19-UNREG and 19x19-GIC$_{\mbox{\scriptsize P}}$ methods.

Finally, we highlight the fact that all 19 of the kernel solution methods that we have tested show complicated (and different) functional dependencies of the
model performance metrics on PSF FWHM and S/N in each of the reference and target images. This has made it far more difficult to identify the
best performing methods than was originally anticipated.

\subsection{Optimal Values Of $\lambda$}
\label{sec:opt_lam}

\begin{figure*}
\centering
\epsfig{file=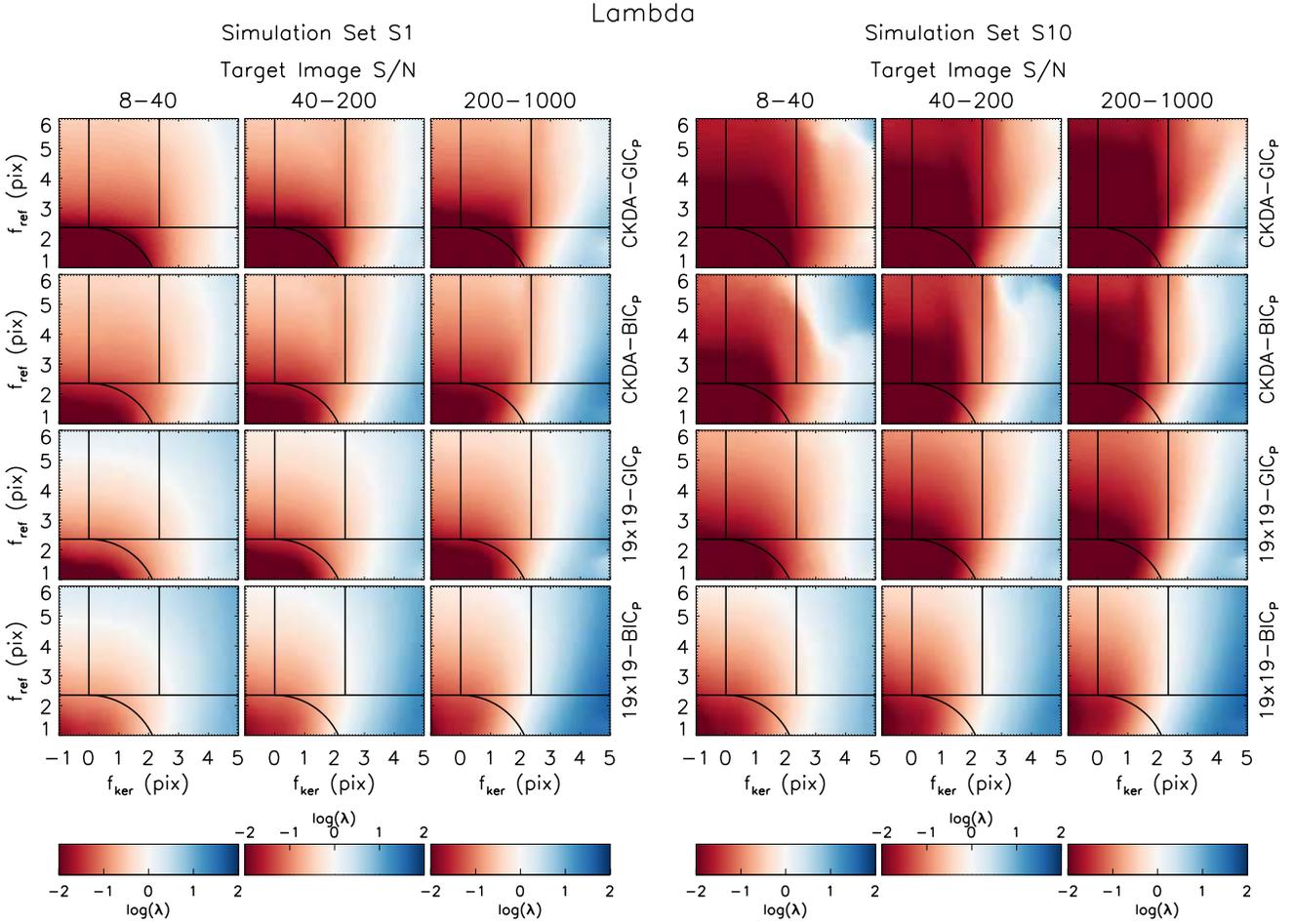,angle=0.0,width=\linewidth}
\caption{Plots of surfaces representing the log of the median $\lambda$ values for simulation sets S1 (left-hand side) and S10 (right-hand side) as a
         function of the reference image and kernel FWHM. Each plot corresponds to a specific kernel solution method (labelled on the right-hand side of each row of three plots)
         and target-image S/N regime (labelled at the top of each column of plots). Otherwise, the format of the figure is the same as in Figure~\ref{fig:results_mpb_map}.
         \label{fig:results_lam_map}}
\end{figure*}

Be12 recommend values of $\lambda$ between 0.1 and 1 for regularised 19$\times$19-pixel kernels although they caution that the optimal value
will be PSF FWHM and S/N dependent. Four of the kernel solution methods that we have tested employ regularised DBFs
(CKDA-GIC$_{\mbox{\scriptsize P}}$, CKDA-BIC$_{\mbox{\scriptsize P}}$, 19x19-GIC$_{\mbox{\scriptsize P}}$ and 19x19-BIC$_{\mbox{\scriptsize P}}$)
where the optimal value of lambda is selected using either the GIC$_{\mbox{\scriptsize P}}$ or BIC$_{\mbox{\scriptsize P}}$ criteria.
In Figure~\ref{fig:results_lam_map}, for each of these four kernel solution methods, and for each target-image S/N regime, we plot surfaces
representing the log of the median $\lambda$ values for both simulation sets S1 (left-hand side) and S10 (right-hand side) as a function of the reference image and kernel
FWHM. A circular smoothing region of radius 0.33~pix, which encompasses the results from $\sim$2000 simulations, is used to calculate the
$\log(\lambda)$ surface values. 

These surfaces clearly show that the optimal value of $\lambda$, when selected by either the GIC$_{\mbox{\scriptsize P}}$ or BIC$_{\mbox{\scriptsize P}}$ criteria,
is highly correlated with the PSF FWHM and S/N in each of the reference and target images, and that it is further dependent on the kernel design algorithm employed
(i.e. each plot panel in Figure~\ref{fig:results_lam_map} shows a different non-uniform surface). The BIC$_{\mbox{\scriptsize P}}$ criterion selects values of $\lambda$ that are greater than
those selected by the GIC$_{\mbox{\scriptsize P}}$ criterion (i.e. the BIC$_{\mbox{\scriptsize P}}$ criterion favours stronger kernel regularisation),
and the GIC$_{\mbox{\scriptsize P}}$ or BIC$_{\mbox{\scriptsize P}}$ criteria rarely select values of $\lambda$ that are greater than 10 or 100, respectively.
The general trends for $\lambda$ are that higher S/N reference images require weaker kernel regularisation, higher S/N target images require
weaker regularisation for $f_{\mbox{\scriptsize ker}}~<~2.35$~pix but stronger regularisation for $f_{\mbox{\scriptsize ker}}~>~2.35$~pix,
and the larger the values of $f_{\mbox{\scriptsize ref}}$ and $f_{\mbox{\scriptsize ker}}$, the stronger the required regularisation.
We conclude that the optimal regularisation of the kernel for any particular kernel solution method is highly data set dependent and that it
should be determined independently for each target image. We caution that the optimal value of $\lambda$, at least according to the GIC$_{\mbox{\scriptsize P}}$
or BIC$_{\mbox{\scriptsize P}}$ criteria, may lie in a very large range $0~\le~\lambda~\le~100-1000$.

\subsection{IKDA: Number Of Delta Basis Functions}
\label{sec:ndbfs}

\begin{figure*}
\centering
\epsfig{file=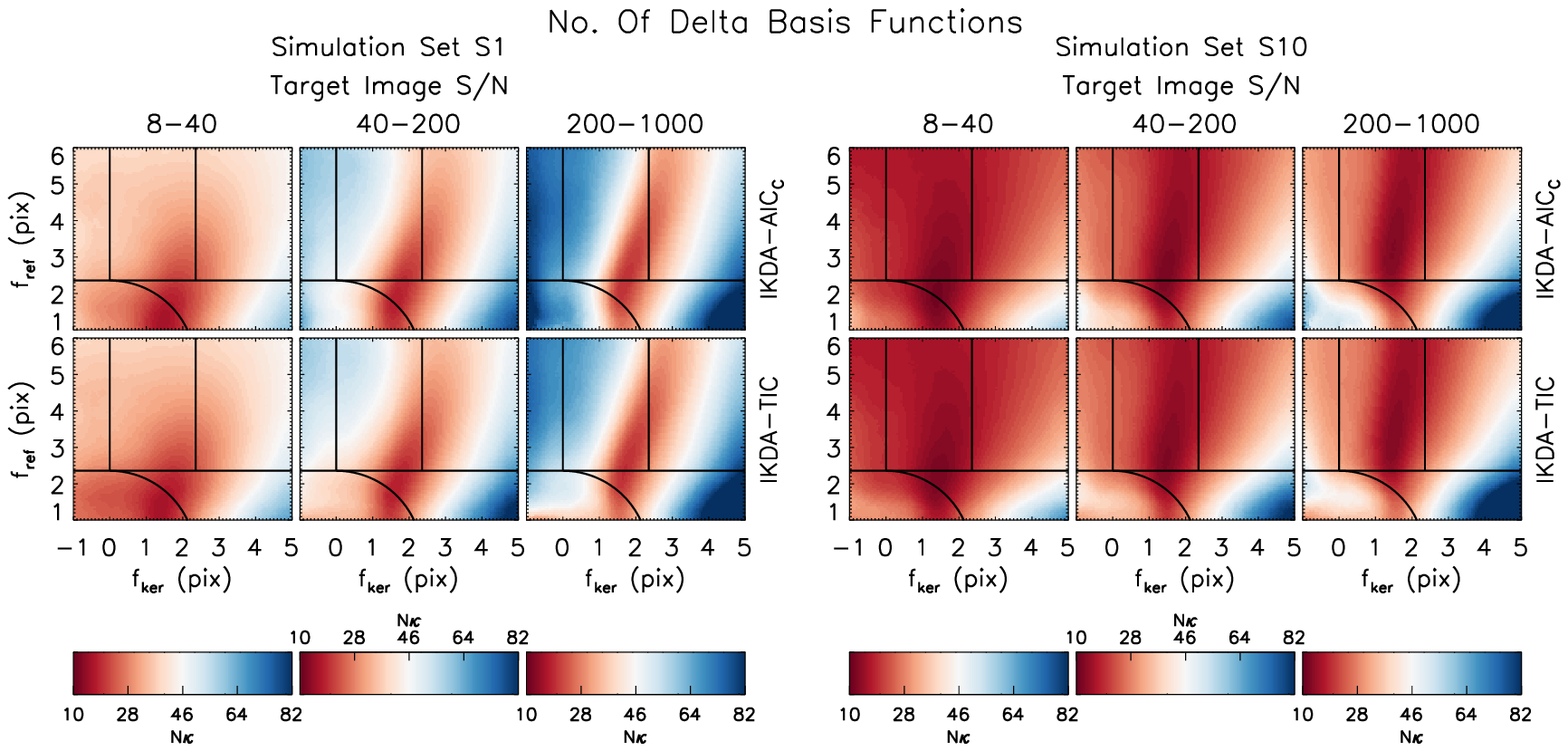,angle=0.0,width=\linewidth}
\caption{Plots of surfaces representing the median number of selected DBFs for simulation sets S1 (left-hand side) and S10 (right-hand side) as a
         function of the reference image and kernel FWHM. Each plot corresponds to a specific kernel solution method (labelled on the right-hand side of each row of three plots)
         and target-image S/N regime (labelled at the top of each column of plots). Otherwise, the format of the figure is the same as in Figure~\ref{fig:results_mpb_map}.
         \label{fig:results_nk_map}}
\end{figure*}

In Section~\ref{sec:sim_fur_inv}, we concluded that the IKDA-AIC$_{\mbox{\scriptsize C}}$ and IKDA-TIC methods are equally the best kernel solution methods in
terms of photometric accuracy. Therefore we are interested in the properties of the kernel designs that they generate, one of which is
the number of delta basis functions $N_{\kappa}$ that are selected. In Figure~\ref{fig:results_nk_map}, for each of these two kernel solution methods,
and for each target-image S/N regime, we plot surfaces
representing the median $N_{\kappa}$ values for both simulation sets S1 (left-hand side) and S10 (right-hand side)
as a function of the reference image and kernel FWHM. As usual, a circular smoothing region of radius 0.33~pix is used to calculate the $N_{\kappa}$ surface values.

These surfaces show a reasonably complicated dependence on PSF FWHM and S/N in each of the reference and target images, although they are very
similar between the two kernel solution methods. The general trends for $N_{\kappa}$ are that higher S/N reference images require less DBFs,
higher S/N target images show larger variations in $N_{\kappa}$ as a function of $f_{\mbox{\scriptsize ref}}$ and $f_{\mbox{\scriptsize ker}}$,
and that all of the surfaces show a minimum trough of approximately the same shape at a similar position. Similar to $\lambda$, we find that the number of selected DBFs
is highly data set dependent and varies over a large range ($10~\la~N_{\kappa}~\la~150$).

\section{Application To Real Images}
\label{sec:realsec}

So far we have employed simulated image data to explore the performance of the proposed kernel solution methods for a
wide range of reference and target image properties, and we have identified the best performing methods in terms
of the photometry that they yield (IKDA-AIC$_{\mbox{\scriptsize C}}$, IKDA-TIC, and 19x19-GIC$_{\mbox{\scriptsize P}}$).
However, the task remains to test the kernel solution methods on real image data to check the validity of the conclusions
from the simulations. While it is not possible to cover the full range of reference and target image properties using real data
in the same systematic and uniform way as it is possible to do with the simulations, we may certainly use real data to
validate the results of the simulations for the small ranges of image properties that they cover. For this purpose,
we will use two independent data sets.

\subsection{Time-Series Observations Of The Open Cluster NGC~7789}
\label{sec:ngc7789}

\subsubsection{Data And Reductions}
\label{sec:ngc7789_data_red}

The first data set comes from a transit survey of the open cluster NGC~7789 (\citealt{bra2005}). The data
were observed using the Wide Field Camera (WFC) on the 2.5-m Isaac Newton Telescope (INT) of the
Observatorio del Roque de los Muchachos, La Palma, Canary Islands. We selected the data from chip 2
(2048$\times$4096 pix; 0.33~arcsec~pix$^{-1}$) for the eleven nights of observations taken
between September 10th-20th in 2000. The exposure time was 300~s for each of the 691 selected images, and each
image covers the same field in NGC~7789. The images had already been calibrated (i.e. bias
subtraction and flat fielding), and the readout noise and gain determined as $\sigma_{0}~=~3.1$~ADU and
$G~=~1.44$~e$^{-}$/ADU, in the work of \citet{bra2005}.

We used the {\tt DanDIA}\footnote{{\tt DanDIA} is built from the {\tt DanIDL} library of IDL
routines available at http://www.danidl.co.uk.} pipeline (\citealt{bra2011}) to create a high-S/N stacked reference image 
and an associated star list. Firstly, stars were detected on, and matched between, the 13~best-seeing images
(observed during a $\sim$2.8~h window on the 10th September 2000). Using the matching stars, a set of linear
transformations were derived between the images, and each image was registered to the pixel grid of the
first image using cubic O-MOMS resampling (\citealt{blu2001}). The stacked reference image was then created by summing
the registered images and dividing by 13. The PSF FWHM of the reference image was measured to be
$f_{\mbox{\scriptsize ref}}~\sim$3.44~pix.

Secondly, we measured the fluxes and positions of the stars in the reference image by extracting a spatially-variable
empirical PSF with polynomial degree 3 from the image and fitting this PSF to each detected object.
Deblending of very close objects was attempted.
From this analysis, we derived a list of 7604 stars with known fluxes and positions in the reference image.

\subsubsection{Applying The Kernel Solution Methods}
\label{sec:ngc7789_apply}

We measured the PSF FWHM $f_{\mbox{\scriptsize tar}}$ of each image and retained only those 587 images such that
$-1 < f_{\mbox{\scriptsize ker}} < 5$ (see equation~\ref{eqn:fwhm_tar}) in order to match the range for $f_{\mbox{\scriptsize ker}}$
employed in our simulations. We then selected 250 random stars from the reference image star list avoiding stars within 200 pixels
of the image edges. The selected stars are approximately uniformly distributed across the image area and cover the range of brightest to faintest
detected stars. For each star, we cut a 141$\times$141 pixel region from the (parent) reference image, and a 101$\times$101 pixel region
from each of the 587 (parent) target images in the time-series, such that the star in question lies at the centre of each region. This
effectively registers each target image region with the relevant reference image region to the nearest integer
pixel without performing image resampling. This process yielded 146750 reference and target image pairs (along with associated
bad pixel masks from the data reduction).

The S/N for each target image region was calculated as follows.
The target-image pixel values were used in place of the $M_{ij}$ in equation~\ref{eqn:noise_model} and 
the $\sigma_{ij}$ were calculated using the known readout noise, gain, and flat-field
image from the data reduction. We then calculated $\mbox{SN}_{\mbox{\scriptsize tar}}$ from equation~\ref{eqn:sn_tar}
using the target-image pixel values in place of the $I_{\mbox{\scriptsize noiseless},ij}$, the previously computed $\sigma_{ij}$ values
in place of the $\sigma_{\mbox{\scriptsize in},\mbox{\scriptsize tar},ij}$, and the estimate of the local sky background level
from the data reduction in place of $S_{\mbox{\scriptsize tar}}$.

To flag any outlier pixel values in the target images, we first ran the 19x19-GIC$_{\mbox{\scriptsize P}}$ method on each
reference and target image pair without applying sigma-clipping. The model image fit and its noise model were used to clip pixel values
with $\left| \varepsilon_{ij} \right| \ge 4$ and the corresponding pixels were included in the bad pixel mask for the target image.
Then, for each kernel solution method, we computed kernel and differential background solutions for all of the reference and target
image pairs, ignoring bad pixels and without using sigma-clipping. In all cases, we used three iterations for each solution.
The optimisation of $\lambda$ for the GIC$_{\mbox{\scriptsize P}}$ and BIC$_{\mbox{\scriptsize P}}$ model selection criteria was
performed in the same way as for the simulations (see Section~\ref{sec:sim_results}).

We are unable to assess the performance of each kernel solution method with regards to model error since the true model image
is unknown for real data. Hence we do not calculate the MSE metric. However, the photometric scale factor for each solution may be
compared on a relative scale. For each of the 587 parent images, we compute the median
value of $P$ from the 19x19-GIC$_{\mbox{\scriptsize P}}$ fits for the 250 corresponding target images, and, for comparison
purposes only, we use these median $P$ values to normalise the values of $P$ estimated by each kernel solution method for each
target image. Hence the median of the normalised $P$ values for each parent image for the 19x19-GIC$_{\mbox{\scriptsize P}}$
method is always unity.

The remaining model performance metrics require a reliable noise model for their computation.
For the simulations, this noise model was precisely known since it was used to generate
the simulated data. However, for the real data, we may only estimate the noise in each pixel. For a fair comparison
between the different kernel solution methods for each reference and target image pair,
we employ a single noise model corresponding to the 19x19-GIC$_{\mbox{\scriptsize P}}$ fit
when calculating the model performance metrics. In other words, we use the pixel values from the 19x19-GIC$_{\mbox{\scriptsize P}}$ model image fit
in equation~\ref{eqn:noise_model} to calculate the $\sigma_{ij}$ values which we then use in place of the $\sigma_{\mbox{\scriptsize in},\mbox{\scriptsize tar},ij}$
in equations~\ref{eqn:fit_bias}~and~\ref{eqn:fit_var} for the purpose of calculating the MFB and MFV metrics.

To assess the photometric accuracy of each kernel solution method, we again perform
PSF fitting on the difference image at the position of the central star in the reference image region
(which is one of the 250 stars selected randomly from the reference image star list). The method is the
same as that used to perform PSF fitting in the simulations, except that we employ the empirical 
PSF at the measured star position on the reference image determined by shifting the empirical PSF
model corresponding to the nearest pixel by the appropriate subpixel shift using cubic O-MOMS resampling.
The noise model used for the PSF fitting is the noise model corresponding to the model image fit for the
kernel solution method under consideration.
The computation of the MPB and MPV measures also requires the calculation
of a reasonable and consistent normalisation factor for the difference fluxes.
Again we use an estimate of the theoretical minimum variance $\sigma_{\mbox{\scriptsize min}}^2$ in the
difference flux $\mathcal{F}_{\mbox{\scriptsize meas}}$ on the photometric scale of the reference image.
We calculate $\sigma_{\mbox{\scriptsize min}}^2$
from equation~\ref{eqn:minsig} by setting $\mathcal{P}_{\mbox{\scriptsize tar}}$ to the empirical PSF for the star
convolved with the 19x19-GIC$_{\mbox{\scriptsize P}}$ kernel solution and normalised to a sum of unity,
and by again setting the $\sigma_{\mbox{\scriptsize in},\mbox{\scriptsize tar},ij}$ to the $\sigma_{ij}$ values
obtained using the 19x19-GIC$_{\mbox{\scriptsize P}}$ model image fit.
Finally, we take the appropriate normalisation factor calculated earlier for the photometric scale factors
(which also use the 19x19-GIC$_{\mbox{\scriptsize P}}$ fits) as the best available estimate of the true
photometric scale factor $P_{\mbox{\scriptsize true}}$. The MPB and MPV measures are then calculated for a set of
$N_{\mbox{\scriptsize set}}$ flux measurements $\mathcal{F}_{\mbox{\scriptsize meas},k}$ indexed by $k$
via equations~\ref{eqn:mean_fdiff} and~\ref{eqn:var_fdiff}.

\subsubsection{Results And Discussion}
\label{sec:ngc7789_results}

\begin{figure*}
\centering
\epsfig{file=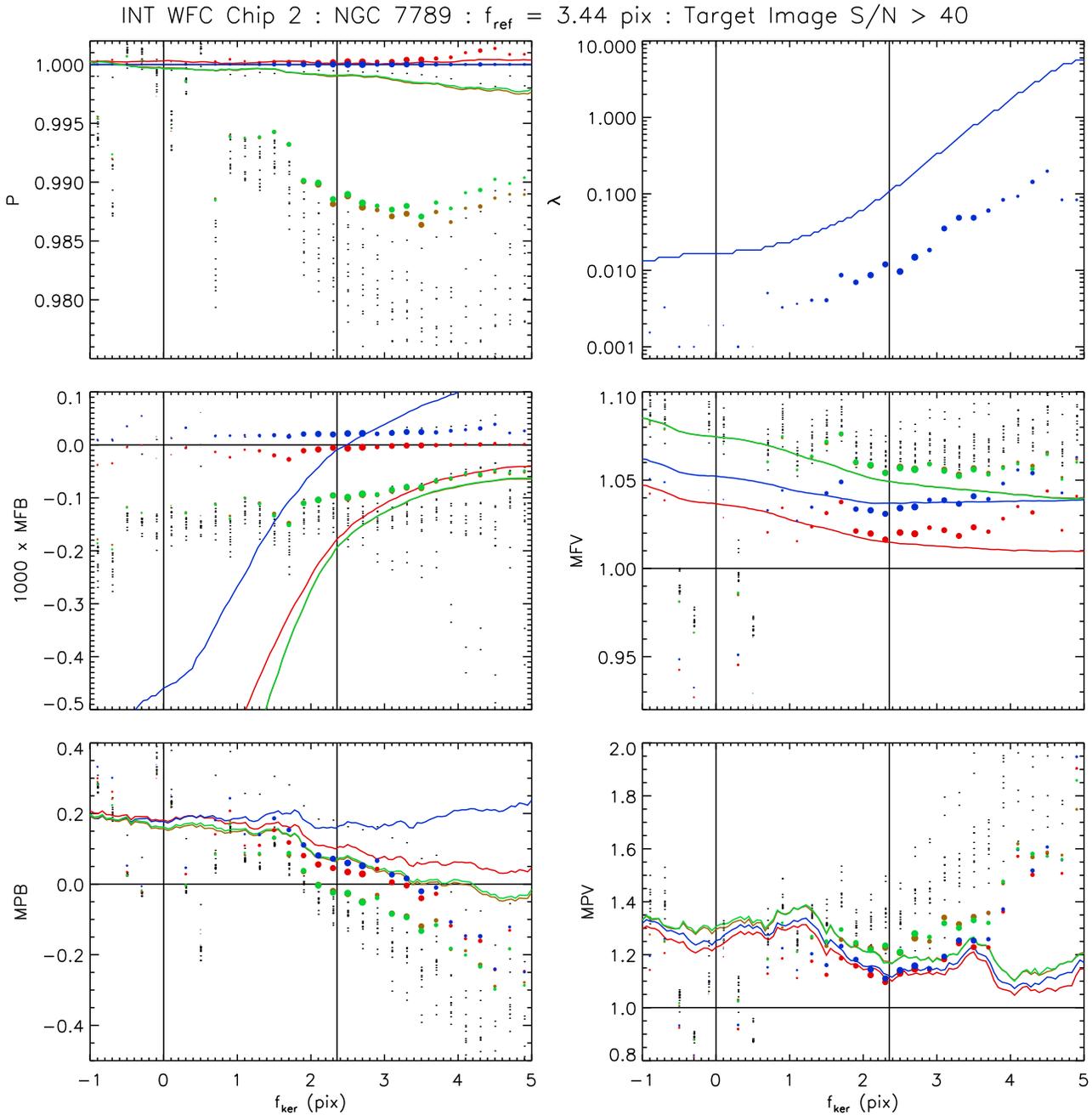,angle=0.0,width=\linewidth}
\caption{Plots of the median $P$, MFB and MFV values (equations~\ref{eqn:phot_scale}, \ref{eqn:fit_bias} and \ref{eqn:fit_var}),
         and the MPB and MPV measures (equations~\ref{eqn:mean_fdiff}~and~\ref{eqn:var_fdiff}), for each kernel solution method
         as a function of $f_{\mbox{\scriptsize ker}}$. The plots correspond to the results for the INT target images
         with SN$_{\mbox{\scriptsize tar}}~>~40$. The data are binned in $f_{\mbox{\scriptsize ker}}$ with bins of size 0.2~pix.
         Coloured solid circles correspond to the IKDA-AIC$_{\mbox{\scriptsize C}}$ (light brown), IKDA-TIC (green), 19x19-GIC$_{\mbox{\scriptsize P}}$ (blue)
         and 19x19-UNREG (red) methods such that the area of each circle is proportional to the number of data values used in the estimation of the central value.
         The largest circle corresponds to 9793 data values. Black dots represent the results for the remaining kernel solution methods.
         The values of $P$ displayed in the top-left panel have been normalised using the results from the 19x19-GIC$_{\mbox{\scriptsize P}}$ method as described in
         Section~\ref{sec:ngc7789_apply}. The median $\lambda$ values for the 19x19-GIC$_{\mbox{\scriptsize P}}$ method are plotted in the top-right panel.
         The curves correspond to the results from simulation set S10 (see text in Section~\ref{sec:ngc7789_results} for details).
         \label{fig:results_data_int}}
\end{figure*}

\begin{figure*}
\centering
\epsfig{file=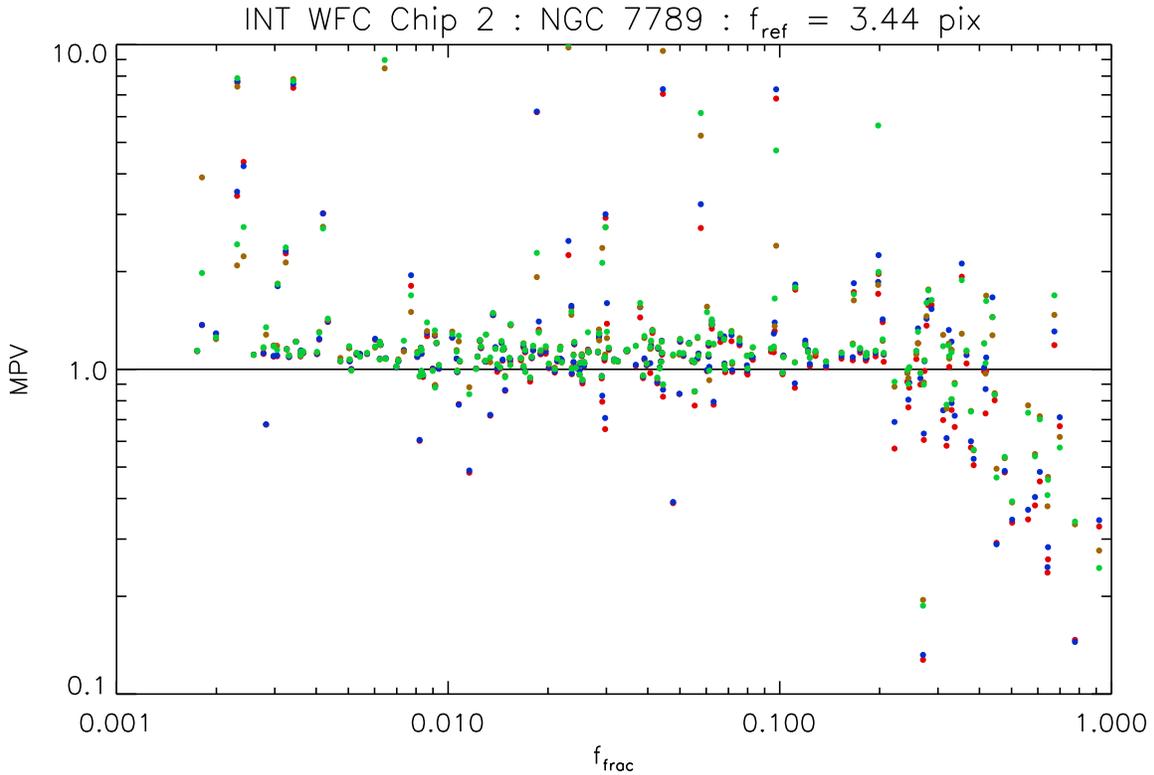,angle=0.0,width=0.9\linewidth}
\caption{Plot of the MPV measure (equation~\ref{eqn:var_fdiff}) for each star light curve as a function of the ratio of the star flux
         to the total object flux within the target image region ($f_{\mbox{\scriptsize frac}}$).
         Coloured solid circles correspond to the
         IKDA-AIC$_{\mbox{\scriptsize C}}$ (light brown), IKDA-TIC (green), 19x19-GIC$_{\mbox{\scriptsize P}}$ (blue) and 19x19-UNREG (red) methods.
         This plot corresponds to the results for the INT target images.
         \label{fig:results_data_int_lc}}
\end{figure*}

\begin{figure*}
\centering
\epsfig{file=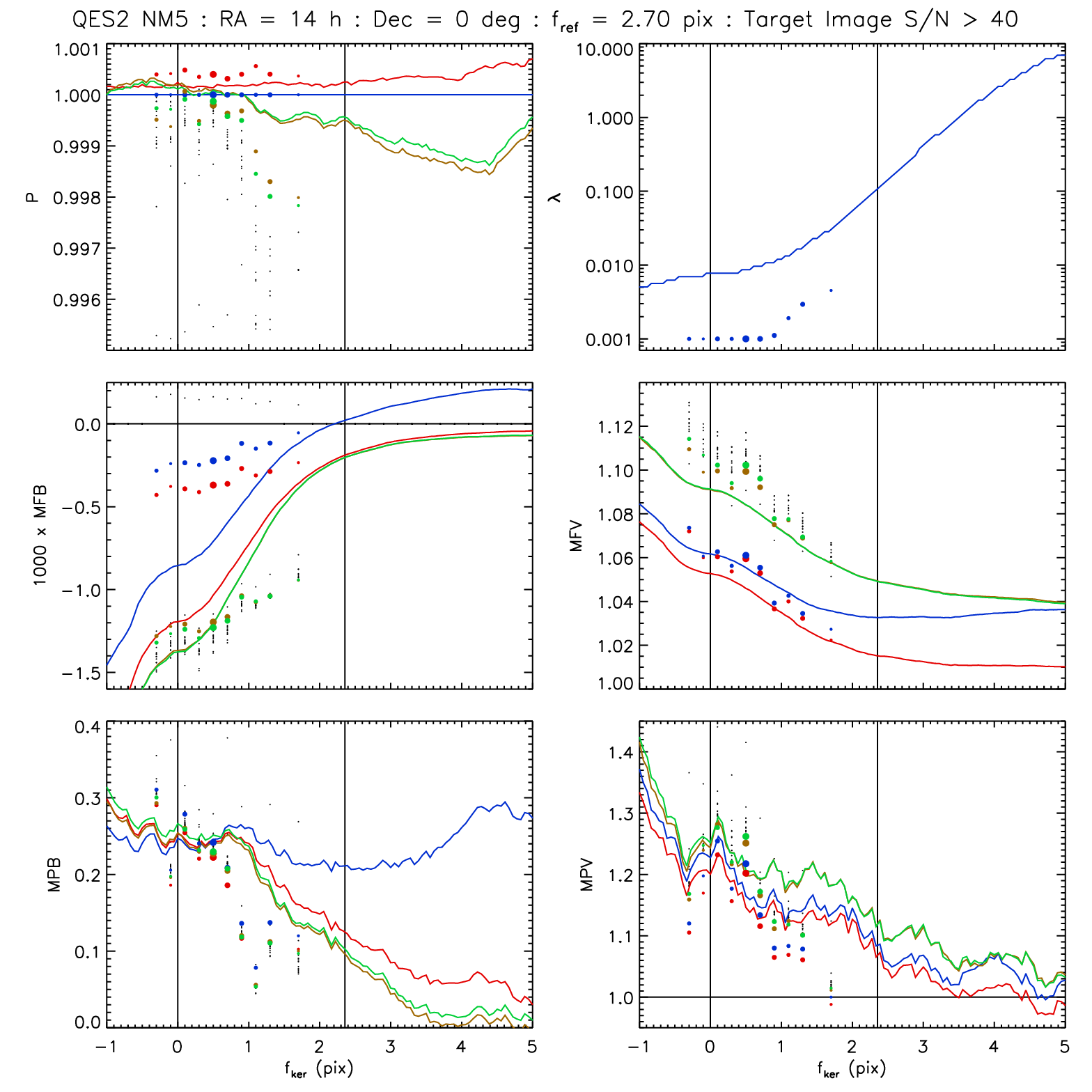,angle=0.0,width=\linewidth}
\caption{Same as Figure~\ref{fig:results_data_int} for the QES data. The largest circle corresponds to 5273 data values.
         \label{fig:results_data_qes}}
\end{figure*}

In Figure~\ref{fig:results_data_int}, for each kernel solution method we plot the median $P$, MFB and MFV values, and the MPB and MPV measures,
as a function of $f_{\mbox{\scriptsize ker}}$ (determined from $f_{\mbox{\scriptsize ref}}$ and $f_{\mbox{\scriptsize tar}}$
using equation~\ref{eqn:fwhm_tar}) for the results from the target images with SN$_{\mbox{\scriptsize tar}}~>~40$.
The data are binned in $f_{\mbox{\scriptsize ker}}$ with bins of size 0.2~pix. The results for the main methods of interest are
plotted with coloured solid circles (light brown for IKDA-AIC$_{\mbox{\scriptsize C}}$; green for IKDA-TIC; blue for 19x19-GIC$_{\mbox{\scriptsize P}}$; red for 19x19-UNREG)
such that the area of each circle is proportional to the number of data values used in the estimation of the central value. The largest
circle corresponds to 9793 data values.
The results for the remaining kernel solution methods are plotted with black dots. Note that the values of $P$ displayed in the top-left plot panel
have been normalised using the results from the 19x19-GIC$_{\mbox{\scriptsize P}}$ method as described in the previous section.
We also produce a similar style plot for the median $\lambda$ values for the 19x19-GIC$_{\mbox{\scriptsize P}}$ method in the top-right plot panel.

To facilitate a comparison with our simulations, we also plot curves in each panel for the IKDA-AIC$_{\mbox{\scriptsize C}}$ (light brown), IKDA-TIC (green),
19x19-GIC$_{\mbox{\scriptsize P}}$ (blue) and 19x19-UNREG (red) methods representing the results from simulation set S10 (high-S/N reference image) for the relevant model performance metric
as a function of $f_{\mbox{\scriptsize ker}}$. Each point on a curve is calculated as the median value (for $P$, $\lambda$, MFB and MFV) of the results from $\sim$4000 simulations,
or via equations ~\ref{eqn:mean_fdiff} and~\ref{eqn:var_fdiff} (for MPB and MPV), by using a smoothing radius of
0.33~pix in the $(f_{\mbox{\scriptsize ker}},f_{\mbox{\scriptsize ref}})$ plane ($f_{\mbox{\scriptsize ref}}~=~3.44$~pix in this case) and
considering only SN$_{\mbox{\scriptsize tar}}~>~40$. For an easier comparison to the results from the real data,
the curves for the MFV and MPV metrics have been scaled by factors of 1.045 and 1.7, respectively.

The plots in Figure~\ref{fig:results_data_int} nicely demonstrate that the results for the real data follow the same patterns as those for the simulations.
For instance, the results for the IKDA-AIC$_{\mbox{\scriptsize C}}$ and IKDA-TIC methods are virtually indistinguishable, and the order of the methods
from high to low values for each of the five model performance metrics are the same (e.g. MPB goes blue-red-brown/green for both the points and the curves).
Furthermore, the trends in the results for the real data as a function of $f_{\mbox{\scriptsize ker}}$ are similar to those seen in the simulation results
(e.g. see $P$ and MPB) even if the actual values do not match up so closely (e.g. see $P$ for the two IKDA methods). Finally, our recommended kernel solution
methods from Section~\ref{sec:sim_fur_inv} outperform the remaining methods (except 19x19-UNREG) with respect to each of the five model performance metrics.
These are comforting results given the fact that the simulated data are generated based on approximations to the properties of real data (e.g. adopting
circularly symmetric Gaussian PSF profiles in the simulations when real data exhibit PSFs that deviate from circular symmetry and Gaussian functions).

A few points about the plots in Figure~\ref{fig:results_data_int} are worth considering in more detail. We find that $P$ is equally under-estimated for these
real data by $\sim$1\% for the two IKDA methods relative to the 19x19-GIC$_{\mbox{\scriptsize P}}$ and 19x19-UNREG methods, although we do not know which method provides the best
estimate of the true value of $P$ for the real data. Also, this fact does not seem to have had a detrimental impact
on the photometric bias for the two IKDA methods. The median MFB values for the real data do not follow the shape of the MFB curves derived from the simulations, but their
absolute values are generally even smaller than those from the simulations ($<$2$\times$10$^{-4}~\sigma_{\mbox{\scriptsize min}}$). The scaling of the simulation results to match the results from
the real data for the MFV metric is most likely necessary because the pixel uncertainties are somewhat under-estimated for the real data due to unmodelled
sources of error (e.g. flat-field errors, error in the gain, errors in the empirical PSF, etc.). However, for the MPV metric, the scale factor between the simulation
results and those from the real data is much larger, and we suggest an alternative explanation for this below. Once the simulation results for the MFV and MPV metrics have been scaled,
they match very satisfactorily with the results from the real data.

Focussing on the results for the real data with regards to photometric accuracy, we find that the gradients in the MPB measure as a function of $f_{\mbox{\scriptsize ker}}$
for the IKDA-AIC$_{\mbox{\scriptsize C}}$, IKDA-TIC, 19x19-GIC$_{\mbox{\scriptsize P}}$ and 19x19-UNREG
methods are very similar and cover a range of $\sim$0.40~$\sigma_{\mbox{\scriptsize min}}$.
However, the MPB values are closest to zero for the two IKDA methods indicating a smaller photometric bias. None of the remaining methods perform as well as these four methods
in terms of MPB. In Figure~\ref{fig:results_data_int_lc}, we plot the MPV measure calculated for each star light curve (i.e. from 587 photometric measurements in each case)
as a function of the ratio of the star flux to the total object flux within the target image region. The quantity on the x-axis ($f_{\mbox{\scriptsize frac}}$) indicates by how much
the flux from the star on which the photometry is performed dominates the total object flux, and hence the kernel solution, in the target image region. For $f_{\mbox{\scriptsize frac}} \ga 0.2$, it is
clear that the four methods under consideration tend to over-fit the real data in the same manner as we found for the simulations, with the 19x19-UNREG and 19x19-GIC$_{\mbox{\scriptsize P}}$
methods doing the most over-fitting. For $f_{\mbox{\scriptsize frac}}~\la~0.2$, the MPV values scatter nicely around $\sim$1.2 for the IKDA-AIC$_{\mbox{\scriptsize C}}$ and IKDA-TIC methods.
This plot also explains the large scale factor for the MPV metric between the results for the simulations and those for the real data if one considers that $\sim$70\% of the simulations
have $f_{\mbox{\scriptsize frac}}~>~0.1$ (where over-fitting mainly occurs), compared to only $\sim$24\% of the reference and target image pairs for the real data.
Referring back to the MPV plot panel in Figure~\ref{fig:results_data_int}, if we ignore the 19x19-UNREG and 19x19-GIC$_{\mbox{\scriptsize P}}$
methods because of their excessive over-fitting, then we can see that the IKDA-AIC$_{\mbox{\scriptsize C}}$ and IKDA-TIC methods consistently attain the best MPV values.
Hence our conclusions from the simulations are fully validated by the application of the kernel solution methods to the INT image data; namely that the IKDA-AIC$_{\mbox{\scriptsize C}}$
and IKDA-TIC methods are equally the best in terms of the
photometry that they yield, and that the 19x19-GIC$_{\mbox{\scriptsize P}}$ method is a close second best.

Finally, it is interesting to note that the values of $\lambda$ selected by the 19x19-GIC$_{\mbox{\scriptsize P}}$ method for the real data are $\sim$10~times smaller than
the values selected for the simulations, while the variation of $\lambda$ as a function of $f_{\mbox{\scriptsize ker}}$ has the same form.

\subsection{Time-Series Observations Of A Galactic Field}
\label{sec:qes}

\subsubsection{Data And Reductions}
\label{sec:qes_data_red}

The second data set comes from a commissioning run for the Qatar Exoplanet Survey (QES; \citealt{als2013}).
The data were observed using camera 5 of the second QES observing station at the New Mexico Skies observatory, New Mexico, USA.
We selected a block of 27 images (4096$\times$4096~pix; 3.1~arcsec pix$^{-1}$) of the same field (R.A.$\sim$14~h; Dec.$\sim$0~deg)
taken on the night of 11th May 2015 during a $\sim$2.2~h period. Each image has an exposure time of 30~s.

We used the {\tt DanDIA} pipeline to calibrate the images (bias level subtraction and flat fielding) and to measure
the chip readout noise ($\sim$7.85~ADU) and gain ($\sim$1.65~e$^{-}$/ADU). We also used the pipeline to create
a high-S/N stacked reference image from a block of ten 30~s images of the same field taken later during the night
and to produce an associated star list with 84069~stars (see Section~\ref{sec:ngc7789_data_red} for the method).
The PSF FWHM of the reference image was measured to be $f_{\mbox{\scriptsize ref}}~\sim$2.70~pix.

We selected 1000 random stars from the reference image star list avoiding stars within 200 pixels of the image edges.
Following the same steps as those used for the INT data, we created 1000 reference and target image pairs for
each of the 27 parent target images in the time-series, yielding 27000 image pairs in total.
The procedures described in Section~\ref{sec:ngc7789_apply} for the INT data were then applied to the QES data
to compute the kernel and differential background solutions for each kernel solution method, and to calculate the
model performance metrics.

\subsubsection{Results And Discussion}
\label{sec:qes_results}

The results for the QES data are plotted in Figure~\ref{fig:results_data_qes}, which has been constructed in exactly the same
way as Figure~\ref{fig:results_data_int} for the INT data. Note that the QES data are more limited in that they
only cover the range $-0.4~<~f_{\mbox{\scriptsize ker}}~<~1.8$~pix. In this case, the curves for the MFV and MPV metrics have been
scaled by factors of 1.045 and 1.5, respectively.

The conclusions for the QES data are the same
as those for the INT data. The few exceptions for these data are that $P$ is only under-estimated by up to $\sim$0.2\% for the
IKDA-AIC$_{\mbox{\scriptsize C}}$ and IKDA-TIC methods relative to the 19x19-GIC$_{\mbox{\scriptsize P}}$ and 19x19-UNREG methods,
the absolute values of the MFB metric are larger,
and there is more scatter in the MPB and MPV measures (probably because fewer data have been analysed).
Therefore the results for the QES data add a further independent validation of the results from our simulations.

\section{Conclusions And Recommendations}
\label{sec:conclusions}

The key achievement in this work is the elaboration of a framework for automatically
constructing a kernel model (or, equivalently, a model image) for DIA where the user is only required
to specify very few external parameters to control the kernel design (e.g. the maximum extent of the kernel).
The framework requires the definition of a kernel solution method that consists of
two components; namely, a kernel design algorithm to generate a set of candidate kernel models, and a
model selection criterion to select the simplest kernel model from the candidate models that provides a sufficiently good fit to the
target image (i.e. an implementation of the Principle of Parsimony). The framework also requires
the definition of an appropriate detector noise model with associated parameters such as readout noise and gain.
It is crucial that this noise model is accurate since the model selection criteria depend heavily
on the pixel uncertainties provided by the noise model.
We developed and tested 18 automatic kernel solution methods using
comprehensive image simulations and real data, and we compared their performance to that of a fixed unregularised kernel design (i.e. the 19x19-UNREG method).

The main conclusion from the image simulations (Section~\ref{sec:simsec}) is that the IKDA-AIC$_{\mbox{\scriptsize C}}$ and IKDA-TIC
methods are equally the best kernel solution methods in terms of photometric accuracy. The 19x19-GIC$_{\mbox{\scriptsize P}}$ method
also performs very well and is a good second choice. This conclusion is also supported by considering the performance of these
methods with regards to model error and fit quality. The 19x19-UNREG method gives excellent estimates of the
photometric scale factor (Figure~\ref{fig:results_p_map}) with what appear to be some of the most uniform
and least biased model image fits (Figure~\ref{fig:results_mfb_map}) and PSF photometry (Figure~\ref{fig:results_mpb_map}).
However, this is somewhat misleading since the model performance metrics that measure the fit variance (Figure~\ref{fig:results_mfv_map})
and the photometric variance (Figure~\ref{fig:results_mpv_map})
reveal that this method is the worst offender for over-fitting the target image. Hence we concur with the findings from Be12 that
the unregularised 19x19-pixel kernel design brings too many parameters to the model image. Moreover, we have shown that
kernel regularisation (via Tikhonov regularisation) is not the only way, or even necessarily the best way, to control the over-fitting.
The IKDA-AIC$_{\mbox{\scriptsize C}}$ and IKDA-TIC methods achieve a better performance than the regularised kernel designs via
a parsimonious choice of unregularised delta basis functions. Taking this further by combining the IKDA with regularisation
(i.e. the IKDA-GIC$_{\mbox{\scriptsize P}}$ and IKDA-BIC$_{\mbox{\scriptsize P}}$ methods) was unfortunately not possible as we found that
the corresponding processing time was prohibitive. From the simulations, we also discovered that the AIC-type model selection criteria work better
than the BIC-type criteria for DIA, which we explain by considering that the true model image is not included in the set
of candidate model images generated by our kernel design algorithms.

We also analysed two independent sets of real image data covering different regions in the reference and target image PSF FWHM and S/N parameter space
(Section~\ref{sec:realsec}). The results
for the real data were found to follow the same patterns and trends as the results from the simulations. Most importantly, the IKDA-AIC$_{\mbox{\scriptsize C}}$,
IKDA-TIC and 19x19-GIC$_{\mbox{\scriptsize P}}$ methods were also found to be the best kernel solution methods in terms of photometric accuracy for the
real data.

In practical terms, the AIC$_{\mbox{\scriptsize C}}$ model selection criterion is much easier to implement (trivial in fact), and various orders of magnitude faster to calculate, than
the TIC or GIC$_{\mbox{\scriptsize P}}$ criteria. Since the IKDA-AIC$_{\mbox{\scriptsize C}}$ and IKDA-TIC methods yield virtually the same results,
it is clear that the IKDA-AIC$_{\mbox{\scriptsize C}}$ method is the most desirable of the two for implementation. However, the IKDA can be somewhat slower to run than the optimisation
of GIC$_{\mbox{\scriptsize P}}$ over $\lambda$ for the fixed 19x19-pixel kernel design. This is especially true when the IKDA attempts to grow a kernel model with many
delta basis functions.
Hence, if processing time is a concern, then the 19x19-GIC$_{\mbox{\scriptsize P}}$ method may be more desirable than the IKDA-AIC$_{\mbox{\scriptsize C}}$ method, even if the
results are slightly less optimal. One caveat of the 19x19-GIC$_{\mbox{\scriptsize P}}$ method is that a 19x19-pixel grid may not be large enough for an adequate kernel solution
for some DIA problems, and its size may therefore need to be increased as appropriate.

Our work constitutes the first fully systematic and comprehensive attempt to characterise the performance of DIA as a function
of the reference and target image properties using simulated images and with validation on real data.
We have learned some important facts from these experiments which may be translated into the following recommendations:
\begin{itemize}
\item It is vastly advantageous to use a reference image with higher S/N than that of the target image regardless of the image properties or
      the kernel solution method. We therefore recommend that the reference image is constructed either by exposing for longer than the
      target image(s), or by stacking a set of registered images to achieve a longer effective exposure time, while at the same time maintaining the
      requirement that the reference image has a PSF FWHM that is among the smallest PSF FWHMs of the target image(s).
      We realise that this particular advice is already followed for most DIA reductions. However, it is comforting to
      see that our comprehensive simulation work strongly supports this approach.
\item In general, the photometric scale factor between the reference and target image is under-estimated. However, this effect is
      smaller for higher S/N target images. Given the importance of obtaining an accurate estimate of the photometric
      scale factor for accurate photometry, we recommend employing all of the pixels in the target image to solve for the kernel, since this
      maximises the S/N of the data that are being fit.
\item In most of our simulations, we found that all of the kernel solution methods are over-fitting
      the brightest star(s) and under-fitting the faintest star(s) since the brightest star(s) dominate the kernel solution.
      The effect on the photometry of the brightest star in each target image is to yield variances that are impossibly, and therefore
      misleadingly, small. We used target images of size 101$\times$101 pixels both for the simulations and the tests on the real data. We found that by
      increasing the size of the target images, this effect on the photometry
      is mitigated since more bright stars, and therefore more pixels from bright stars, are used to derive the kernel solution.
      Hence we again recommend employing all of the pixels in the target image to solve for the kernel. 
\end{itemize}

These last two recommendations have important implications for some popular DIA software implementations that generate
a spatially varying kernel solution for an image by interpolating a set of spatially-invariant kernel solutions determined
independently from small image regions called ``stamps''
(e.g. ISIS - \citet{ala2000}, HOTPANTS\footnote{http://www.astro.washington.edu/users/becker/v2.0/hotpants.html}). 
The stamps are chosen to be approximately uniformly distributed across the image area, centred on isolated bright stars, and only
slightly larger than the objects they encompass (e.g. in Be12, the stamp size is $\sim$57$\times$57~pix, with only 
$\sim$39$\times$39~pix used for the kernel solution).
We believe that it will be highly beneficial to modify the stamp selection strategy in these algorithms to match our recommendations.
Specifically, image stamps should be defined to be as large
as possible without seriously violating the assumption of a spatially-invariant kernel model within the stamp, and they should be selected
such that each stamp contains a minimum of at least a few bright objects (not necessarily stars, and there is no reason to avoid
blended objects). Our tests on real data suggest the following useful rule-of-thumb: the ratio of the flux of the brightest star to the total flux
of all objects in a stamp should be less than $\sim$0.1.
It is not surprising that Be12 found that the unregularised 19x19-pixel kernel with 361 parameters was over-fitting a
target image stamp with just $\sim$1520 pixels, of which only a small proportion contain signal from the single
object\footnote{Be12 found that the normalised residuals in the difference image for their example have a standard deviation
of $\sim$0.79 for the unregularised 19x19-pixel kernel. However, it should be noted that this is perfectly consistent with a reduced chi-squared of unity.}
(see their section~4). We are confident that if Be12 were to repeat their experiment for a set of image stamps selected following our recommendations,
then the unregularised 19x19-pixel kernel would have been found to be over-fitting the stamps to a much lesser extent
than reported.

Be12 recommend values of $\lambda$ between 0.1 and 1 for regularised 19x19-pixel kernels while cautioning that the ``optimal
value of $\lambda$ will be a function of the S/N in the template and science images, ... and of the respective seeings
in the input images, ...''. We have characterised precisely how the optimal value of $\lambda$, as selected by the
GIC$_{\mbox{\scriptsize P}}$ and BIC$_{\mbox{\scriptsize P}}$ criteria, varies as a function of the reference and target image properties 
for four kernel solution methods (Section~\ref{sec:opt_lam}).
We find that the optimal value of $\lambda$ is highly correlated with the PSF FWHM and S/N in each of the reference
and target images, and that it spans values from $\lambda = 0$ (i.e. no regularisation) up to maximum values of the order of
$\lambda = 10$ and $\lambda = 100$ for GIC$_{\mbox{\scriptsize P}}$ and BIC$_{\mbox{\scriptsize P}}$, respectively. We conclude
that the optimal regularisation of the kernel model for any particular kernel solution method is highly data set dependent and
that it should be determined {\it independently} for each target image.

Looking to the future, we can see much potential for the development and testing of new kernel design algorithms
within our framework that may perform better than those presented in this work. In fact, we believe that there is still
plenty of room for improvement in the kernel solution methods, especially with regards to achieving the best photometry.
For example, the poor performance of the CKDA methods, including those employing kernel regularisation, was a disappointment.
It would be interesting to investigate whether adopting a radial dependence for the strength of the kernel regularisation
can improve the CKDA performance, since we expect the variations in the true kernel model to be smallest in the outer
parts of the kernel. Most intriguingly, the ``spidery'' form of the kernels that are generated by the IKDA methods,
combined with the fact that they perform exceptionally well, implies that sparsity may be the key to the optimal use of
delta basis functions in DIA. Finally, it remains to extend the methods presented here to the case of a spatially varying
convolution kernel.

\section*{Acknowledgements}

I dedicate this work to the twin bright lights in my life - Phoebe and Chloe Bramich Mu\~niz.

We thank the referee for investing the time to carefully study this paper and for providing useful feedback.
This publication was made possible by NPRP grant \# X-019-1-006 from the Qatar National
Research Fund (a member of Qatar Foundation). The statements made herein are solely
the responsibility of the authors. The HPC resources and services used in this work
were provided by the IT Research Computing group in Texas A\&M University at Qatar.
IT Research Computing is funded by the Qatar Foundation for 
Education, Science and Community Development (http://www.qf.org.qa).

\section*{Appendix A}
\label{app:append_A}

\begin{figure}
\centering
\begin{tabular}{cc}
\subfigure[]{\epsfig{file=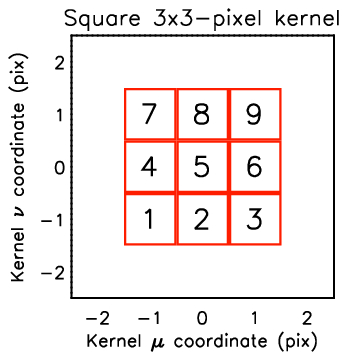,angle=0.0,width=0.48\linewidth} \label{fig:ker_ex1}} &
\subfigure[]{\epsfig{file=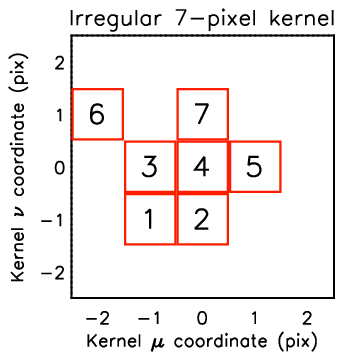,angle=0.0,width=0.48\linewidth} \label{fig:ker_ex2}} \\
\end{tabular}
\caption{Example configurations of sets of kernel DBFs. Individual DBFs are represented by red squares
         positioned at the kernel pixel coordinates where they take the value unity. Each red square displays the value of $q$ for the
         corresponding DBF.
         \label{fig:ker_ex}}
\end{figure}

We provide two examples of the Laplacian matrix $\mathbf{L}$.
For the square 3$\times$3-pixel kernel design shown in Figure~\ref{fig:ker_ex1}, where the value of $q$ is displayed
inside each kernel pixel, we may use equation~(\ref{eqn:L_elements}) to obtain the 10$\times$10 matrix:
\begin{equation}
\mathbf{L} = \left(
\begin{array}{cccccccccc}
 2  &    -1   &    0   &   -1   &    0   &    0   &    0   &    0   &    0  & 0 \\
-1  &     3   &   -1   &    0   &   -1   &    0   &    0   &    0   &    0  & 0 \\
 0  &    -1   &    2   &    0   &    0   &   -1   &    0   &    0   &    0  & 0 \\
-1  &     0   &    0   &    3   &   -1   &    0   &   -1   &    0   &    0  & 0 \\
 0  &    -1   &    0   &   -1   &    4   &   -1   &    0   &   -1   &    0  & 0 \\
 0  &     0   &   -1   &    0   &   -1   &    3   &    0   &    0   &   -1  & 0 \\
 0  &     0   &    0   &   -1   &    0   &    0   &    2   &   -1   &    0  & 0 \\
 0  &     0   &    0   &    0   &   -1   &    0   &   -1   &    3   &   -1  & 0 \\
 0  &     0   &    0   &    0   &    0   &   -1   &    0   &   -1   &    2  & 0 \\ 
 0  &     0   &    0   &    0   &    0   &    0   &    0   &    0   &    0  & 0 \\
\end{array} \right)
\notag
\end{equation}
This matrix is of rank equal to 8, as expected since all of the DBFs are connected to each other. The elements
of the last row and column correspond to the differential background parameter and are consequently all zero.

Kernels may of course be of any shape. For the 7-pixel kernel design shown in Figure~\ref{fig:ker_ex2}, equation~(\ref{eqn:L_elements}) yields:
\begin{equation}
\mathbf{L} = \left(
\begin{array}{cccccccc}
 2 &     -1  &    -1  &     0  &     0  &     0  &     0  & 0 \\
-1 &      2  &     0  &    -1  &     0  &     0  &     0  & 0 \\
-1 &      0  &     2  &    -1  &     0  &     0  &     0  & 0\\
 0 &     -1  &    -1  &     4  &    -1  &     0  &    -1  & 0 \\
 0 &      0  &     0  &    -1  &     1  &     0  &     0  & 0 \\
 0 &      0  &     0  &     0  &     0  &     0  &     0  & 0 \\
 0 &      0  &     0  &    -1  &     0  &     0  &     1  & 0 \\
 0 &      0  &     0  &     0  &     0  &     0  &     0  & 0 \\
\end{array} \right)
\notag
\end{equation}
This matrix is of rank equal to 5, which is explained by the fact that there are two disconnected sets of connected DBFs in the kernel model.

\section*{Appendix B}
\label{app:append_B}

%Sharpening and blurring kernels correspond to the cases $f_{\mbox{\scriptsize ker}} \le 0$~pix   
%and $f_{\mbox{\scriptsize ker}} \ge 0$~pix, respectively. Under-sampled and over-sampled reference/target images correspond to the cases $f_{\mbox{\scriptsize ref}}, f_{\mbox{\scriptsize tar}} \le 2.35$~pix
%and $f_{\mbox{\scriptsize ref}}, f_{\mbox{\scriptsize tar}} \ge 2.35$~pix, respectively.

\begin{figure*}
\centering
\epsfig{file=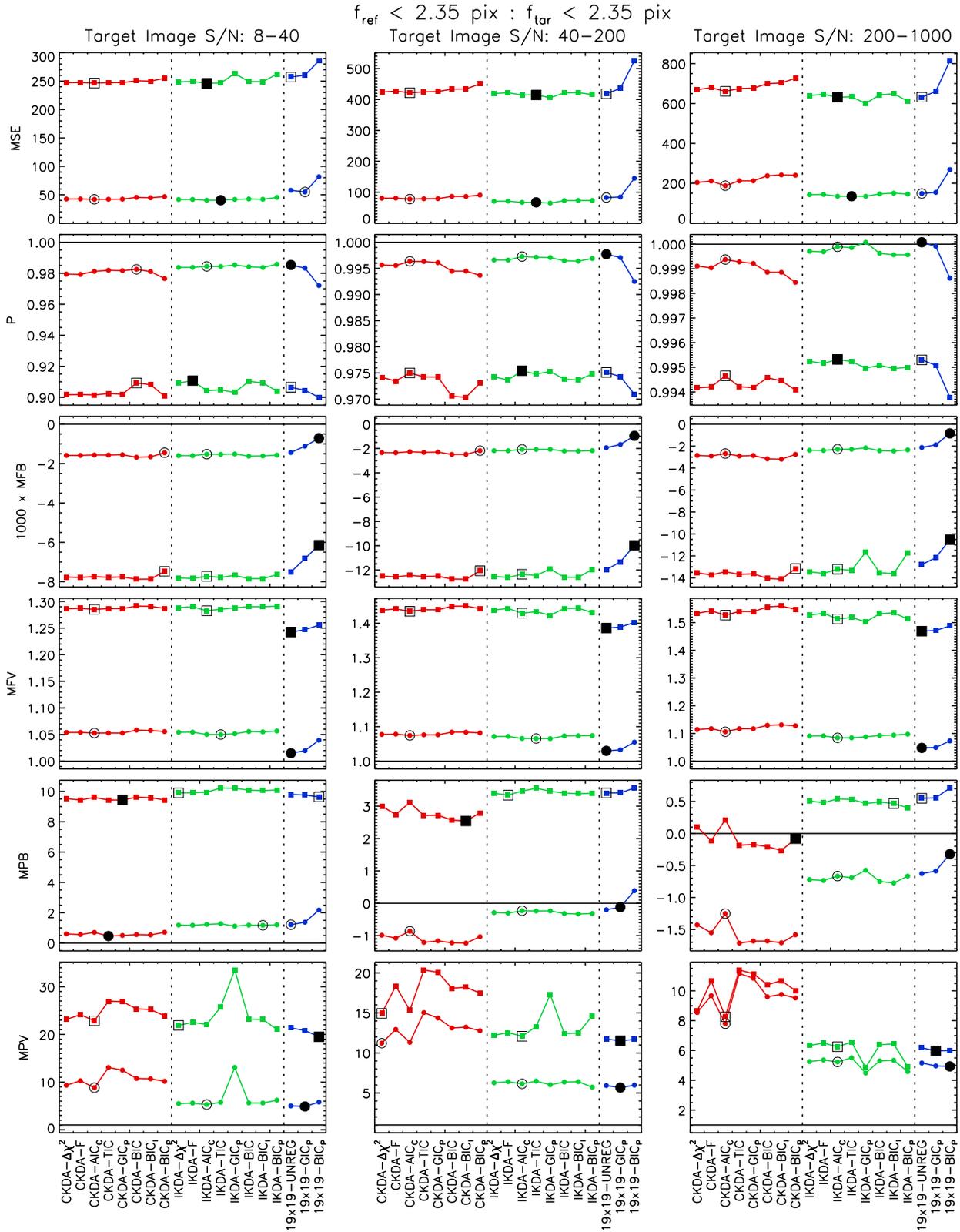,angle=0.0,width=\linewidth}
\caption{Plots of the median MSE, $P$, MFB, and MFV values (equations~\ref{eqn:mod_err}, \ref{eqn:phot_scale}, \ref{eqn:fit_bias} and \ref{eqn:fit_var}),
         and the MPB and MPV measures (equations~\ref{eqn:mean_fdiff}~and~\ref{eqn:var_fdiff}), for each kernel solution method for
         $f_{\mbox{\scriptsize ref}} \le 2.35$~pix and $f_{\mbox{\scriptsize tar}} \le 2.35$~pix.
         The results in each plot have been calculated from $\sim$19000 simulations for each of the simulation sets S1 and S10.
         The layout, symbols, and colours used are the same as in Figure~\ref{fig:results_reg5}.
         The IKDA-GIC$_{\mbox{\scriptsize P}}$ and IKDA-BIC$_{\mbox{\scriptsize P}}$ methods are excluded when determining the best values of the relevant
         model performance metric.
         \label{fig:results_reg1}}
\end{figure*}

\begin{figure*}
\centering
\epsfig{file=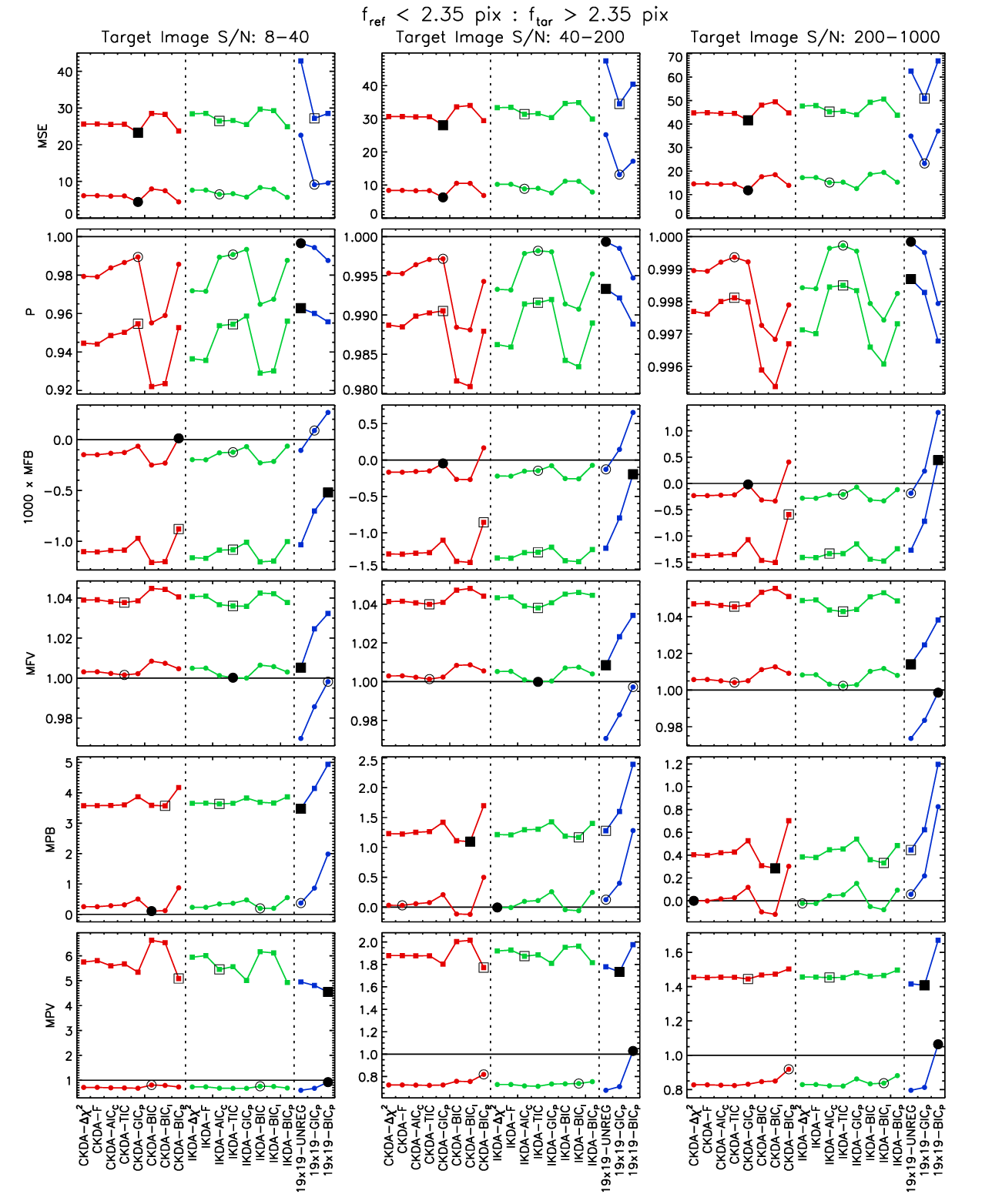,angle=0.0,width=\linewidth}
\caption{Plots of the median MSE, $P$, MFB, and MFV values (equations~\ref{eqn:mod_err}, \ref{eqn:phot_scale}, \ref{eqn:fit_bias} and \ref{eqn:fit_var}),
         and the MPB and MPV measures (equations~\ref{eqn:mean_fdiff}~and~\ref{eqn:var_fdiff}), for each kernel solution method for
         $f_{\mbox{\scriptsize ref}} \le 2.35$~pix and $f_{\mbox{\scriptsize tar}} \ge 2.35$~pix.
         The results in each plot have been calculated from $\sim$29000 simulations for each of the simulation sets S1 and S10.
         The layout, symbols, and colours used are the same as in Figure~\ref{fig:results_reg5}.
         The IKDA-GIC$_{\mbox{\scriptsize P}}$ and IKDA-BIC$_{\mbox{\scriptsize P}}$ methods are excluded when determining the best values of the relevant
         model performance metric.
         \label{fig:results_reg2}}
\end{figure*}

\begin{figure*}
\centering
\epsfig{file=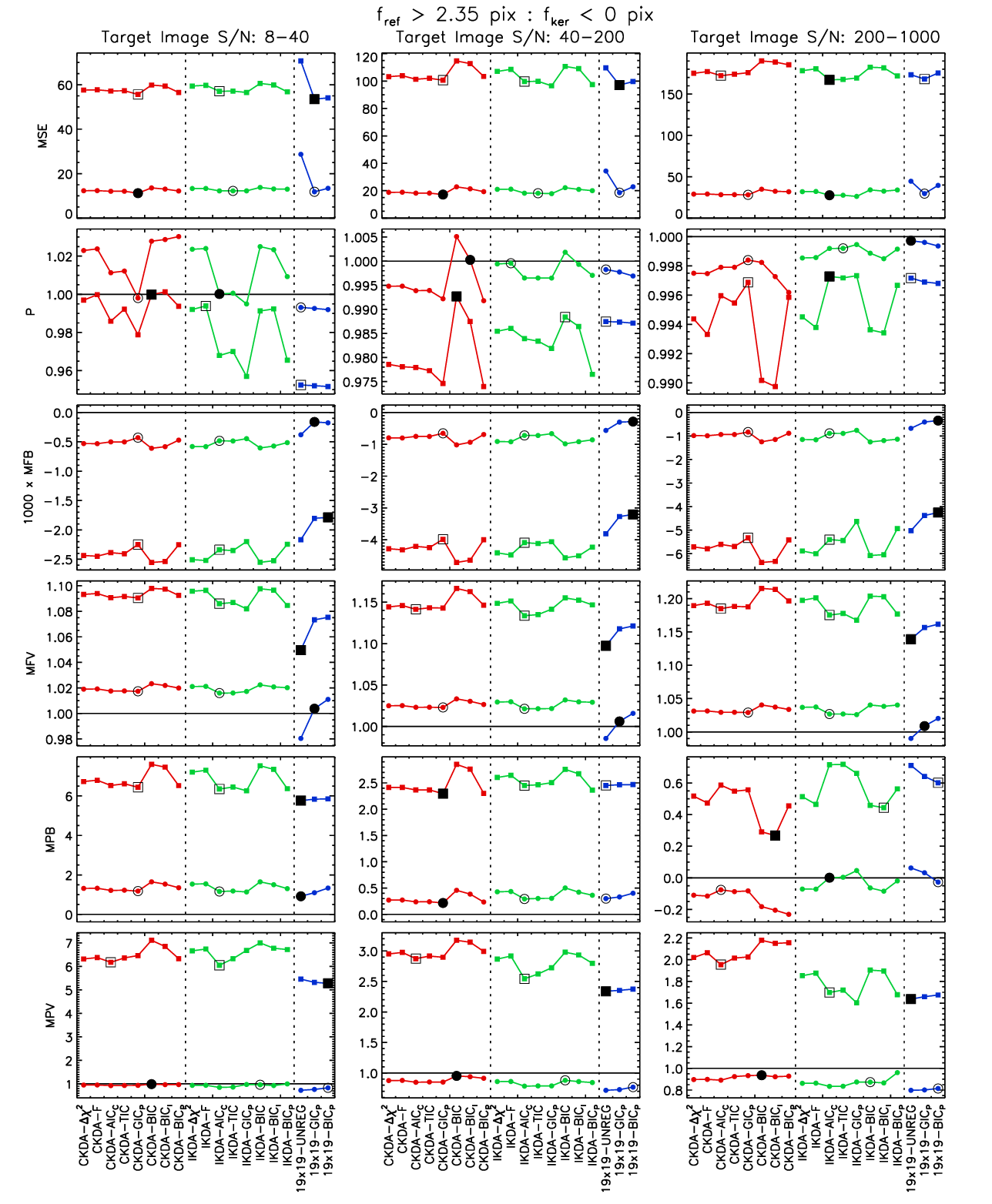,angle=0.0,width=\linewidth}
\caption{Plots of the median MSE, $P$, MFB, and MFV values (equations~\ref{eqn:mod_err}, \ref{eqn:phot_scale}, \ref{eqn:fit_bias} and \ref{eqn:fit_var}),
         and the MPB and MPV measures (equations~\ref{eqn:mean_fdiff}~and~\ref{eqn:var_fdiff}), for each kernel solution method for
         $f_{\mbox{\scriptsize ref}} \ge 2.35$~pix and $f_{\mbox{\scriptsize ker}} \le 0$~pix.
         The results in each plot have been calculated from $\sim$22000 simulations for each of the simulation sets S1 and S10.
         The layout, symbols, and colours used are the same as in Figure~\ref{fig:results_reg5}.
         The IKDA-GIC$_{\mbox{\scriptsize P}}$ and IKDA-BIC$_{\mbox{\scriptsize P}}$ methods are excluded when determining the best values of the relevant
         model performance metric.
         \label{fig:results_reg3}}
\end{figure*}

\begin{figure*}
\centering
\epsfig{file=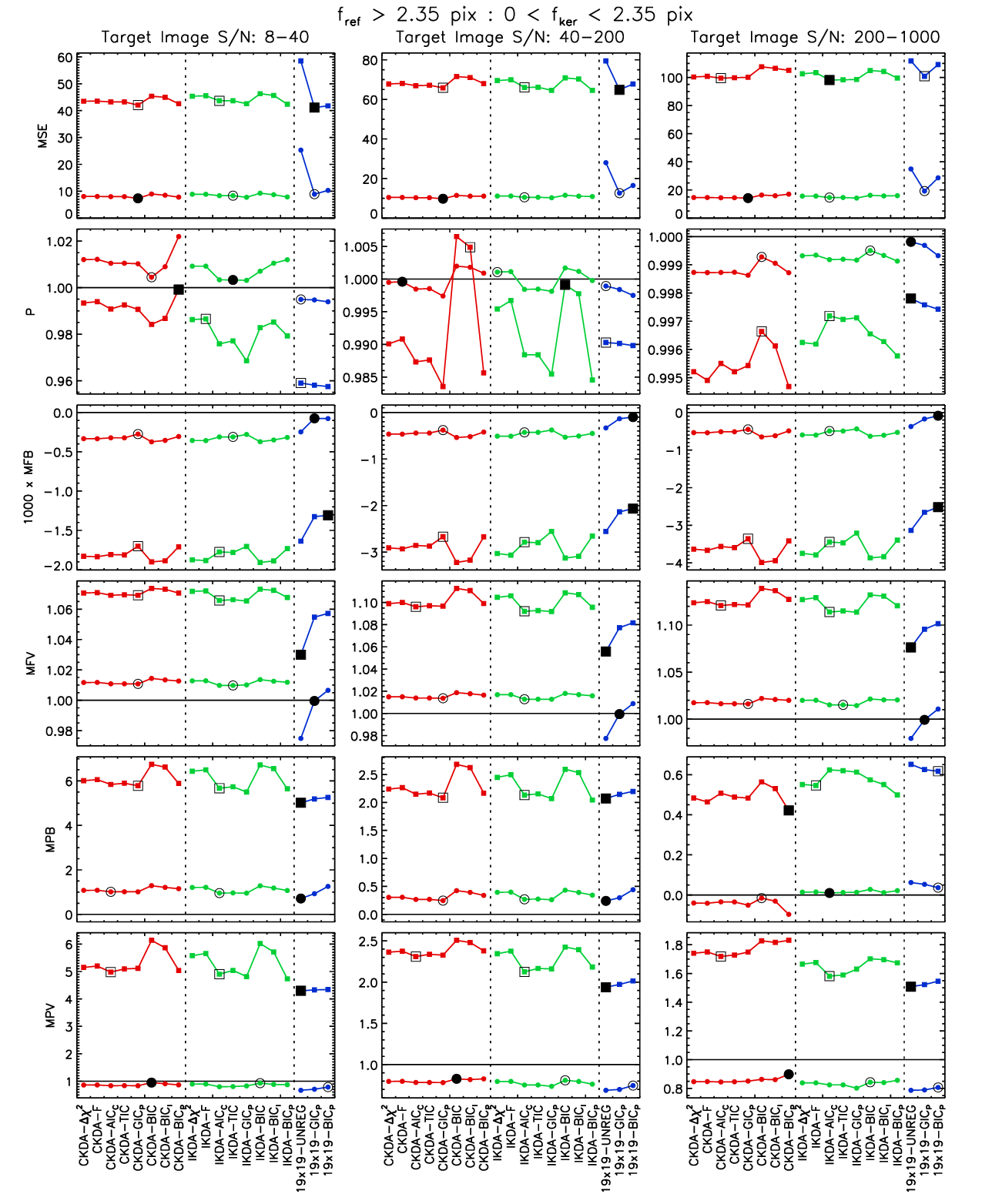,angle=0.0,width=\linewidth}
\caption{Plots of the median MSE, $P$, MFB, and MFV values (equations~\ref{eqn:mod_err}, \ref{eqn:phot_scale}, \ref{eqn:fit_bias} and \ref{eqn:fit_var}),
         and the MPB and MPV measures (equations~\ref{eqn:mean_fdiff}~and~\ref{eqn:var_fdiff}), for each kernel solution method for
         $f_{\mbox{\scriptsize ref}} \ge 2.35$~pix and $0 \le f_{\mbox{\scriptsize ker}} \le 2.35$~pix.
         The results in each plot have been calculated from $\sim$53000 simulations for each of the simulation sets S1 and S10.
         The layout, symbols, and colours used are the same as in Figure~\ref{fig:results_reg5}.
         The IKDA-GIC$_{\mbox{\scriptsize P}}$ and IKDA-BIC$_{\mbox{\scriptsize P}}$ methods are excluded when determining the best values of the relevant
         model performance metric.
         \label{fig:results_reg4}}
\end{figure*}

In this appendix, we present Figures~\ref{fig:results_reg1}~-~\ref{fig:results_reg4} where we plot the median MSE, $P$, MFB, and MFV values,
and the MPB and MPV measures, for each kernel solution method for various subsets of our simulations chosen based on image sampling.
These plots are referred to briefly in Sections~\ref{sec:sim_results}~and~\ref{sec:sim_disc}.

\section*{Appendix C}
\label{app:append_C}

\begin{figure*}
\centering
\epsfig{file=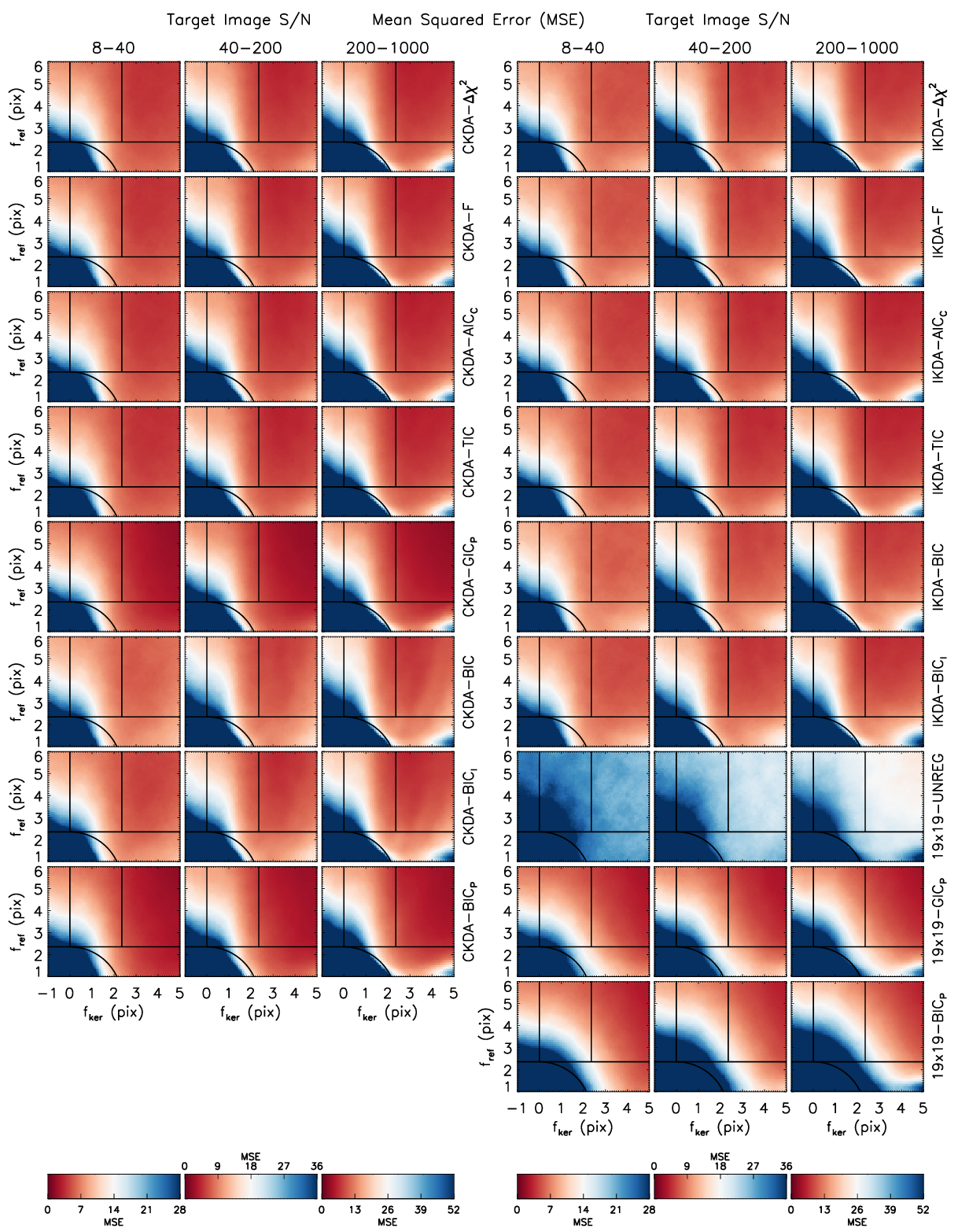,angle=0.0,width=0.99\linewidth}
\caption{Plots of surfaces representing the median MSE values (equation~\ref{eqn:mod_err}) for simulation set S10 as a function of the reference image and kernel FWHM.
         The format of the figure is the same as in Figure~\ref{fig:results_mpb_map}.
         \label{fig:results_mse_map}}
\end{figure*}

\begin{figure*}
\centering
\epsfig{file=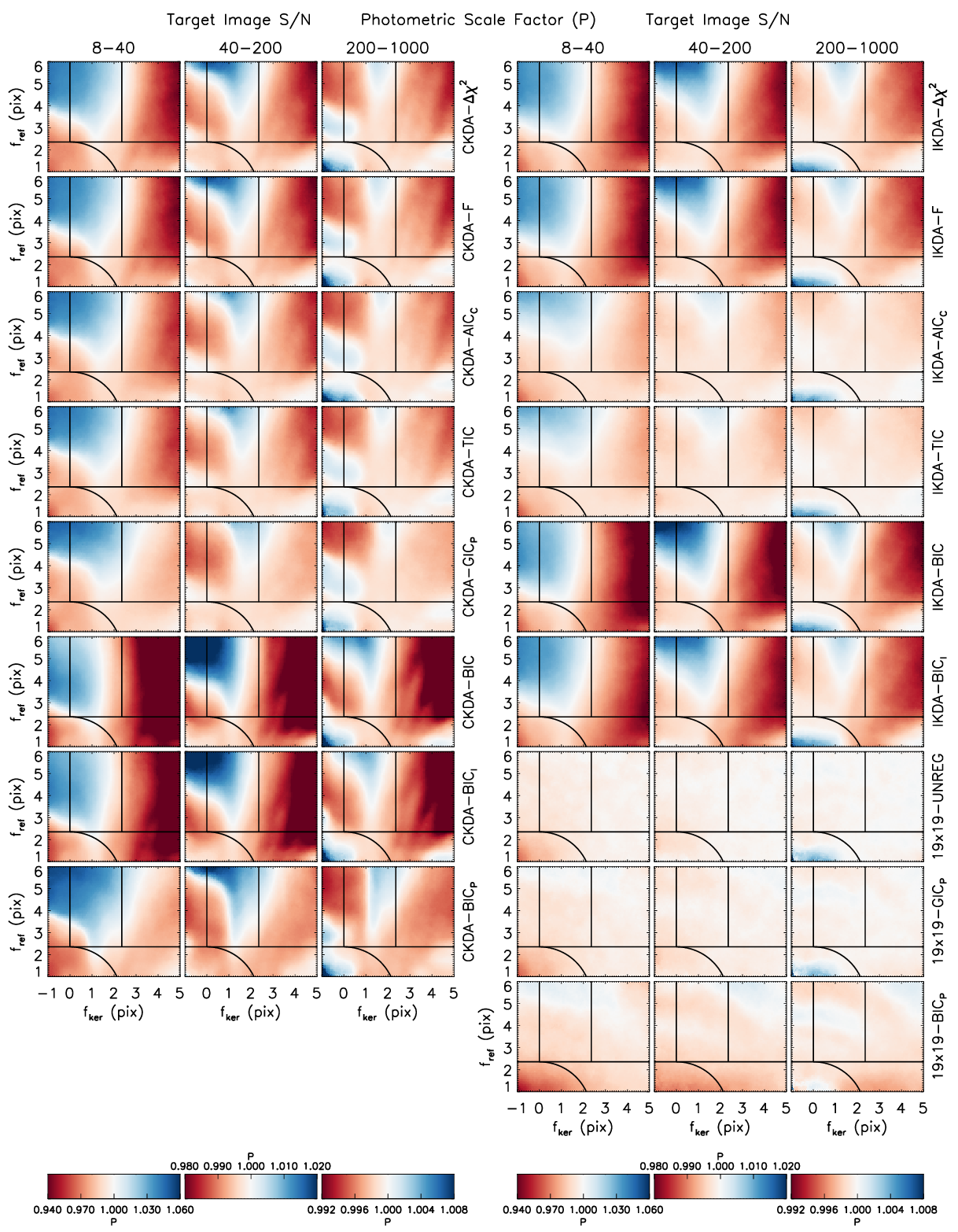,angle=0.0,width=0.99\linewidth}
\caption{Plots of surfaces representing the median $P$ values (equation~\ref{eqn:phot_scale}) for simulation set S10 as a function of the reference image and kernel FWHM.
         The format of the figure is the same as in Figure~\ref{fig:results_mpb_map}.
         \label{fig:results_p_map}}
\end{figure*}

\begin{figure*}
\centering
\epsfig{file=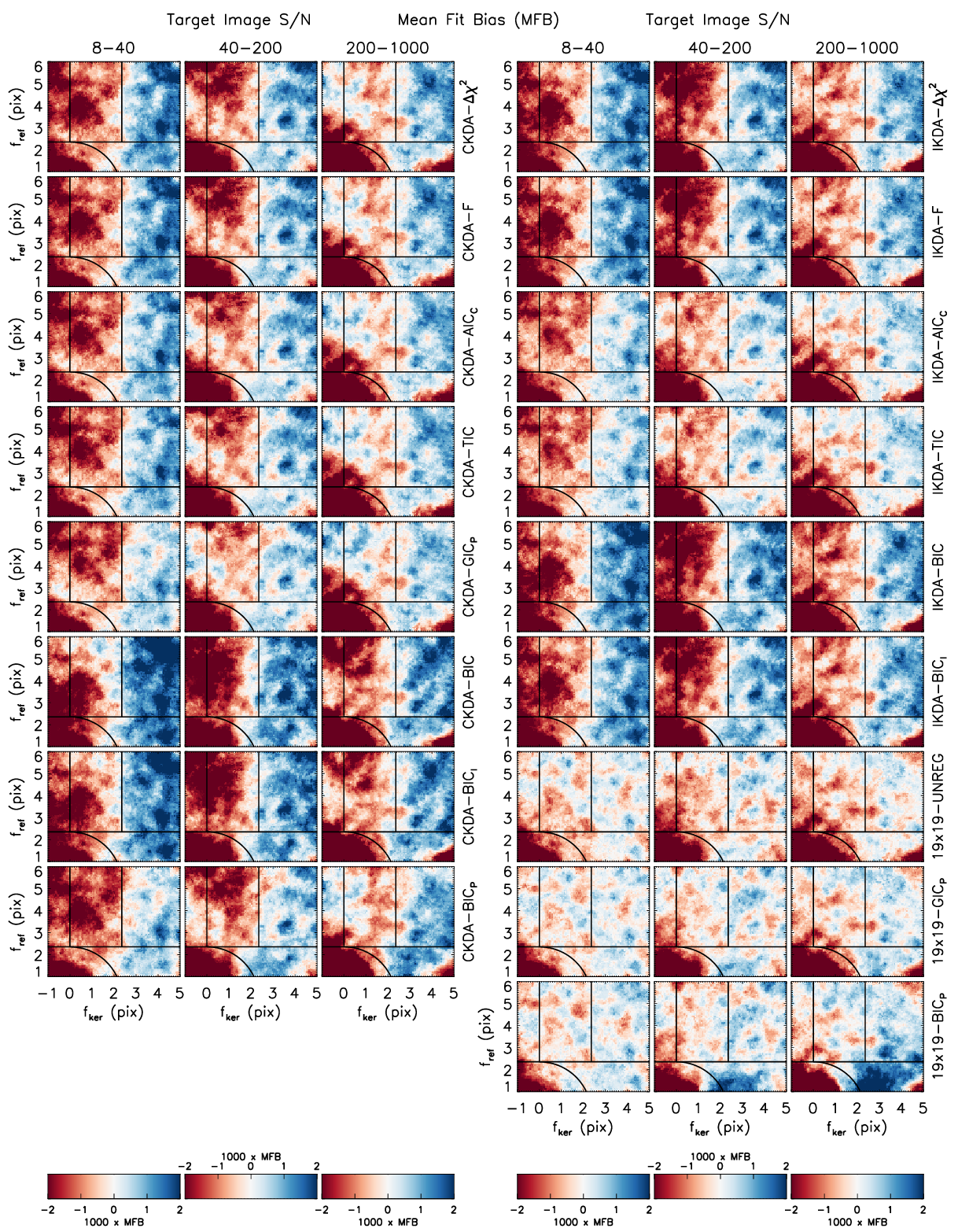,angle=0.0,width=0.99\linewidth}
\caption{Plots of surfaces representing the median MFB values (equation~\ref{eqn:fit_bias}) for simulation set S10 as a function of the reference image and kernel FWHM.
         The format of the figure is the same as in Figure~\ref{fig:results_mpb_map}.
         \label{fig:results_mfb_map}}
\end{figure*}

\begin{figure*}
\centering
\epsfig{file=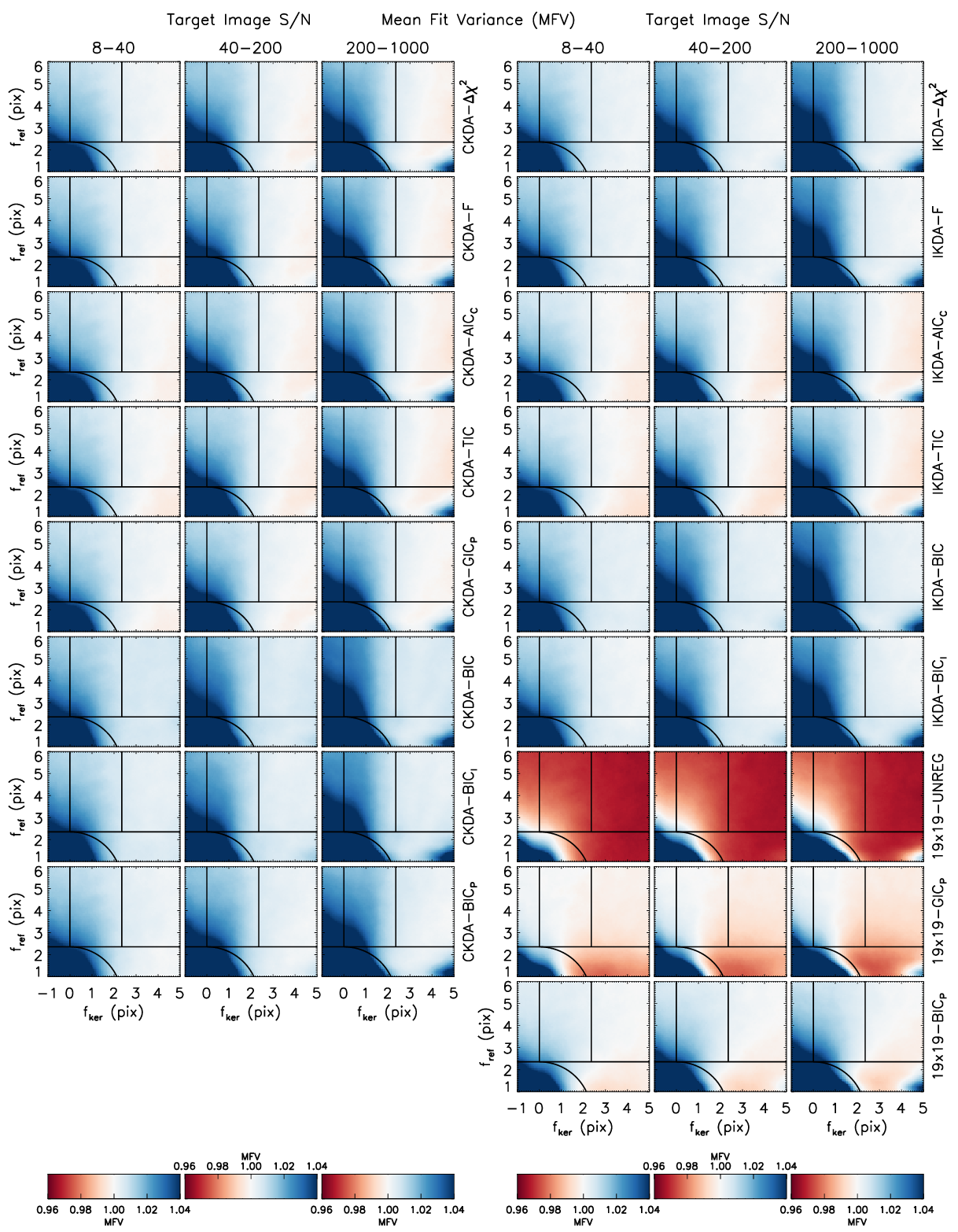,angle=0.0,width=0.99\linewidth}
\caption{Plots of surfaces representing the median MFV values (equation~\ref{eqn:fit_var}) for simulation set S10 as a function of the reference image and kernel FWHM.
         The format of the figure is the same as in Figure~\ref{fig:results_mpb_map}.
         \label{fig:results_mfv_map}}
\end{figure*}

For completeness of this paper, in Figures~\ref{fig:results_mse_map}~-~\ref{fig:results_mfv_map} we plot surfaces representing
the median MSE, $P$, MFB, and MFV values for simulation set S10 as a function of the reference image and kernel FWHM. These plots are
referred to briefly in Section~\ref{sec:sim_fur_inv}.

\label{lastpage}

\end{document}